\documentclass[prb,superscriptaddress, preprintnumbers,twocolumn,showpacs,amsmath,amssymb]{revtex4}
\usepackage{amsmath}
\usepackage{amssymb}
\usepackage{amsthm}
\usepackage{amsfonts}
\usepackage{enumerate}
\usepackage{latexsym}
\usepackage{graphicx}
\usepackage{subfigure}

\def\Im {\mbox{Im}}
\def\be{\begin{equation}}       \def\ee{\end{equation}}
\def\bea{\begin{eqnarray}}      \def\eea{\end{eqnarray}}

\begin{document}

\title{Quasiparticle Scattering Interference in Superconducting Iron-pnictides}
\author{Yan-Yang Zhang}
\affiliation{Department of Physics, Purdue University, West
Lafayette, Indiana 47907, USA}
\author{Chen Fang}
\affiliation{Department of Physics, Purdue University, West
Lafayette, Indiana 47907, USA}
\author{Xiaoting Zhou}
\affiliation{Department of Physics, Purdue University, West
Lafayette, Indiana 47907, USA}
\author{Kangjun Seo}
\affiliation{Department of Physics, Purdue University, West
Lafayette, Indiana 47907, USA}
\author{Wei-Feng Tsai}
\affiliation{Department of Physics, Purdue University, West
Lafayette, Indiana 47907, USA}
\author{B. Andrei Bernevig}
\affiliation{Princeton Center for Theoretical Science, Princeton
University, NJ 08544}
\author{Jiangping Hu}
\affiliation{Department of Physics, Purdue University, West
Lafayette, Indiana 47907, USA}
\date{\today}

\begin{abstract}
 Using both two orbital and five orbital models, we investigate the
quasiparticle interference (QPI) patterns in the superconducting
(SC) state of  iron-based superconductors. We compare the results
for nonmagnetic and magnetic impurities in sign-changed s-wave
$\cos(k_x)\cdot\cos(k_y)$ and sign-unchanged
$|\cos(k_x)\cdot\cos(k_y)|$ SC states. While the patterns strongly
depend on the chosen band structure details, the sensitivity of peaks
around $(\pm\pi,0)$ and $(0,\pm\pi)$ wavevectors on magnetic or
non-magnetic impurities, and  on sign-changed or sign-unchanged SC orders
is a common feature. Our results strongly suggest that the QPI
can provide a direct evidence of the
pairing symmetry in the SC states.
\end{abstract}


\maketitle
\section{introduction}
Recently, the discovery of high temperature superconductivity in
oxypnictide compounds \cite{kamihara2008,takahashi2008,
Chen2008,Chenxh2008,wen2008} stirred great interests in the
condensed matter community. One important problem is to elucidate the
pairing symmetry of the order parameter of the superconducting state. Theoretically
many possible gap pairing symmetries have been proposed. Due to the proximity of the superconducting state to a collinear antiferromagnetic state,  a magnetism-based mechanism has emerged in both the weak and strong coupling
approaches. This mechanism suggests that an extended s-wave pairing symmetry is
 favored\cite{seo2008,Mazin2008a,Wang2009}.

 The weak-coupling
approach favors an s-wave (so called $s_{\pm}$) state\cite{Mazin2008a} in
which the relative sign of order parameters changes between the hole and
electron pockets. However, the
weak-coupling approach does not specify the exact form of  order
parameter. In a recent paper
\cite{seo2008}, we showed that, in strong-coupling, the pairing symmetry is determined
mainly by the next nearest neighbor antiferromagnetic exchange
coupling $J_2$\cite{Fang2008,Yildirim2008,si} and has an explicit
$s_{x^2y^2}$ form in momentum space, $\cos(k_x)\cdot\cos(k_y)$. This
result is completely independent of any model, \emph{as long as} the
dominating interaction is next-nearest neighbor $J_2$ \emph{and} the
Fermi surfaces are located close to the $\Gamma$ and $M$ points in
the Brillouin zone.  The $\cos(k_x) \cdot \cos(k_y)$ changes sign
between the electron and hole pockets in the Brillouin zone. In this
sense, it resembles the order parameter, $s_{\pm}$, proposed through
 general weak-coupling arguments \cite{Mazin2008a}.

The magnitudes of
superconducting gaps measured by angle-resolved photo-emission
spectroscopy (ARPES) on different Fermi surfaces are  in  good
agreement with the simple
$\cos(k_x)\cdot\cos(k_y)$\cite{Ding2008a,Nakayama2008a,Hasan2008}.
The magnetic properties in the SC state have also been shown to be
consistent with the proposed pairing
symmetry\cite{Parish2008,Parker2008b,Laad2009,Maier2008d}.  Although
several theoretical works\cite{Tsai2008,Ghaemi2008,Parker2008a,Wu2009} propose different ways to measure the sign change between the
electron and hole pockets, directly
probing this change is still a fundamental experimental challenge. Without
any detailed calculations, a theoretical suggestion for probing the
sign change through  quasiparticle interference (QPI) in the presence of magnetic and nonmagnetic impurities.
 has been made in \cite{Wang2009}.

The QPI can be probed directly in modern STM
experiments\cite{HOFFMAN2002,Wang2003} and has been intensively
studied in copper-based high temperature superconductors.
In the presence of impurities, elastic scattering mixes two
eigenstates with different momentum $\mathbf{k}_1$ and
$\mathbf{k}_2$ on the same contour of constant energy and a
scattering interference pattern appears as a modulation in the local density states
(LDOS) at wavevector
$\mathbf{q}=\mathbf{k}_2-\mathbf{k}_1$. Such kind of interference
pattern in the wavevector space can be observed in the Fourier
transform scanning tunneling spectroscopy (FT-STS)
\cite{MCELROY2003, MCELROY2003A}. The quasiparticle scattering
between regions in the $\mathbf{k}$ space with high density of
states (DOS) yields peaks or arcs in the FT-STS. For example, in the
d-wave pairing SC state, many QPI dispersive peaks can be
identified; in
the cuprates, they provide details of the band structure, the nature of superconducting gap. or other competing orders\cite{CHEN2004,SEO2007,Seo2008a,CHEN2002,CHEN2004A,
BENA2004,GHOSAL2005,TESANOVIC2004,TESANOVIC2005,SACHDEV2004,KIVELSON2003,ROBERTSON2006,PODOLSKY2003}.

In this paper, we perform a detailed investigation of the QPI in
iron-based superconductors. We use both two orbital and five
orbital models. In general, the QPI strongly depends on the bare
band structure. The QPI patterns change significantly from a two orbital
model to a five-orbital one, which suggests that the QPI can provide
direct information of the detailed band structure and orbital
degrees of freedom. By carefully examining the pattern, we can also
identify common features of the QPI pattern in both models, which
are tied to the symmetry of SC order parameter and the impurity type.
These general features include: (1) The intra-orbital scattering by
impurities always dominates the inter-orbital scattering. The latter
has negligible effect on the QPI (though it breaks discrete $C_4$
symmetry of the patterns); (2) Unlike in  the d-wave SC state of
cuprates where a large  density of states at the banana tips cause
dispersive features in the QPI\cite{MCELROY2003}, the nodeless
s-wave has little density of states inside SC gaps and hence no strong
points dominate the scattering; (3) A magnetic impurity always
causes a broad and large peak near $\mathbf{q}=(0,0)$ in the QPI;
this stems from intra-band scattering. For a non-magnetic
impurity, the intensity around $\mathbf{q}=(0,0)$ is small. This
result can be used to distinguish two types of impurities; (4) The
peaks around $(\pm\pi,0)$ and $(0,\pm\pi)$ are sensitive to both the type
of impurities and to the sign change of the SC orders between the electron and
hole pockets. Magnetic impurities along with sign-unchanged SC orders or
non-magnetic impurity with sign-changed SC orders cause strong
interference peaks. Finally, as in a fully-gapped s-wave SC state the results from a full
T-matrix calculation do not differ considerably from  results of a simple
first order perturbation calculation\cite{Barnea03,Capriotti}, we are able to also provide an analytic derivation of the
above results.

\section{Two-orbital model and single impurity scattering}
We first  investigate the QPI in a simple two orbital model\cite{raghu2008,seo2008,Parish2008}. The mean field Hamiltonian of the model in  SC states is written as $H=\sum_\mathbf{k}
\Psi^{\dagger}(\mathbf{k})B(\mathbf{k})\Psi(\mathbf{k})$ with
\begin{equation}
B(\mathbf{k})=\left(
  \begin{array}{cccc}
    \epsilon_x(\mathbf{k})-\mu & \Delta_1(\mathbf{k}) & \epsilon_{xy}(\mathbf{k}) & 0 \\
    \Delta^*_1(\mathbf{k}) & -\epsilon_x(\mathbf{k})+\mu & 0 & -\epsilon_{xy}(\mathbf{k}) \\
    \epsilon_{xy}(\mathbf{k}) & 0 & \epsilon_y(\mathbf{k})-\mu & \Delta_2(\mathbf{k}) \\
    0 & -\epsilon_{xy}(\mathbf{k}) & \Delta^*_2(\mathbf{k}) & -\epsilon_y(\mathbf{k})+\mu \\
  \end{array}
\right),\label{eq1}
\end{equation}
where
$\Psi^{\dagger}(\mathbf{k})=(c^{\dagger}_{1,\mathbf{k},\uparrow},c_{1,-\mathbf{k},\downarrow},c^{\dagger}_{2,\mathbf{k},\uparrow},c_{2,-\mathbf{k},\downarrow})$
in the Nambu formalism. The single-particle bands read
\begin{eqnarray}
&&\epsilon_x(k_x,k_y)=-2t_1\cos k_x-2t_2\cos k_y-4t_3\cos k_x \cos
k_y \nonumber\\
&&\epsilon_y(k_x,k_y)=\epsilon_x(k_y,k_x),\quad
\epsilon_{xy}(k_x,k_y)=-4t_4\sin k_x \sin k_y,\nonumber
\end{eqnarray}
where $t_1=-1$, $t_2=1.3$, $t_3=t_4=-0.85$ and $\mu$ is chosen in
the electron-doped regime. Hereafter, $|t_1|$ will be used as the
energy unit.   For $s_{x^2y^2}$ -wave pairing, the order parameter
is $\Delta_1(k_x,k_y)=\Delta_2(k_x,k_y)=\Delta_0\cos k_x\cos
k_y$\cite{seo2008,Parish2008}.

\begin{figure} [htbp]
\centering \subfigure[$\omega=-0.20$]{
\includegraphics*[bb=5 5 260 260,width=0.2\textwidth]{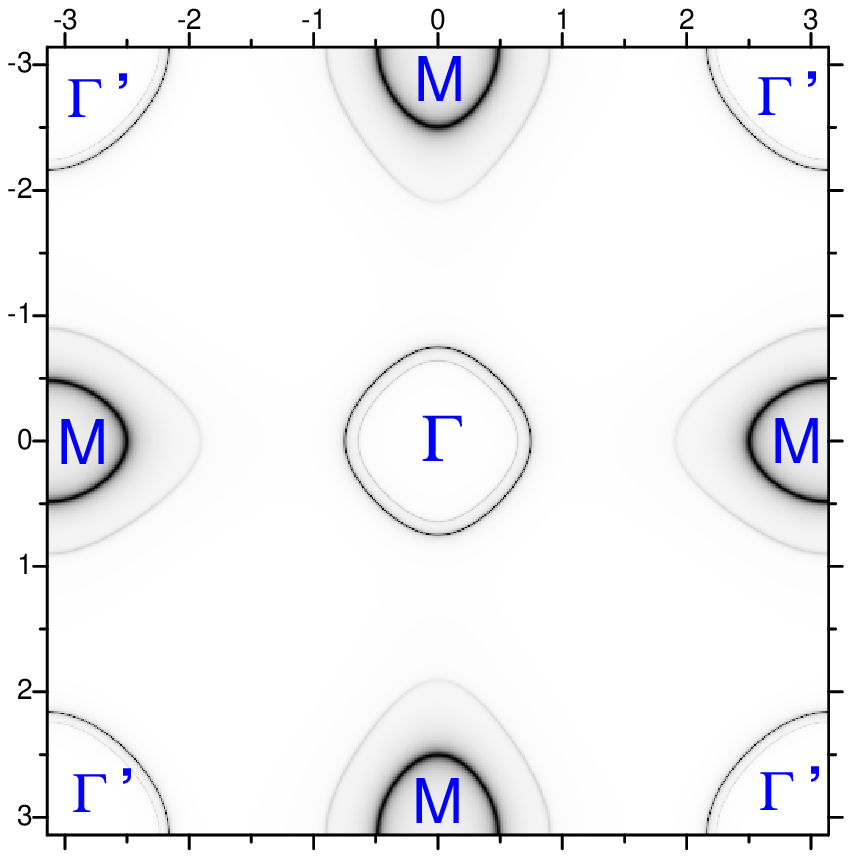}}
\subfigure[$\omega=-0.09$]{
\includegraphics*[bb=5 5 260 260,width=0.2\textwidth]{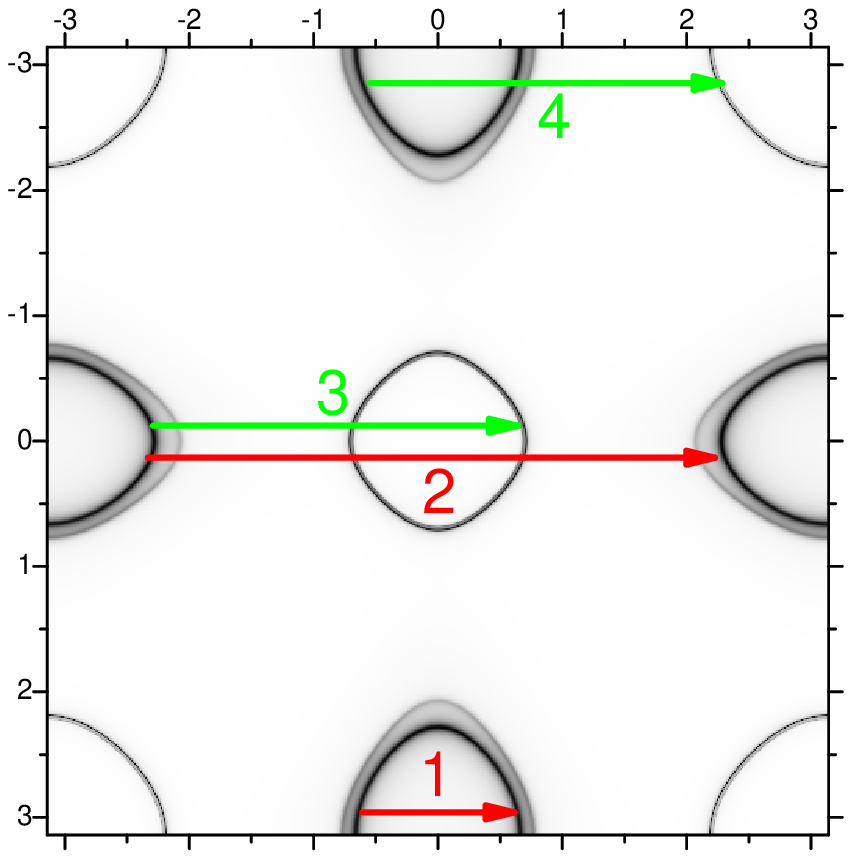}}
\subfigure[$\omega=-0.08$]{
\includegraphics*[bb=5 5 260 260,width=0.2\textwidth]{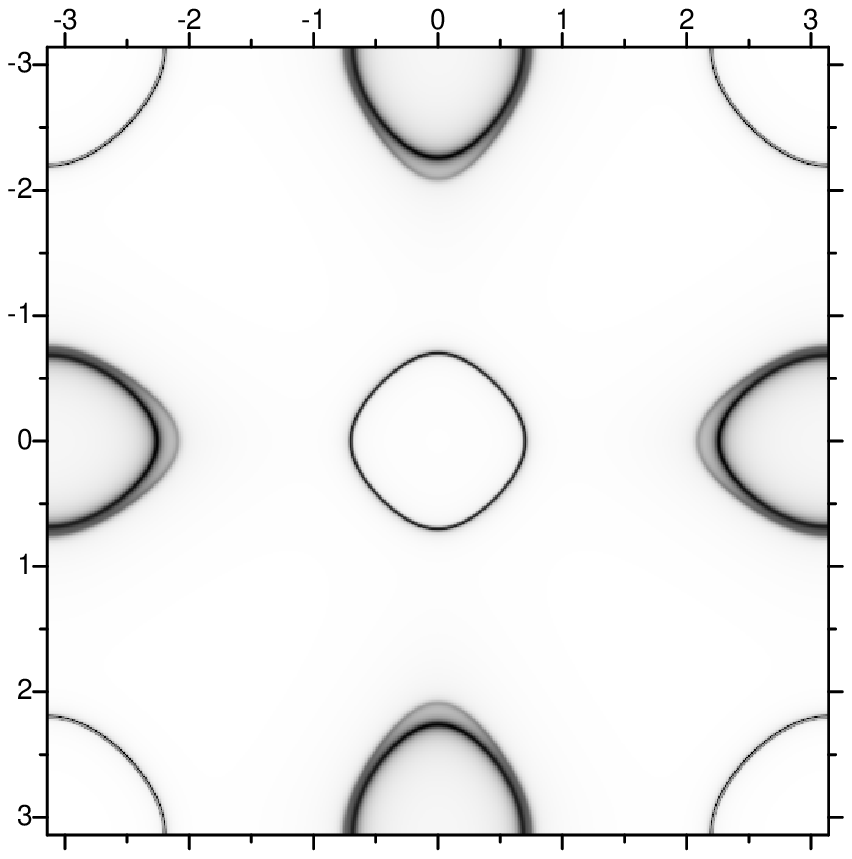}}
\subfigure[$\omega=-0.07$]{
\includegraphics*[bb=5 5 260 260,width=0.2\textwidth]{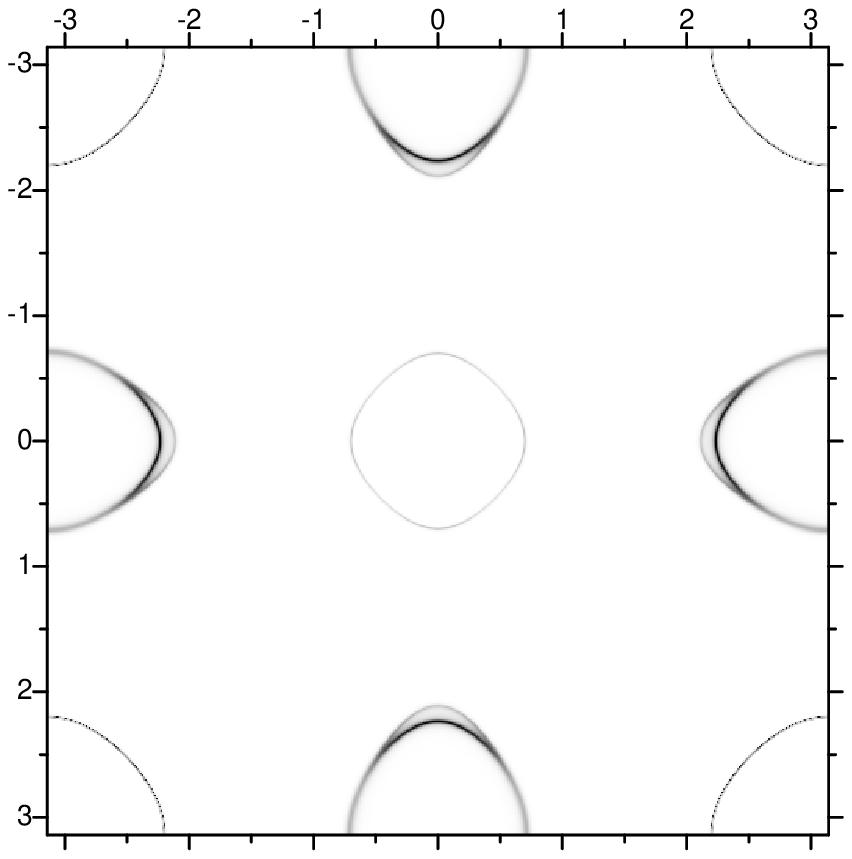}}
\subfigure[$\omega=0.07$]{
\includegraphics*[bb=5 5 260 260,width=0.2\textwidth]{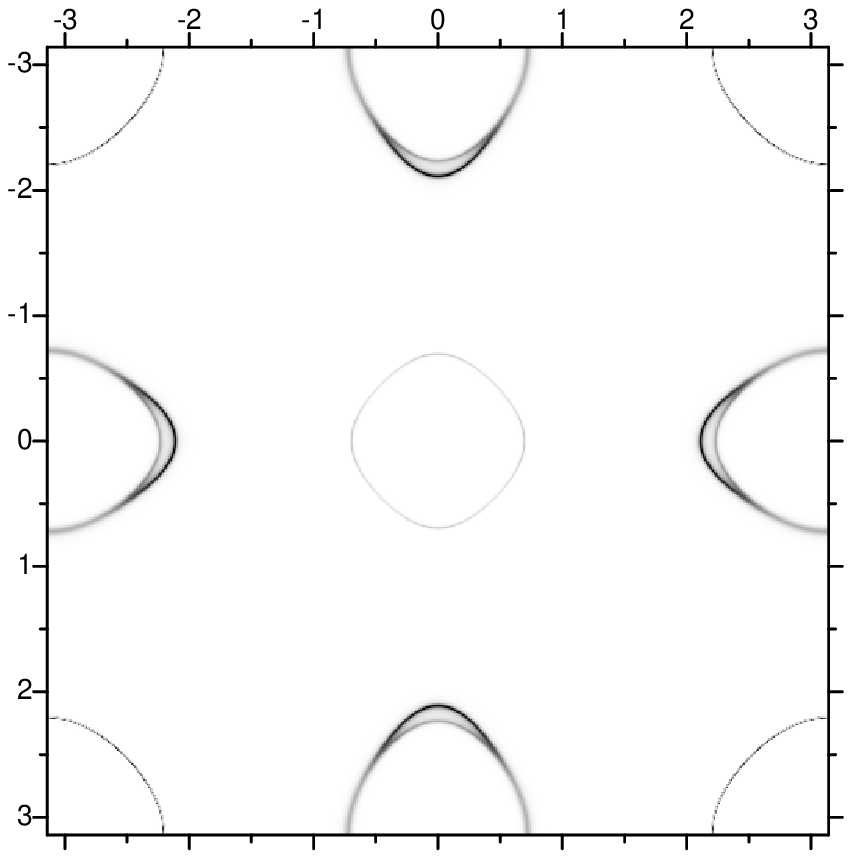}}
\subfigure[$\omega=0.08$]{
\includegraphics*[bb=5 5 260 260,width=0.2\textwidth]{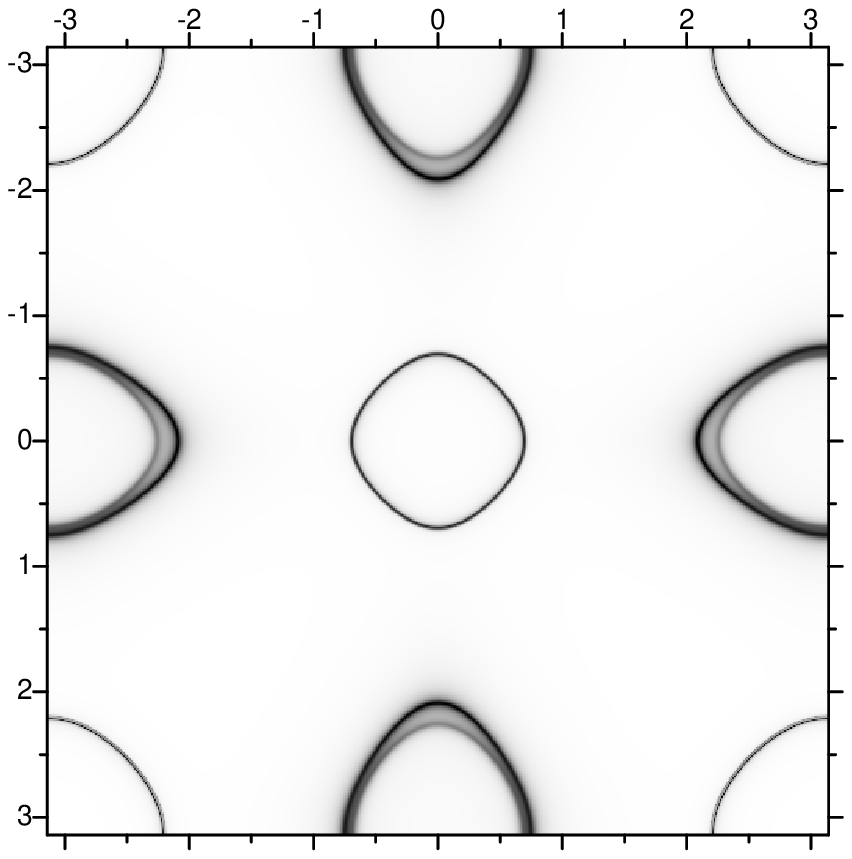}}
\subfigure[$\omega=0.09$]{
\includegraphics*[bb=5 5 260 260,width=0.2\textwidth]{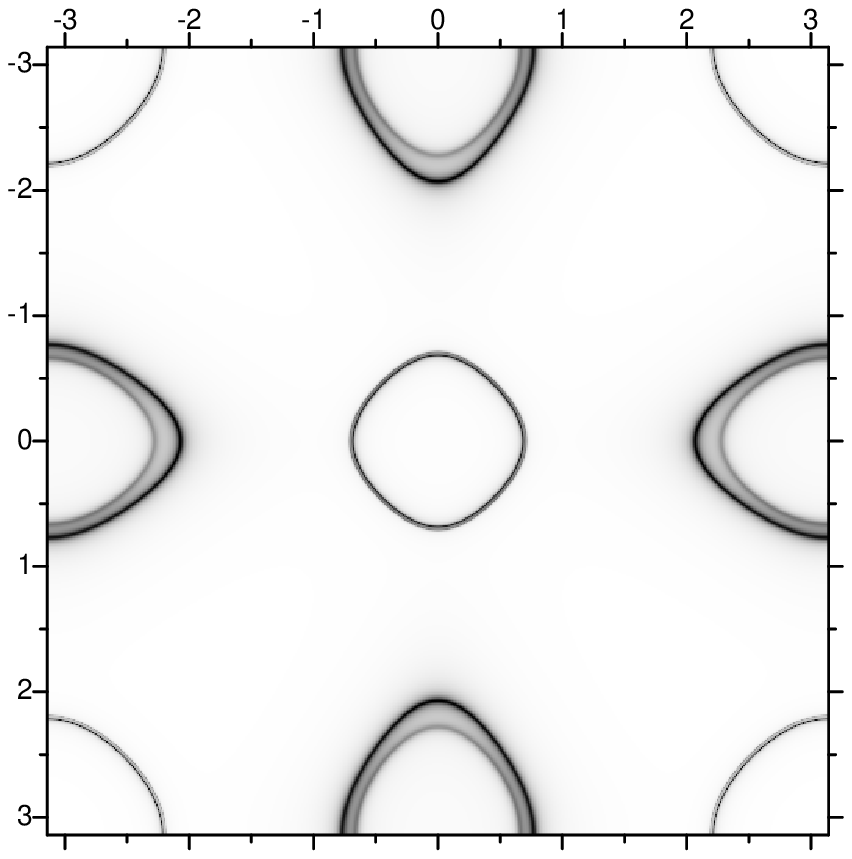}}
\subfigure[$\omega=0.20$]{
\includegraphics*[bb=5 5 260 260,width=0.2\textwidth]{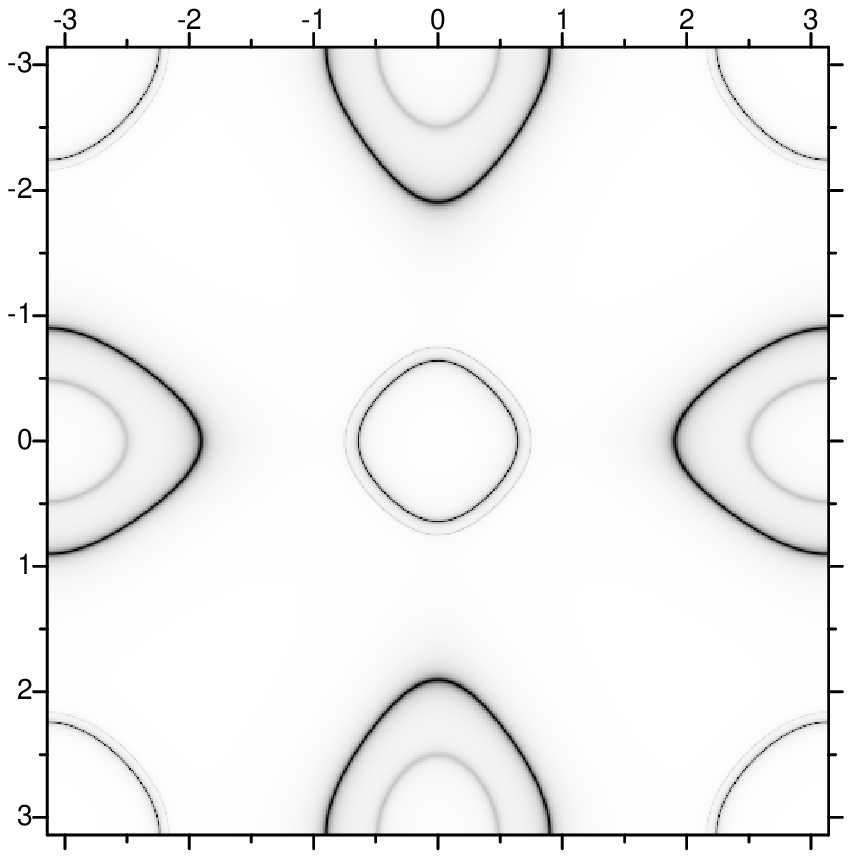}}
\caption{The spectral function $\mathcal{A}(\mathbf{k},\omega)$ in
the unfolded Brillioun zone, for $s_{x^2y^2}$ with $\Delta_0=0.1$.
Darker regions correspond to larger values of $\mathcal{A}$ hence
larger DOS in $\mathbf{k}$ space.} \label{FContour}
\end{figure}

The Green's function for the clean system is
\begin{equation}
G^0(\mathbf{k},\omega)\equiv
G^0(\mathbf{k},\mathbf{k},\omega)=[(\omega+i\delta)I-B(\mathbf{k})]^{-1},\label{eq3}
\end{equation}
where $I$ is the identity matrix and $\delta$ is the energy width broadening. In this work, we only consider a single impurity with
potential $\sim\delta(\mathbf{x})$ so that the impurity matrix
$V(\mathbf{k}_1,\mathbf{k}_2)=V$ is independent of $\mathbf{k}$. The
impurity induced Green's function is expressed as
\begin{equation}
\delta
G(\mathbf{k}_1,\mathbf{k}_2,\omega)=G^0(\mathbf{k}_1,\omega)T(\mathbf{k}_1,\mathbf{k}_2,\omega)G^0(\mathbf{k}_2,\omega).
\label{eq4}
\end{equation}
Standard perturbation theory gives
\begin{eqnarray}
T(\omega)&=&V+V\Gamma^0(\omega)V\nonumber+V\Gamma^0(\omega)V\Gamma^0(\omega)V+\ldots\label{eq5a}\\
&=&[I-V\Gamma^0(\omega)]^{-1}V,\label{eq5b}
\end{eqnarray}
where
\begin{equation}
\Gamma^0(\omega)=\int\frac{d^2k}{(2\pi)^2}G^0(\mathbf{k},\omega).\label{eq6}
\end{equation}
Consequently, the Fourier transform of the (induced) local density
of states is
\begin{equation}
\delta\rho(\mathbf{q},\omega)=\frac{i}{2\pi}\int\frac{d^2k}{(2\pi)^2}g(\mathbf{k},\mathbf{q},\omega),
\label{eq7}
\end{equation}
where $\mathbf{q}=\mathbf{k}'-\mathbf{k}$ and
\begin{eqnarray}
&&g(\mathbf{k},\mathbf{q},\omega)=\delta G_{11}(\mathbf{k},\mathbf{k}',\omega)-\delta G^*_{11}(\mathbf{k}',\mathbf{k},\omega)+\nonumber\\
&&\delta G_{33}(\mathbf{k},\mathbf{k}',\omega)-\delta
G^*_{33}(\mathbf{k}',\mathbf{k},\omega).\label{eq8}
\end{eqnarray}

\begin{figure} [htbp]
\begin{center}
\includegraphics[bb=20 20 300 250,width=0.4\textwidth]{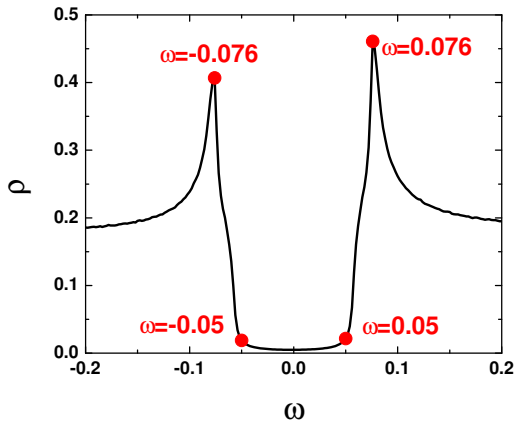}
\end{center}
\caption{ Bulk density of states $\rho$ for $s_{x^2y^2}$ as a
function of $\omega$, $\Delta_0=0.1$ and no impurities. Some special
values are marked by red dots. The energy broadening width is
$\delta=0.002$.} \label{FDOS}
\end{figure}

Due to the multi-orbital nature of the band model, we distinguish
different types of impurities. They are
\begin{equation}
V= V_{\mathrm{intra}}=I\otimes\left(
  \begin{array}{cc}
    V_0 & 0  \\
    0 & \pm V_0  \\
  \end{array}
\right), \label{eq9a}
\end{equation}
for an impurity with only intra-orbital scattering, and
\begin{equation}
V= V_{\mathrm{inter}} =\sigma_x\otimes\left(
  \begin{array}{cc}
     V_0 & 0  \\
    0 & \pm V_0  \\
  \end{array}
\right),\label{eq9b}
\end{equation}
for inter-orbital scattering, where the upper (down) sign
corresponds to magnetic (non-magnetic) impurity. Since it was argued
that for cuprate superconductors this $T$-matrix method is valid
when impurity scattering strength is much larger than the maximal
pairing gap\cite{Wang2003}, we take $V_0=4\Delta_0$ in our
calculation. Our following results do not depend on $V_0$ as long as
$V_0$ is much larger than $\Delta_0$.

\subsection{Numerical results}

 We first calculate  electronic properties in an impurity-free
system. Most of the results in this section have been already
computed in less detail in \cite{Parish2008}. In Fig.
\ref{FContour}, we plot the spectral function
\begin{equation}
\mathcal{A}(\mathbf{k},\omega)=-\frac{1}{\pi}\Im
[G^0_{11}(\mathbf{k},\omega)+G^0_{33}(\mathbf{k},\omega)]\label{eq10}
\end{equation}
of the clean system at different $\omega$. The typical value of
order parameter $\Delta_0=0.1$ will be used throughout this paper.
It should be noted that for $\omega$ and $-\omega$, the shapes
(topology) of the contours of constant energy (CCE) are almost
identical, but the numerical values of $\mathcal{A}$ on these
contours are remarkably different. As an example, let us focus on
the regions near one of the M points ($\pi,0$), where the CCE
consists of two semi-oval-circles, or two complete oval-circles due
to the periodicity of the Brillioun zone (BZ). As can be seen from
Fig. \ref{FContour} (a) and (b), for negative $\omega$, the spectral
weight on the inner circle is larger than that on the outer one. The
situation is opposite for positive $\omega$ (Fig. \ref{FContour} (g)
and (h)). This leads to different scattering interference patterns
at $\pm\omega$, as we will see later on.

\begin{figure} [htbp]
\begin{center}
\includegraphics[bb=20 20 460 220,width=0.5\textwidth]{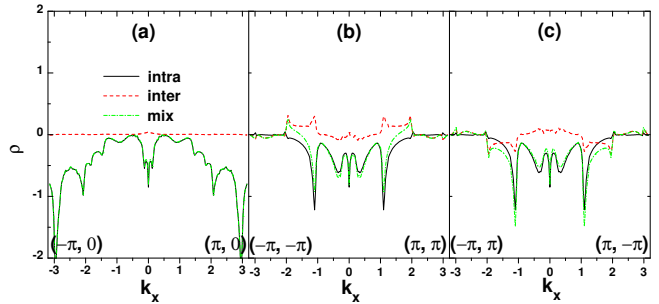}
\end{center}
\caption{(color online) $\delta\rho(\mathbf{q},\omega=0.07)$ for
non-magnetic impurity with $V_{\mathrm{intra}}$ (black solid),
$V_{\mathrm{inter}}$ (red dash) and $V_{\mathrm{mix}}\equiv
V_{\mathrm{intra}}+V_{\mathrm{inter}}$ (green dash dot) along three
directions: (a) $(-\pi,0)\rightarrow(\pi,0)$, (b)
$(-\pi,-\pi)\rightarrow(\pi,\pi)$ and (c)
$(-\pi,\pi)\rightarrow(\pi,-\pi)$.} \label{FIntraInter}
\end{figure}

The bulk density of states $\rho(\omega)=\sum_{\mathbf{k}}
\mathcal{A}(\mathbf{k},\omega)$ is plotted in Fig. \ref{FDOS}. It is
fully gaped within $\sim(-0.05,0.05)$ and the coherent peak occurs
at $\sim\pm0.076$ \cite{Parish2008}.

\begin{figure} [htbp]
\centering \subfigure[$\omega=-0.2$]{
\includegraphics*[bb=5 5 260 260,width=0.2\textwidth]{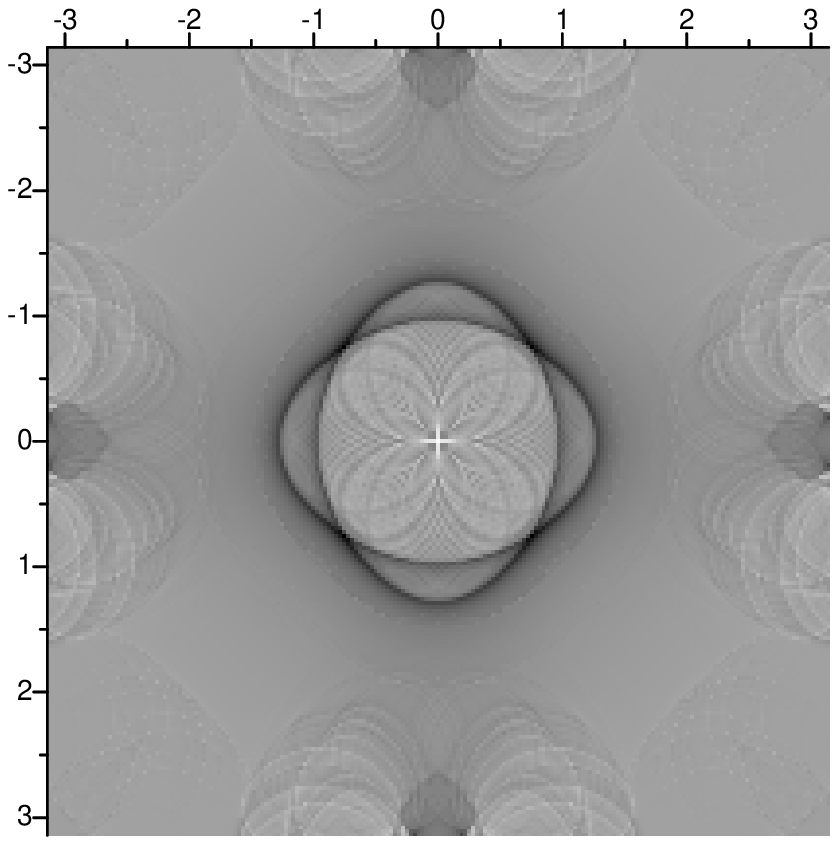}}
\subfigure[$\omega=-0.09$]{
\includegraphics*[bb=5 5 260 260,width=0.2\textwidth]{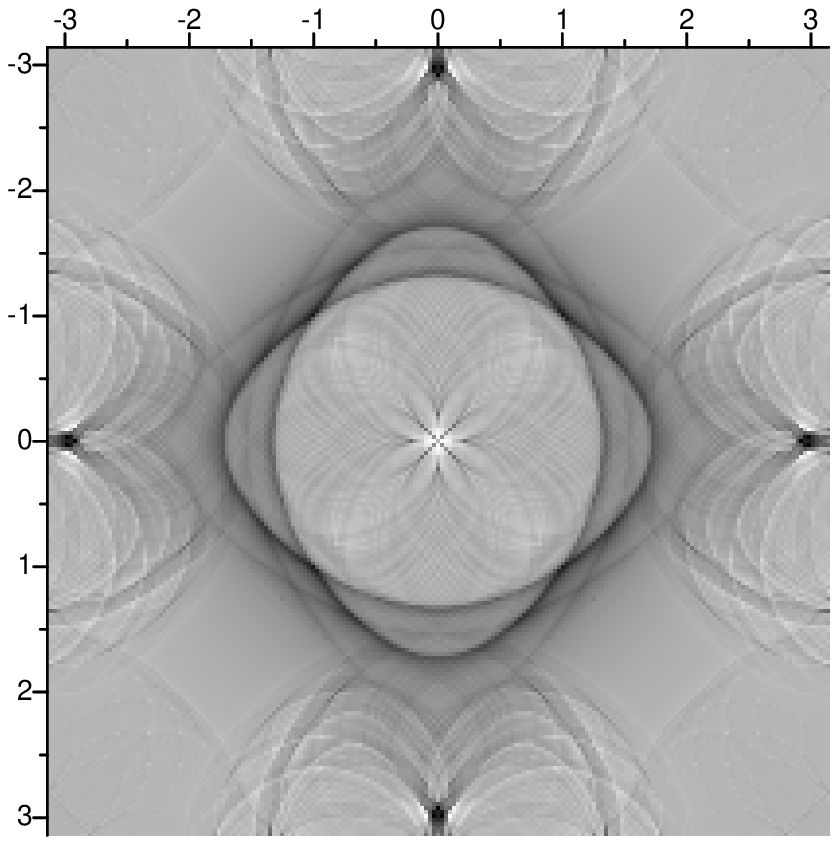}}
\subfigure[$\omega=-0.08$]{
\includegraphics*[bb=5 5 260 260,width=0.2\textwidth]{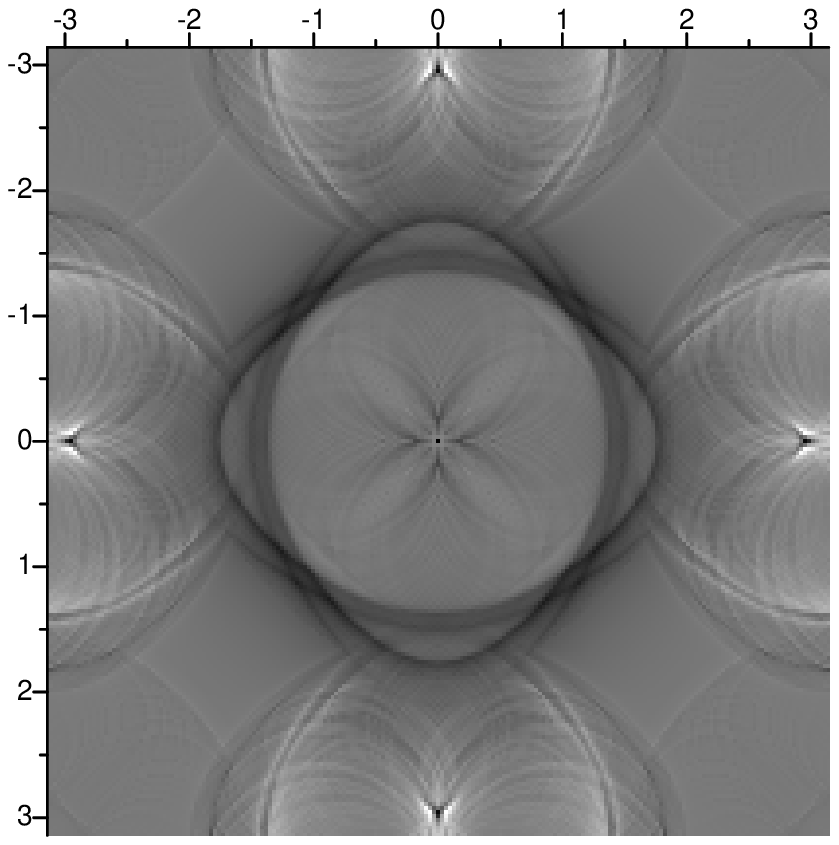}}
\subfigure[$\omega=-0.07$]{
\includegraphics*[bb=5 5 260 260,width=0.2\textwidth]{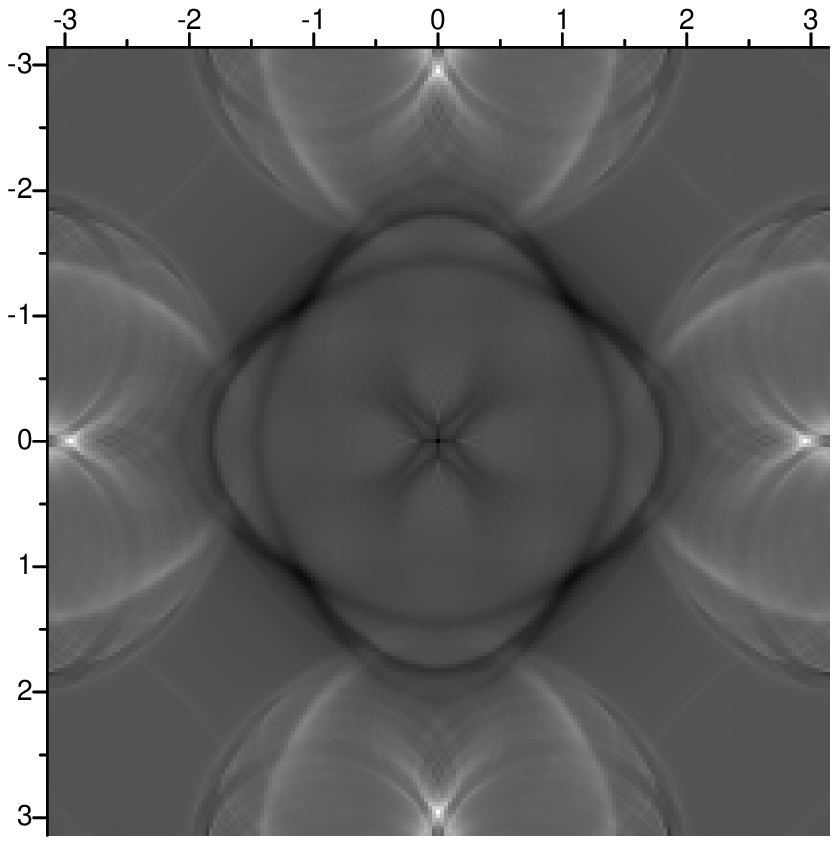}}
\subfigure[$\omega=0.07$]{
\includegraphics*[bb=5 5 260 260,width=0.2\textwidth]{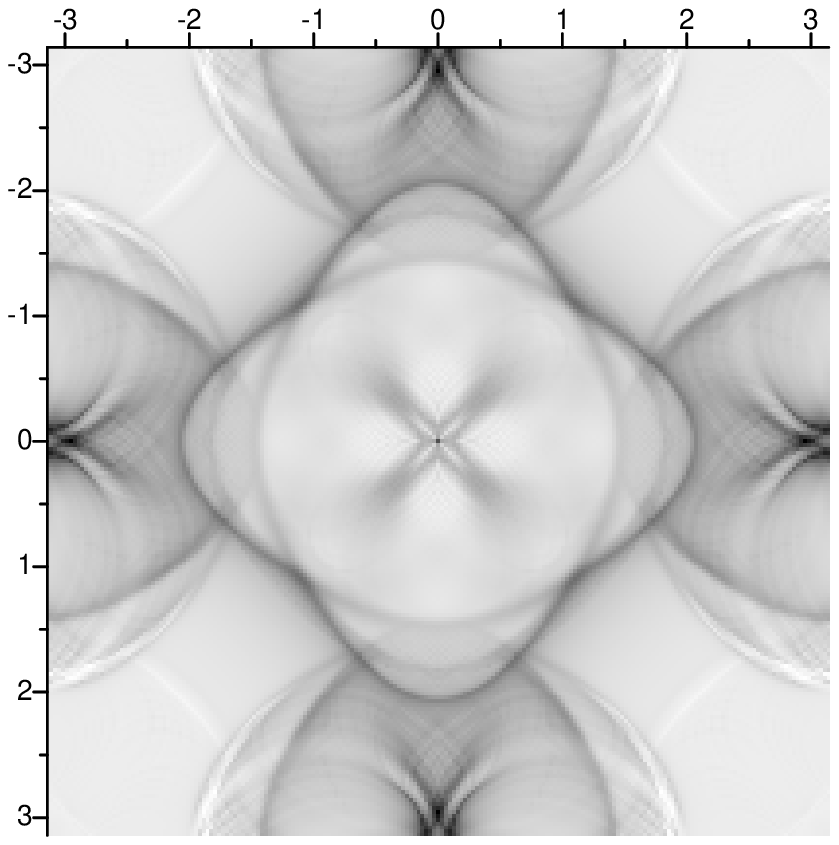}}
\subfigure[$\omega=0.08$]{
\includegraphics*[bb=5 5 260 260,width=0.2\textwidth]{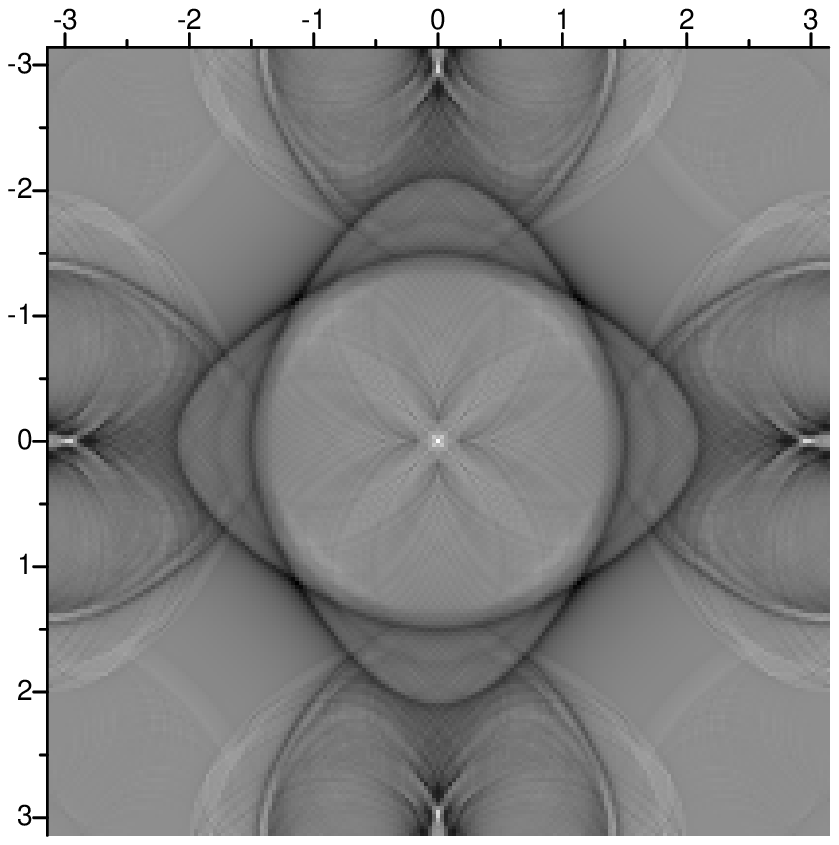}}
\subfigure[$\omega=0.09$]{
\includegraphics*[bb=5 5 260 260,width=0.2\textwidth]{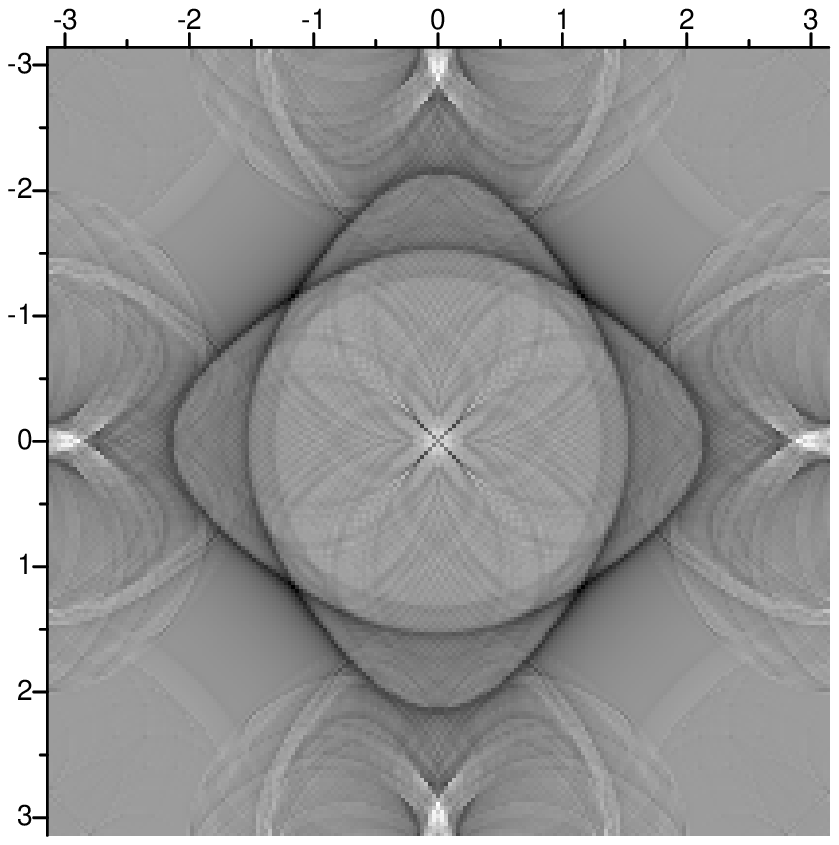}}
\subfigure[$\omega=0.2$]{
\includegraphics*[bb=5 5 260 260,width=0.2\textwidth]{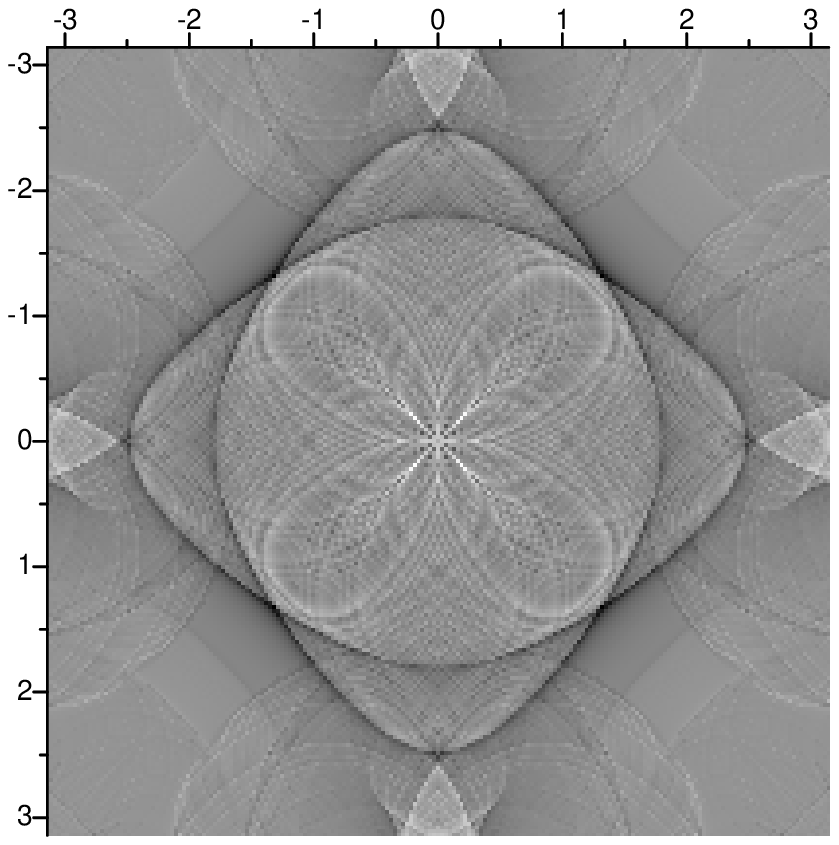}}
\caption{$\delta\rho(\mathbf{q})$ for non-magnetic impurity with
intra-orbital scattering, $V_0=0.4$. A $200\times200$ lattice in
$\mathbf{k}$-space is used in numerical integration of equation
(\ref{eq6}) and the energy broadening width $\delta=0.005$.}
\label{FNonM2D}
\end{figure}

\begin{figure} [htbp]
\centering \subfigure[$\omega=-0.2$]{
\includegraphics*[bb=5 5 260 260,width=0.2\textwidth]{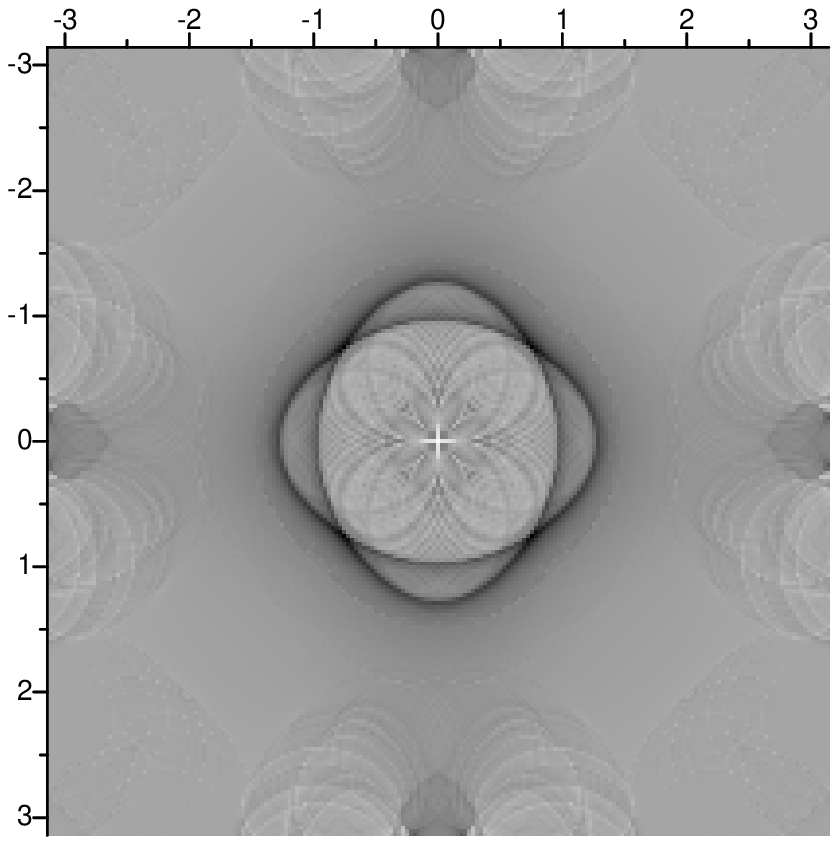}}
\subfigure[$\omega=-0.09$]{
\includegraphics*[bb=5 5 260 260,width=0.2\textwidth]{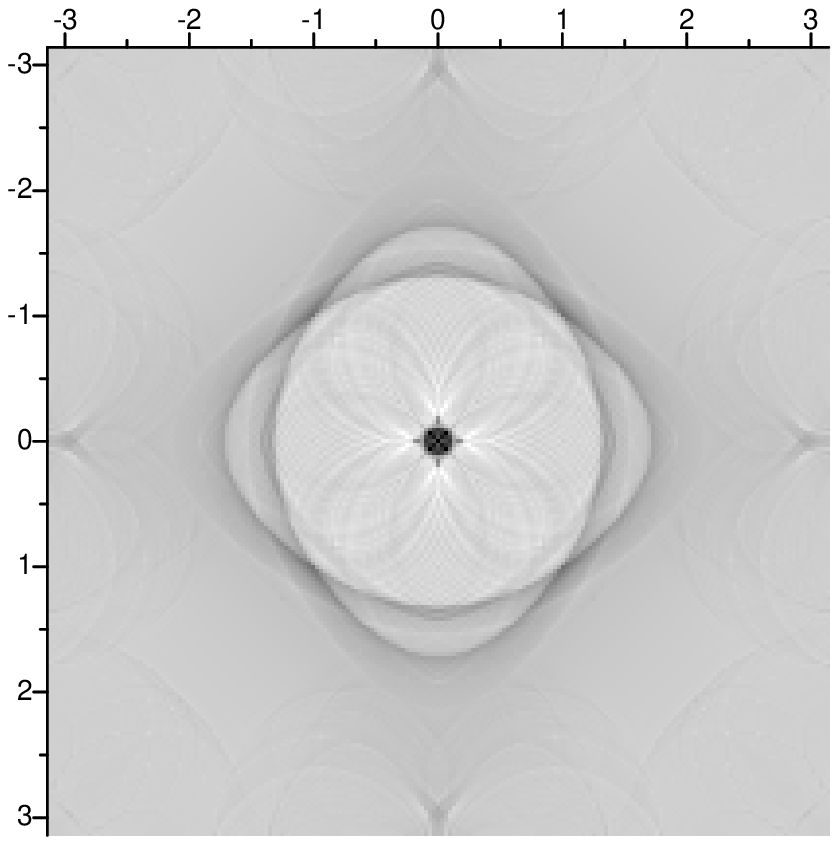}}
\subfigure[$\omega=-0.08$]{
\includegraphics*[bb=5 5 260 260,width=0.2\textwidth]{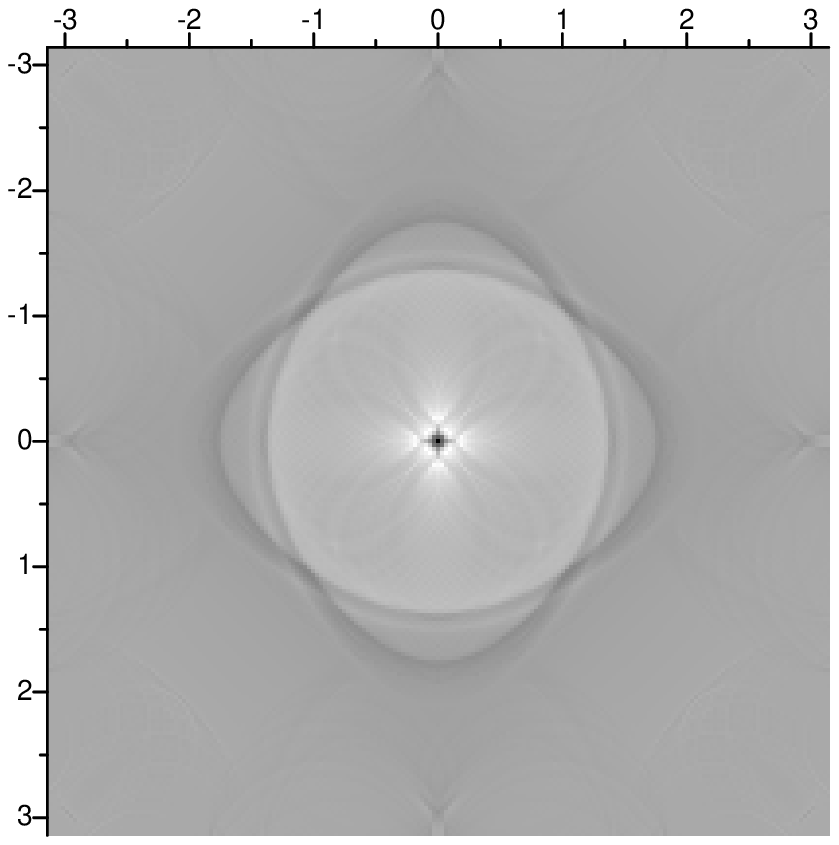}}
\subfigure[$\omega=-0.07$]{
\includegraphics*[bb=5 5 260 260,width=0.2\textwidth]{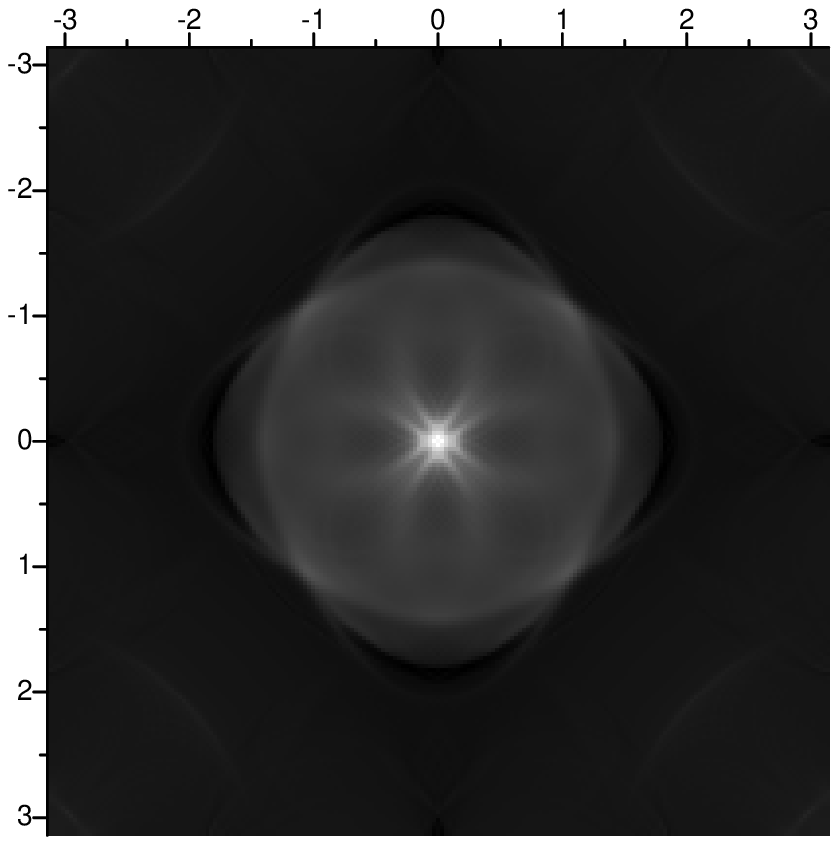}}
\subfigure[$\omega=0.07$]{
\includegraphics*[bb=5 5 260 260,width=0.2\textwidth]{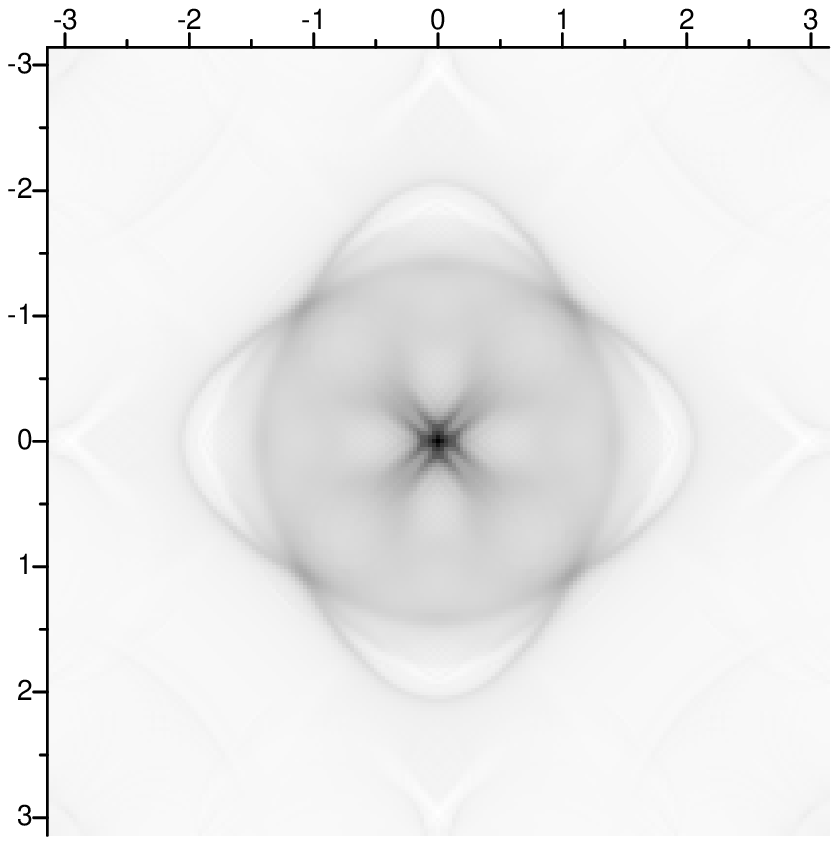}}
\subfigure[$\omega=0.08$]{
\includegraphics*[bb=5 5 260 260,width=0.2\textwidth]{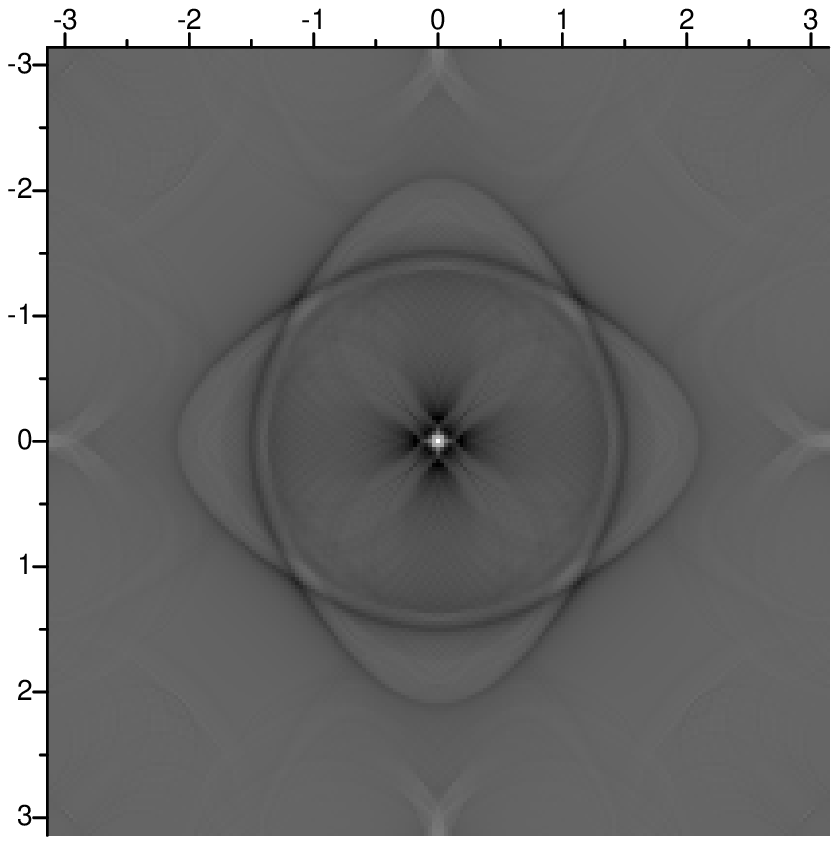}}
\subfigure[$\omega=0.09$]{
\includegraphics*[bb=5 5 260 260,width=0.2\textwidth]{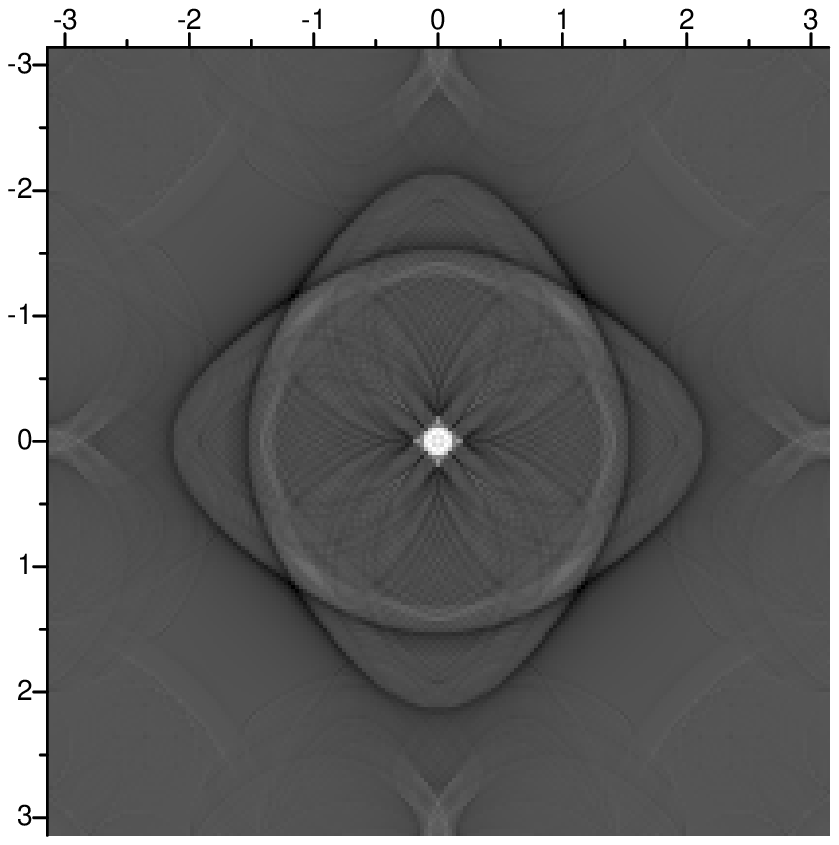}}
\subfigure[$\omega=0.2$]{
\includegraphics*[bb=5 5 260 260,width=0.2\textwidth]{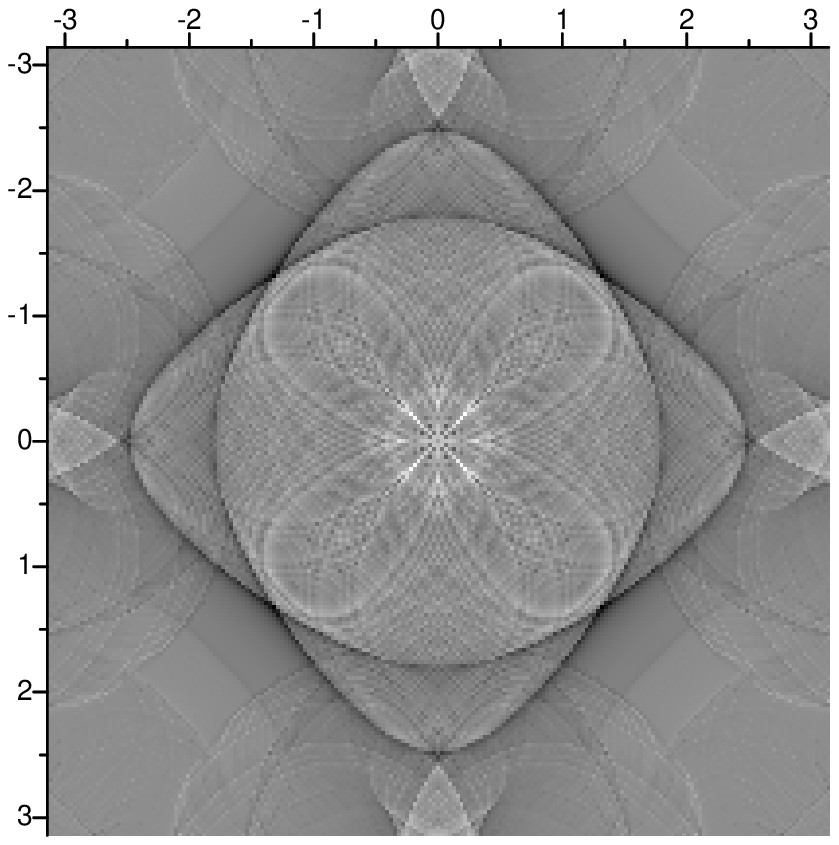}}\caption{The same with Fig. \ref{FNonM2D} but for magnetic impurity, $V_0=0.4$.}
\label{FM2D}
\end{figure}

To exemplify some scattering amplitudes, we plot
$\delta\rho(\mathbf{q})$ for a non-magnetic impurity near the edge
of the gap ($\omega=0.07$) along three special directions in Fig.
\ref{FIntraInter}. Two observations in these figures are common to
all our results. First, $\rho(\mathbf{q})$ for inter-orbital
scattering in the two diagonal directions are quite different
(compare the red dash lines in Fig. \ref{FIntraInter} (b) and (c)).
This is not surprising since the inter-orbital scattering such as
$c^{\dagger}_{1,\mathbf{k},\uparrow}c_{2,\mathbf{k},\uparrow}$
breaks the symmetry between directions
$(-\pi,-\pi)\rightarrow(\pi,\pi)$ and
$(-\pi,\pi)\rightarrow(\pi,-\pi)$. Second, the amplitude of
intra-orbital impurity is stronger than that of the inter-orbital
one; therefore the intra-orbital scattering is dominant when both
are present. As such, in the following, we present only numerical
results for intra-orbital impurities.

\begin{figure} [htbp]
\begin{center}
\includegraphics*[width=0.4\textwidth]{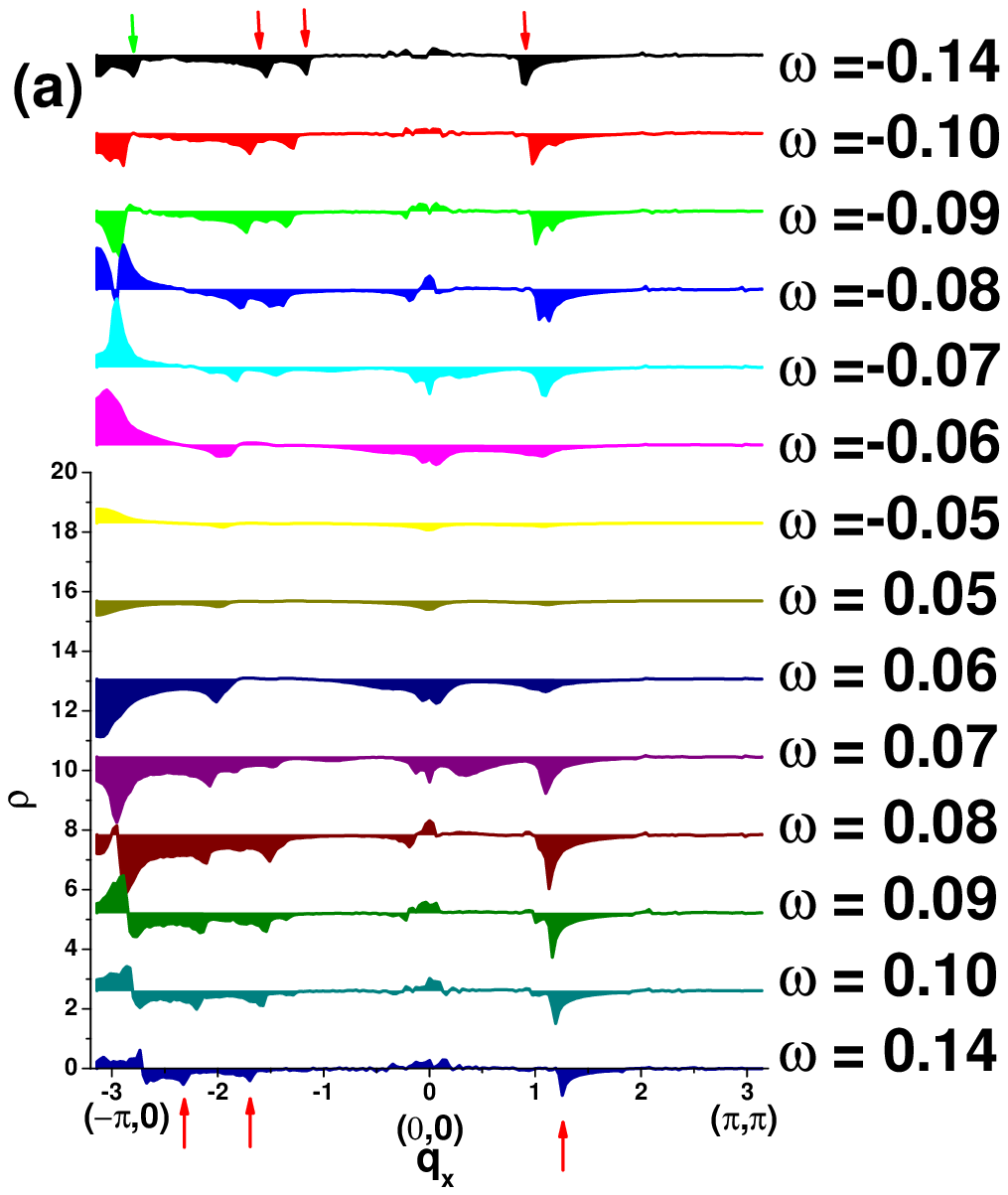}
\includegraphics*[width=0.4\textwidth]{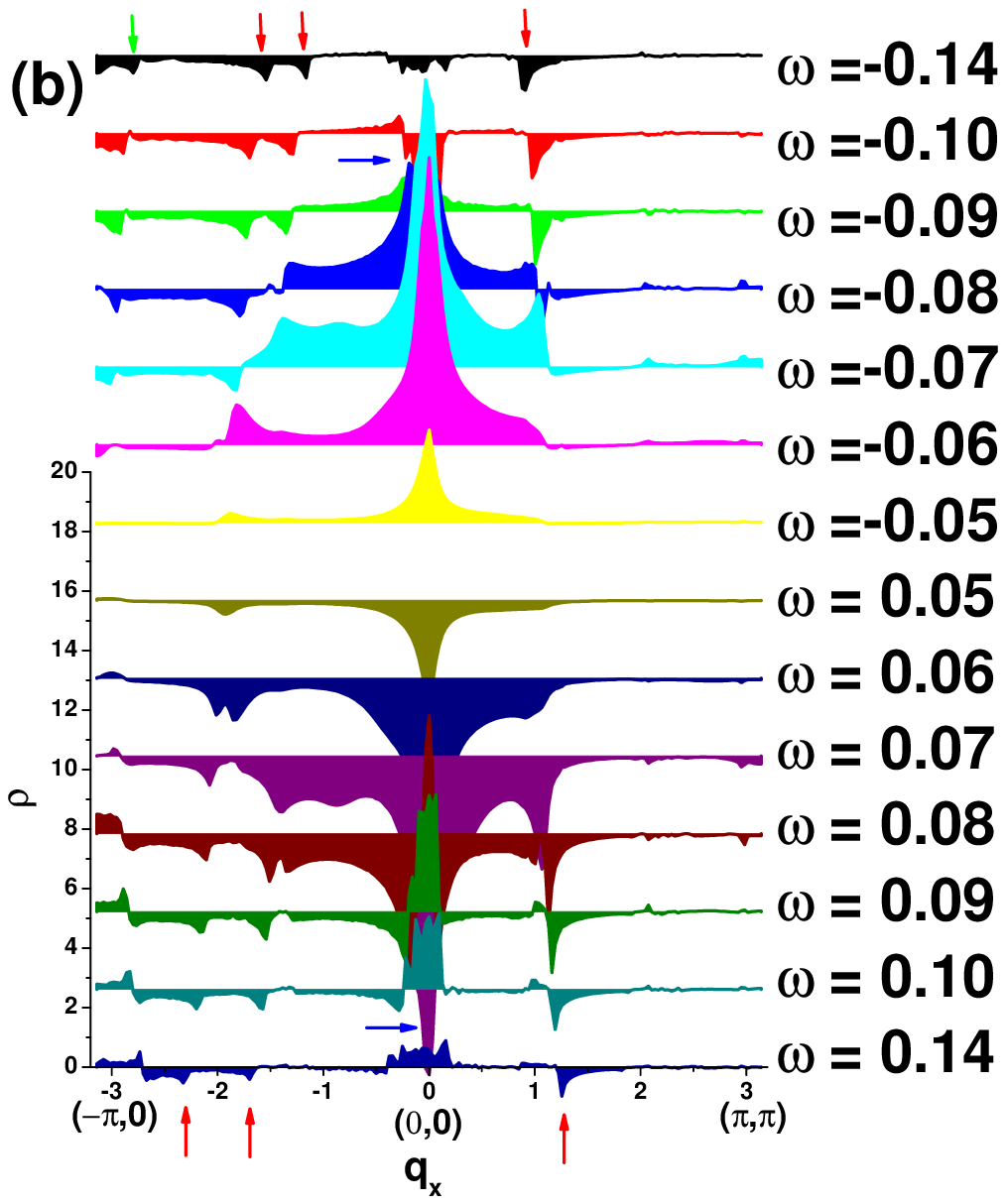}
\end{center}
\caption{(color online) Profiles of $\delta\rho(\mathbf{q},\omega)$
along $\mathrm{M}\rightarrow\Gamma\rightarrow\Gamma'$ for (a)
non-magnetic impurity and (b) magnetic impurity. The data are
shifted vertically relative to each other for clarity.} \label{F1D}
\end{figure}

In Fig. \ref{FNonM2D} and Fig. \ref{FM2D}, we show the
two-dimensional contour of the scattering pattern
$\delta\rho(\mathbf{q})$ for non-magnetic and magnetic impurities,
respectively. The $\delta\rho(\mathbf{q})$ profiles along special
directions are plotted in Fig \ref{F1D}. In Figs. \ref{FNonM2D} and
\ref{FM2D}, the most prominent features for all $\omega$ values are
two intersecting ovals (see also the peaks directed by red arrows in
Fig. \ref{F1D}), reflecting the strong intra-pocket scattering
between equal-energy curves near M points with the largest
DOS. These scattering processes are labeled by red arrows (1 and 2)
in Fig. \ref{FContour} (b), where scattering wavevectors outside the
first BZ (e.g., arrow 2) should be understood as their equivalent
counterparts in the first BZ.  By increasing $|\omega|$ on the
negative (positive) energy side, the size of the ovals decreases
(increases) because the important scattering takes place between the
inner (outer) circle of CCE near M with the largest DOS. At a
definite energy $\omega$, this gives two peaks along direction
$\mathrm{M}\rightarrow \Gamma$, corresponding to the scattering
along the major and minor axis of the oval CCE respectively. Due to
the congruence of these CCE oval circles, they always intersect on
the diagonal line ( $\Gamma'\rightarrow \Gamma$), therefore only one
peak can be observed along this direction. For $\omega$ far outside
the gap, there is no noticeable difference between the magnetic and
non-magnetic impurities (compare Fig. \ref{FNonM2D} (a) and (h) with
Fig. \ref{FM2D} (a) and (h)). This is understandable since  the
tendency of the (Cooper) pair breaking due to a magnetic impurity is
most significant near the Fermi level. Within the single-particle
scattering regime considered here, for $|\omega|<0.05$ inside the
gap, no interference pattern is expected due to lack of scattering states. This is confirmed
(but not shown here) by the fact that, when decreasing the imaginary part of the
energy $\delta$ in Green's functions, the peaks of
$\delta\rho(\mathbf{q},|\omega|>0.05)$ are sharper, while
$\delta\rho(\mathbf{q},|\omega|<0.05)$ vanish trivially for all
$\mathbf{q}$.

There are additional peaks around the M points, as directed by green
arrows in Fig. \ref{F1D}, which play an important role in
distinguishing different types of impurities. When decreasing
$|\omega|$, they move steadily toward M . These originate from the
inter-pocket scattering as demonstrated by green arrows (3 and 4) in
Fig. \ref{FContour} (b). The differences of these peaks between
non-magnetic and magnetic impurities are clear near the gap edges,
suggesting strong dependence on coherent factors due to impurities,
as well as on DOS contour of the clean
system\cite{Wang2003,Barnea03}. For non-magnetic impurity (Fig.
\ref{F1D} (a)), the peaks are much more sharper. Moreover, a large
peak appears around $\Gamma$ (blue arrows in (Fig. \ref{F1D} (b)).
This suggests that the magnetic impurity's ability to localize the
quasiparticle is weaker than that of the non-magnetic one.

\begin{figure} [htbp]
\begin{center}
\includegraphics*[width=0.4\textwidth]{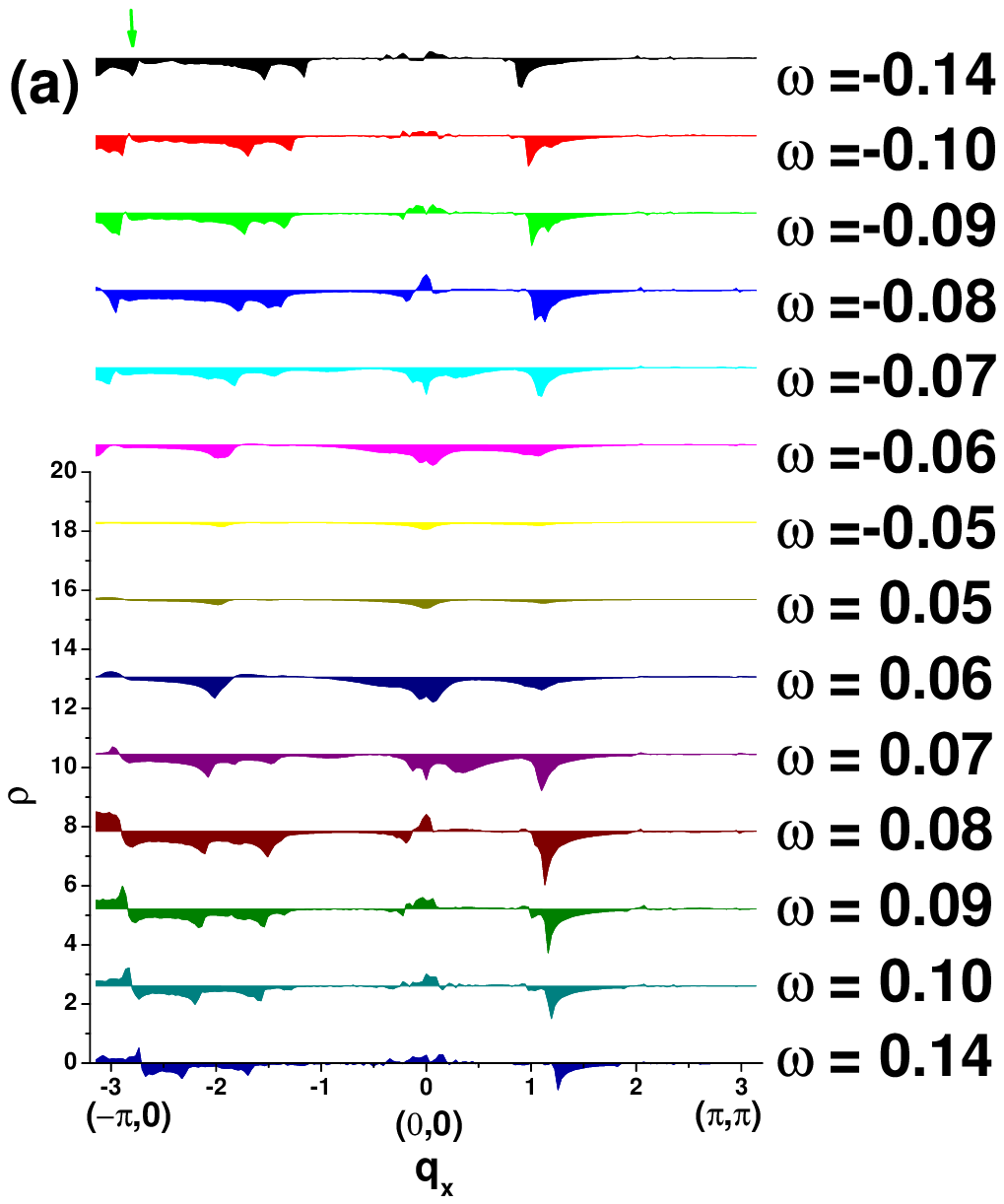}
\includegraphics*[width=0.4\textwidth]{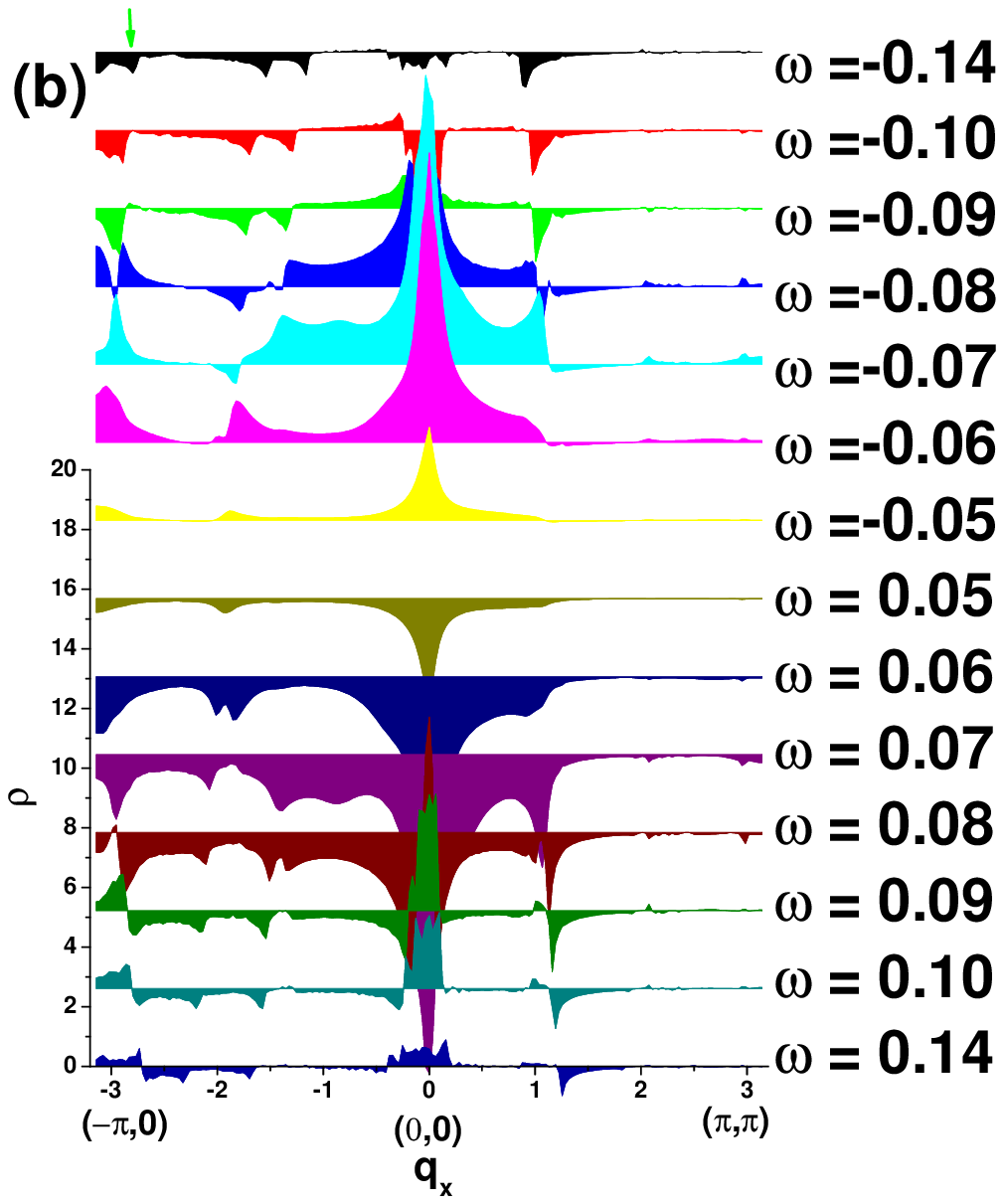}
\end{center}
\caption{(color online) Same as Fig. \ref{F1D} but for the case
without sign change,
$\Delta_1(k_x,k_y)=\Delta_2(k_x,k_y)=|\Delta_0\cos k_x\cos k_y|$.
(a) Non-magnetic impurity and (b) magnetic impurity. }
\label{F1DNoSign}
\end{figure}

\subsection{The effect of sign change}
The $s_{x^2y^2}$ pairing differs from the conventional $s$-wave
pairing that it changes sign in the BZ. To investigate the physical
consequence of this effect, we artificially prohibit this sign
change by letting $\Delta_1(k_x,k_y)=\Delta_2(k_x,k_y)=|\Delta_0\cos
k_x\cos k_y|$ in Eq.~(\ref{eq1}). This may not correspond to any
realistic physical system, but can reveal, by comparison, the
effects of the sign change of the order parameter. We show the
profile of $\delta\rho(\mathbf{q})$ In Fig. \ref{F1DNoSign}. The
most observable feature is that, contrary to the sign-change case,
the inter-pocket peak around the M point is now sharper for the
magnetic impurity.

\subsection{The QPI in other pairing symmetry}
\begin{figure} [htbp]
\centering \subfigure[$\omega=-0.09$]{
\includegraphics*[bb=5 5 260 260,width=0.2\textwidth]{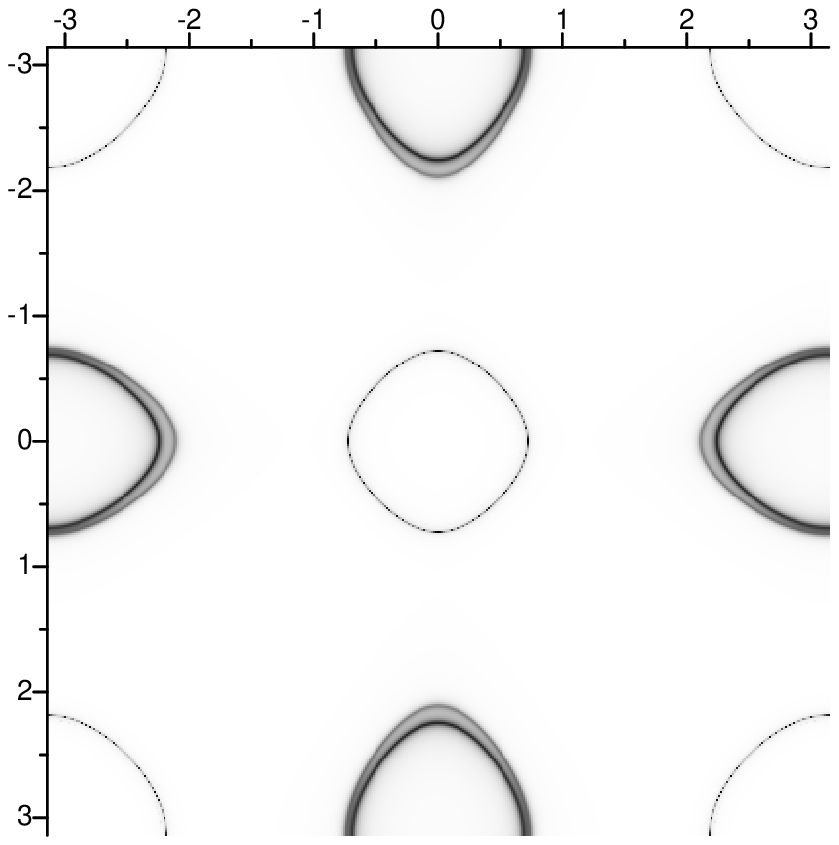}}
\subfigure[$\omega=-0.08$]{
\includegraphics*[bb=5 5 260 260,width=0.2\textwidth]{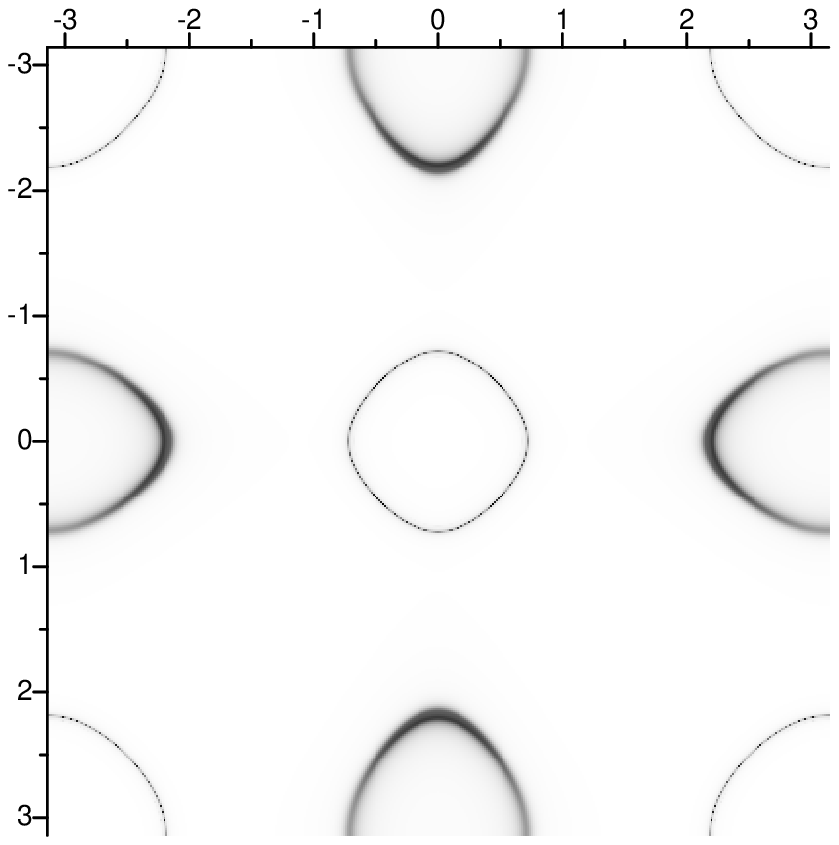}}
\subfigure[$\omega=-0.07$]{
\includegraphics*[bb=5 5 260 260,width=0.2\textwidth]{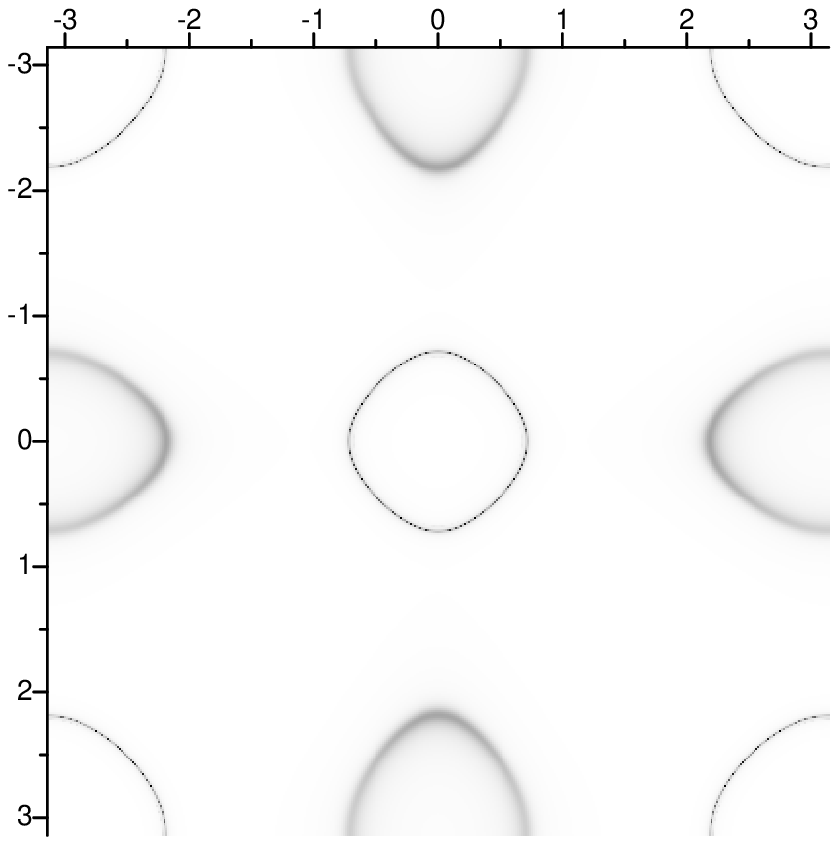}}
\subfigure[$\omega=-0.03$]{
\includegraphics*[bb=5 5 260 260,width=0.2\textwidth]{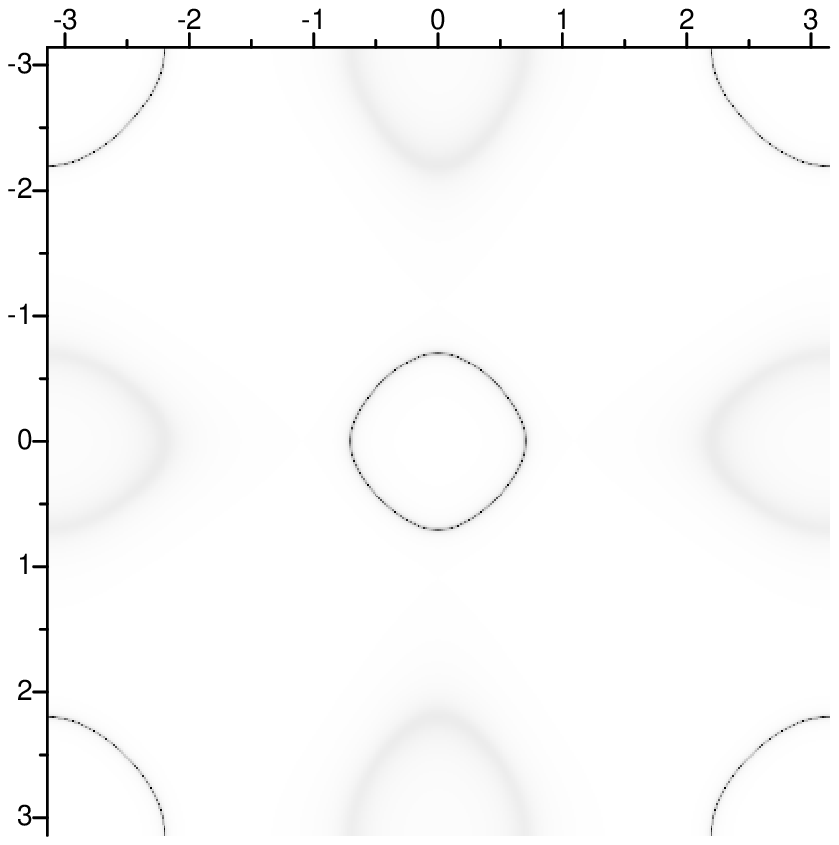}}
\subfigure[$\omega=0$]{
\includegraphics*[bb=5 5 260 260,width=0.2\textwidth]{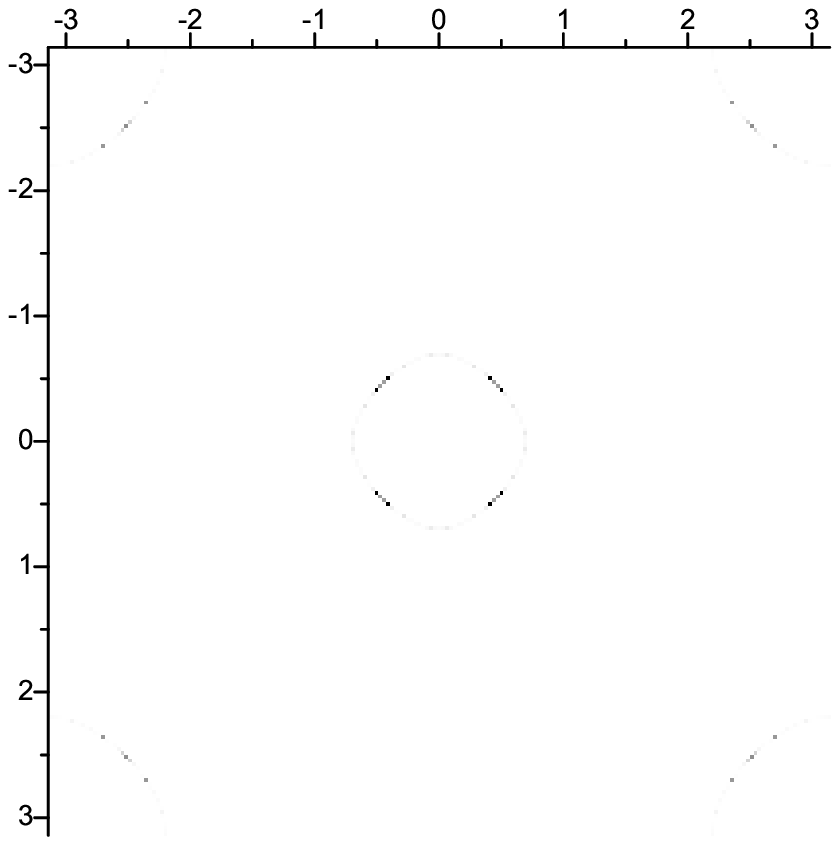}}
\subfigure[$\omega=0.03$]{
\includegraphics*[bb=5 5 260 260,width=0.2\textwidth]{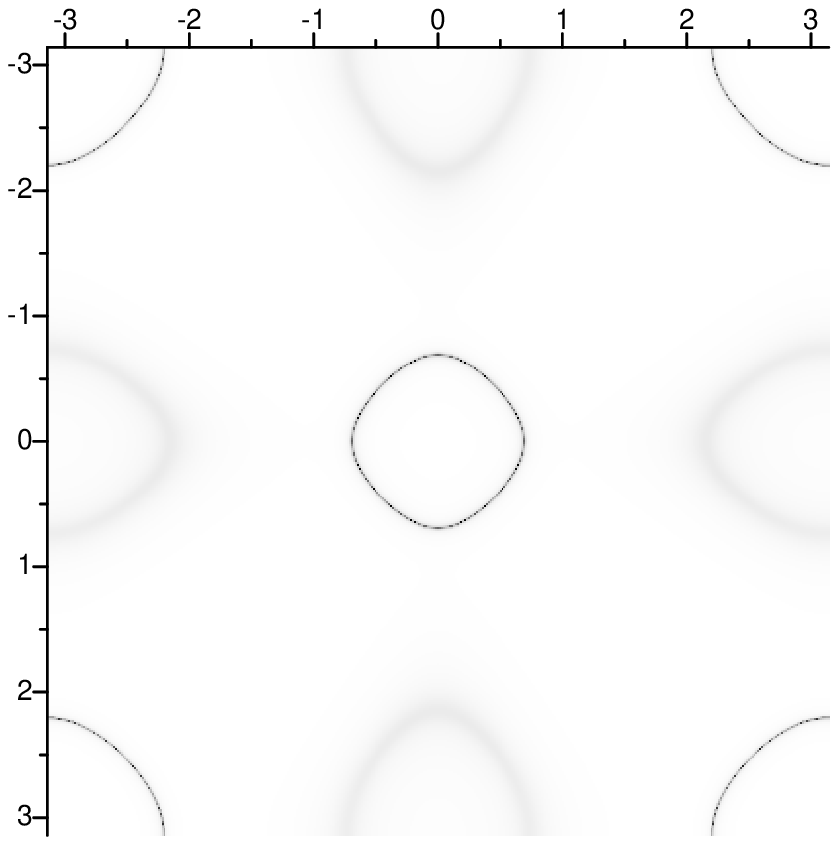}}
\subfigure[$\omega=0.07$]{
\includegraphics*[bb=5 5 260 260,width=0.2\textwidth]{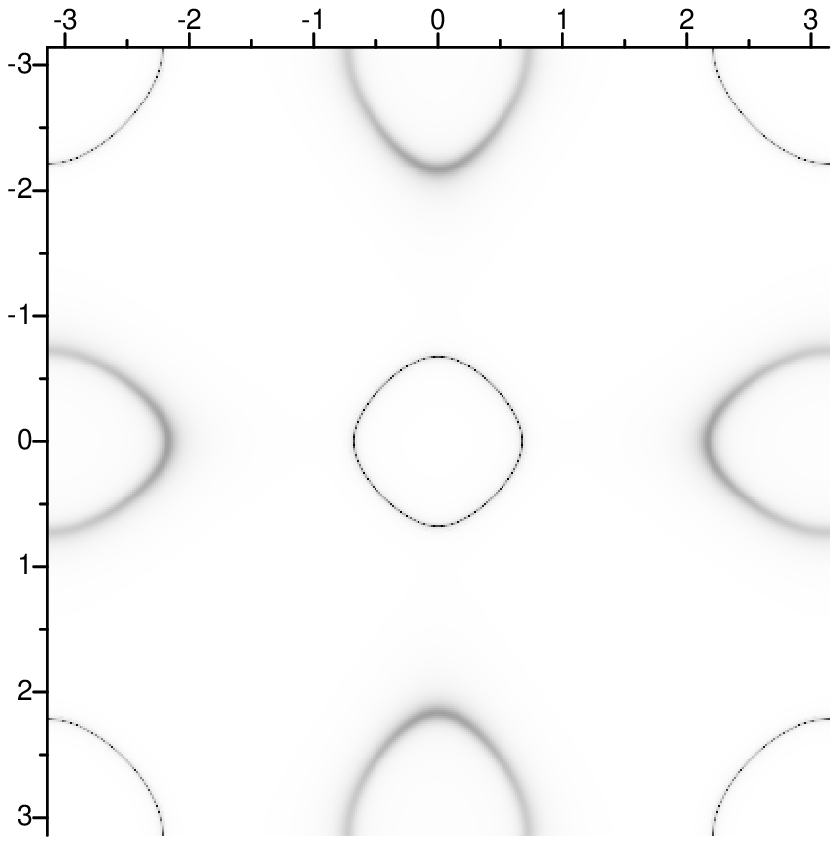}}
\subfigure[$\omega=0.08$]{
\includegraphics*[bb=5 5 260 260,width=0.2\textwidth]{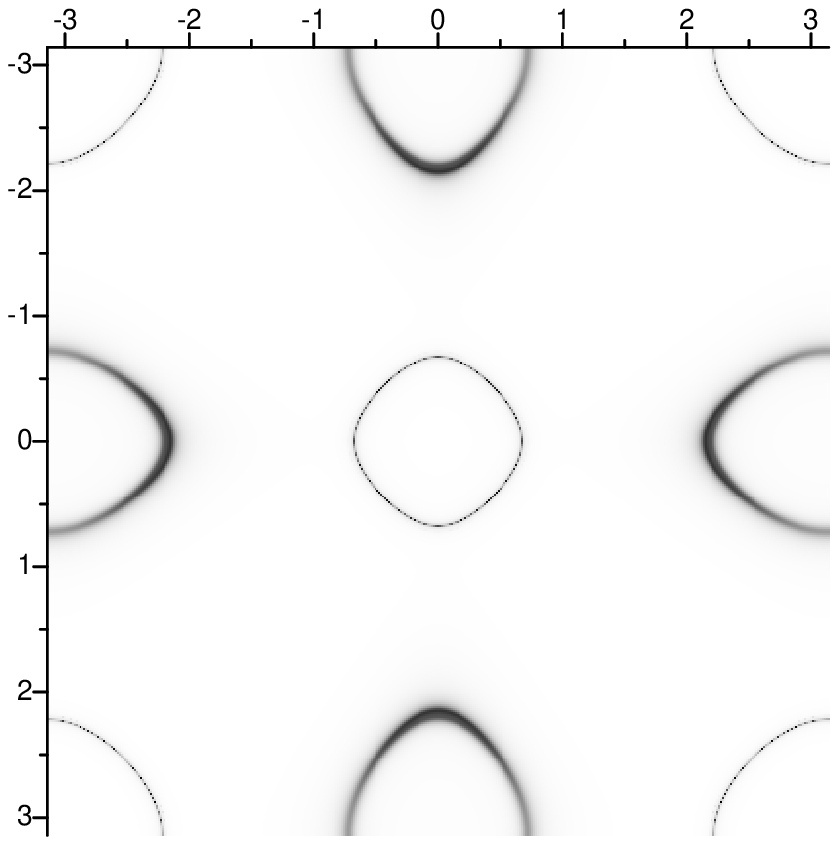}}
\subfigure[$\omega=0.09$]{
\includegraphics*[bb=5 5 260 260,width=0.2\textwidth]{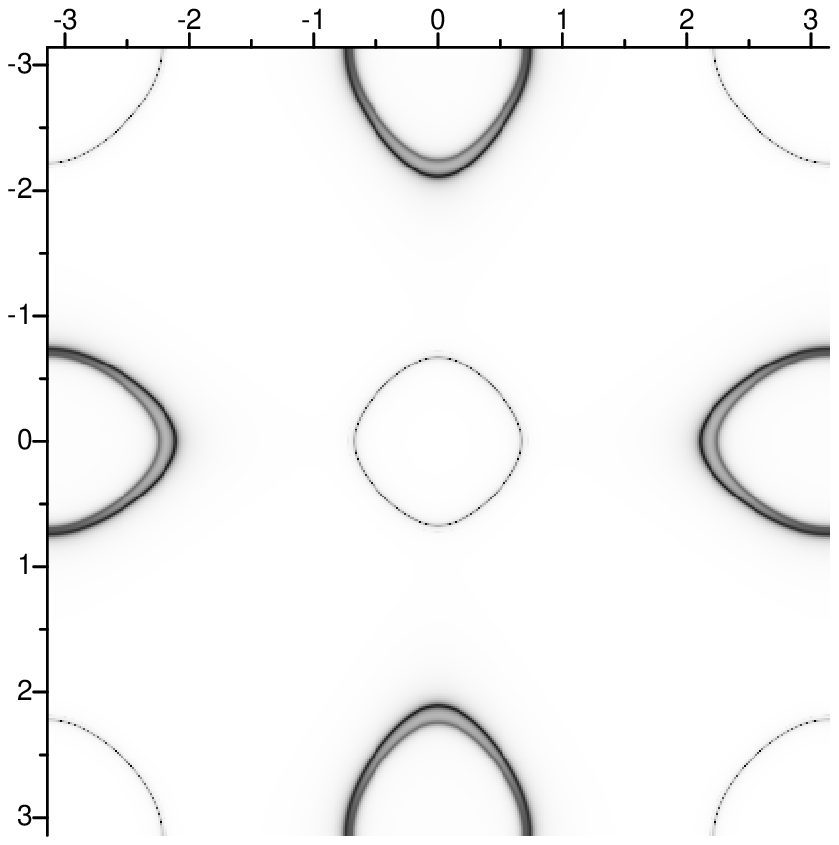}}
\caption{The spectral function $\mathcal{A}(\mathbf{k},\omega)$ for
$d_{x^2-y^2}$ pairing symmetry with $\Delta_0=0.1$. }
\label{FdContour}
\end{figure}

\begin{figure} [htbp]
\centering \subfigure[$\omega=-0.09$]{
\includegraphics*[bb=5 5 260 260,width=0.2\textwidth]{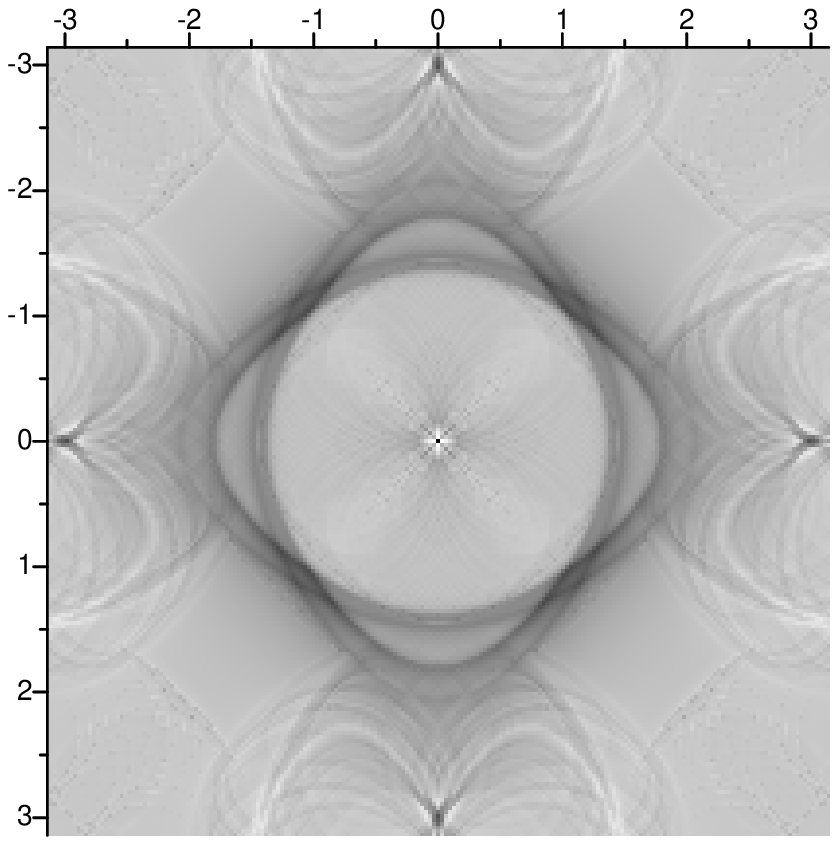}}
\subfigure[$\omega=-0.08$]{
\includegraphics*[bb=5 5 260 260,width=0.2\textwidth]{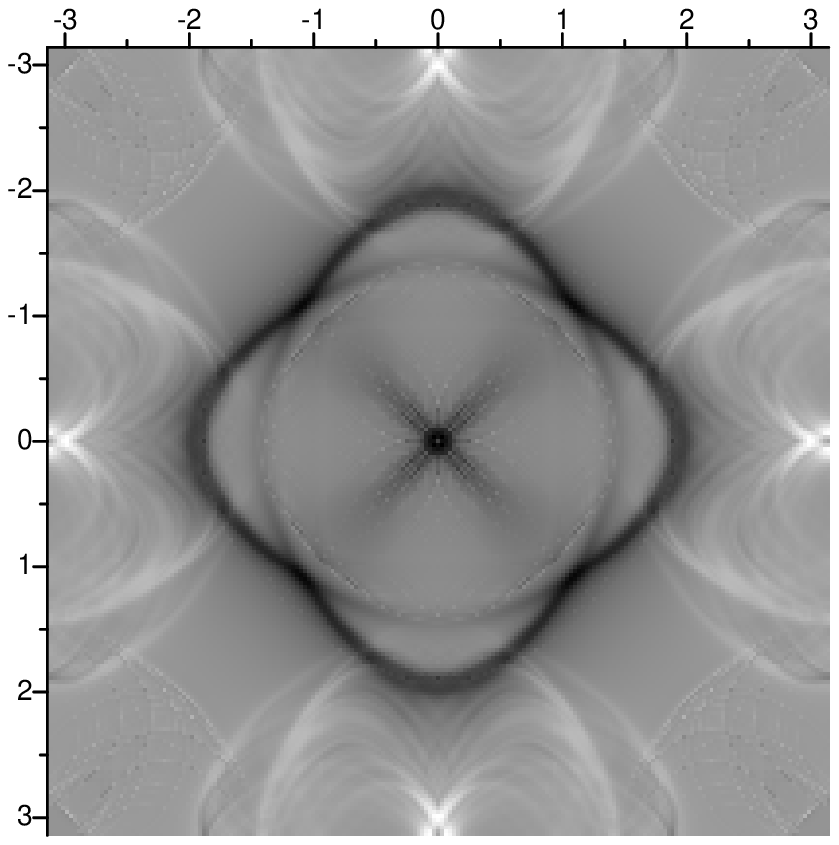}}
\subfigure[$\omega=-0.07$]{
\includegraphics*[bb=5 5 260 260,width=0.2\textwidth]{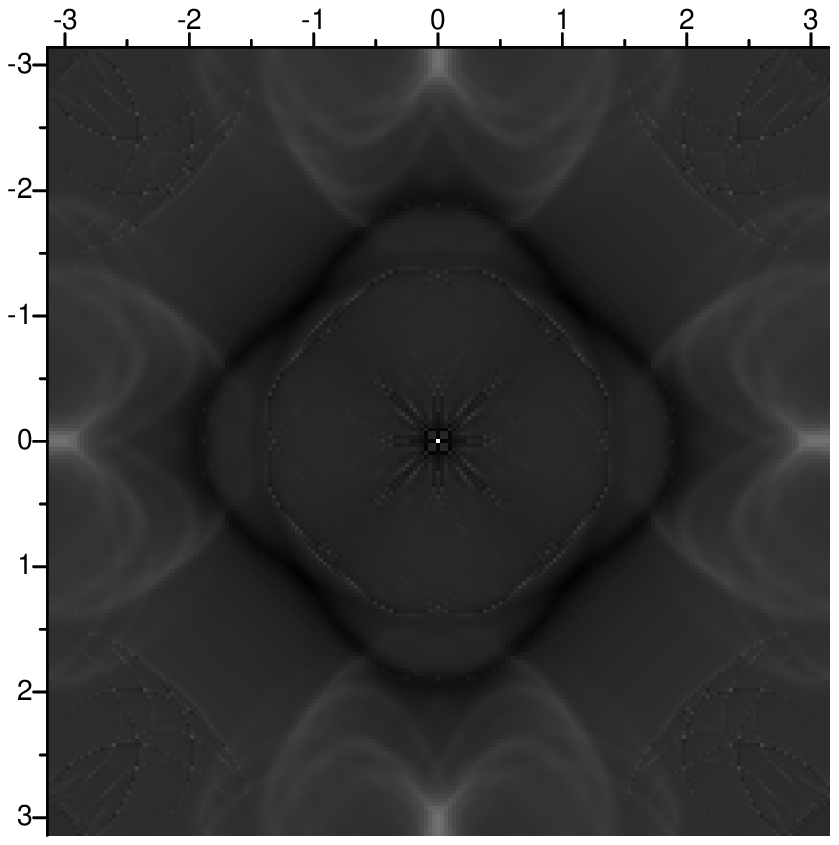}}
\subfigure[$\omega=-0.03$]{
\includegraphics*[bb=5 5 260 260,width=0.2\textwidth]{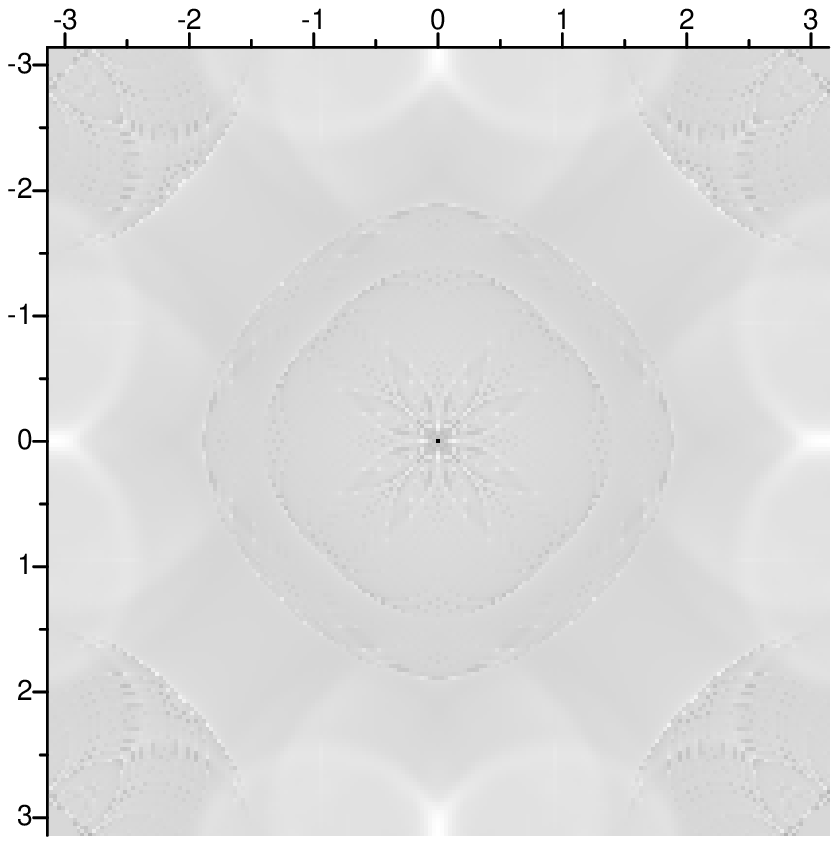}}
\subfigure[$\omega=0$]{
\includegraphics*[bb=5 5 260 260,width=0.2\textwidth]{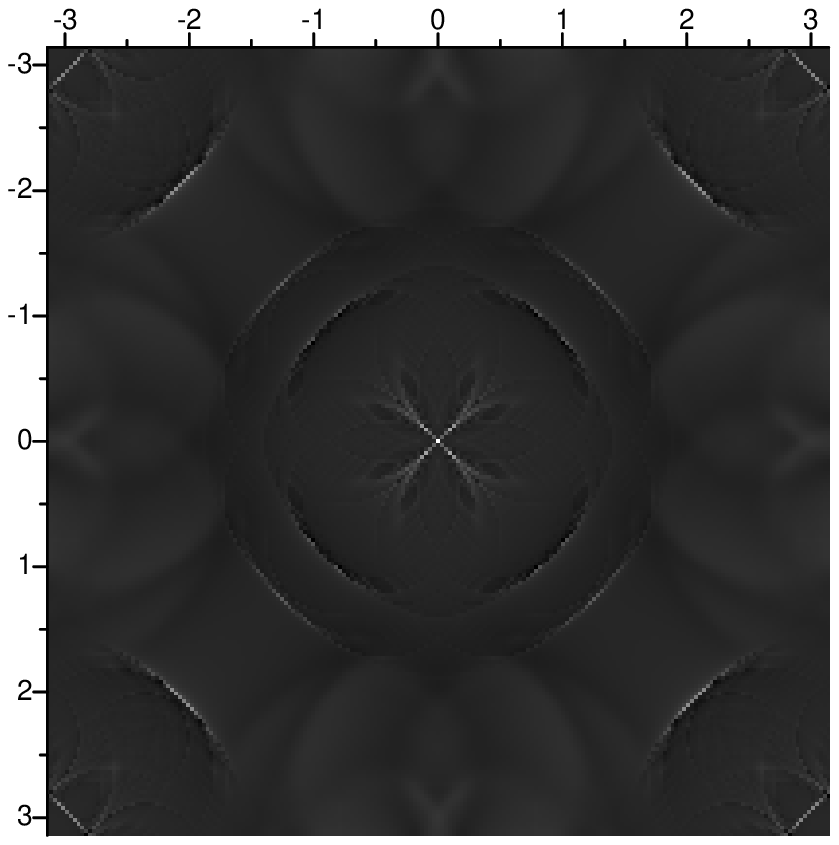}}
\subfigure[$\omega=0.03$]{
\includegraphics*[bb=5 5 260 260,width=0.2\textwidth]{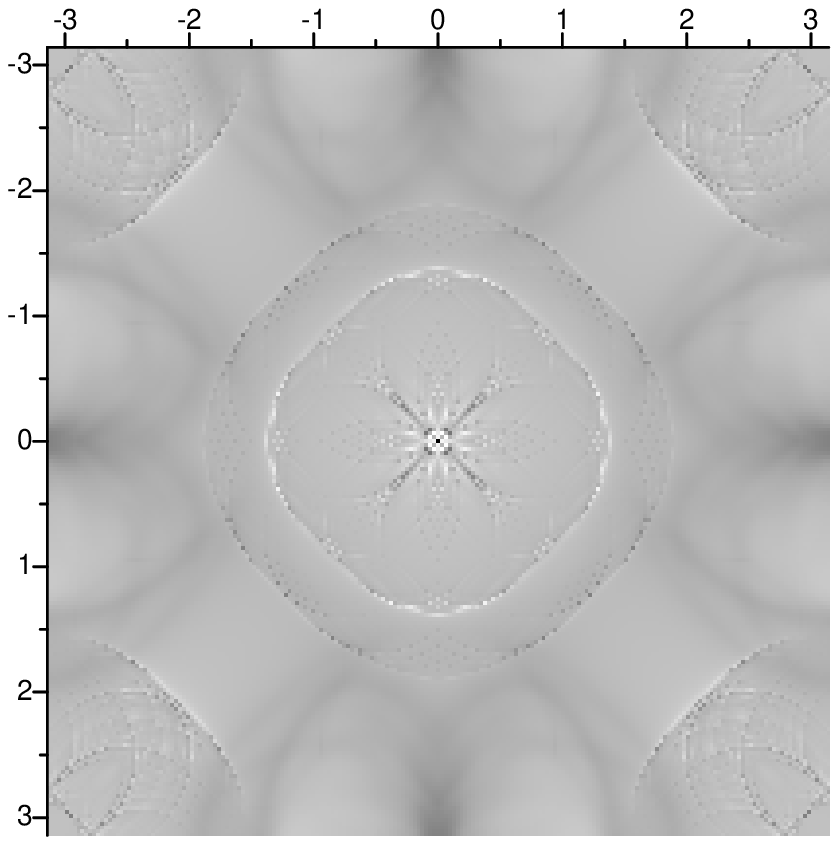}}
\subfigure[$\omega=0.07$]{
\includegraphics*[bb=5 5 260 260,width=0.2\textwidth]{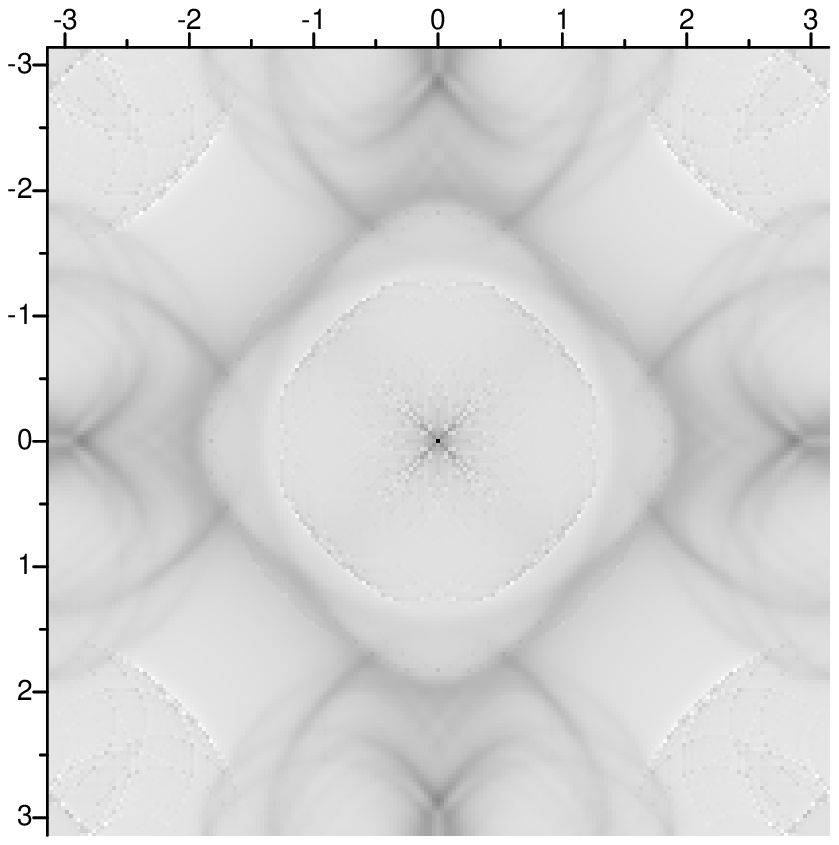}}
\subfigure[$\omega=0.08$]{
\includegraphics*[bb=5 5 260 260,width=0.2\textwidth]{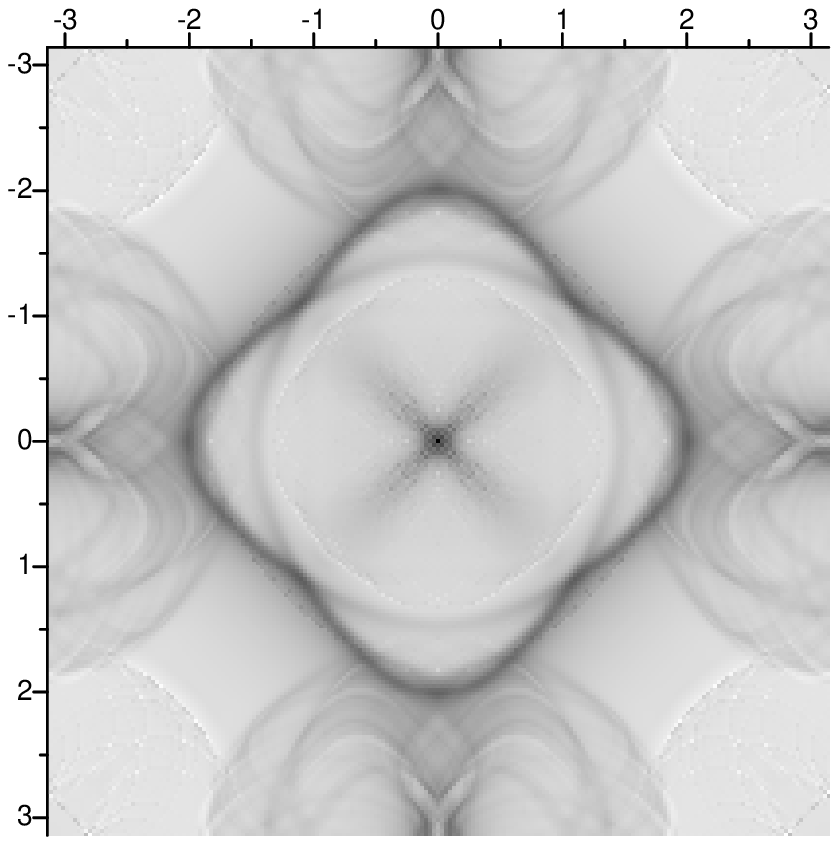}}
\subfigure[$\omega=0.09$]{
\includegraphics*[bb=5 5 260 260,width=0.2\textwidth]{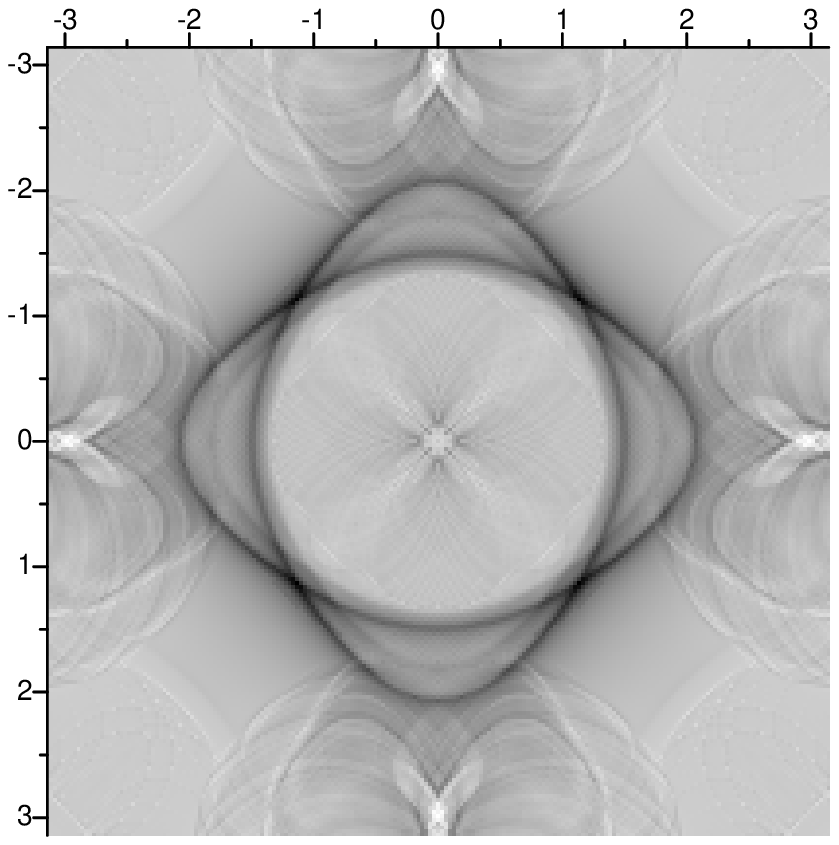}}
\caption{$\delta\rho(\mathbf{q})$ for non-magnetic impurity for
$d_{x^2-y^2}$ pairing symmetry with intra-orbital scattering,
$V_0=0.4$. } \label{FdNonM2D}
\end{figure}

\begin{figure} [htbp]
\centering \subfigure[$\omega=-0.09$]{
\includegraphics*[bb=5 5 260 260,width=0.2\textwidth]{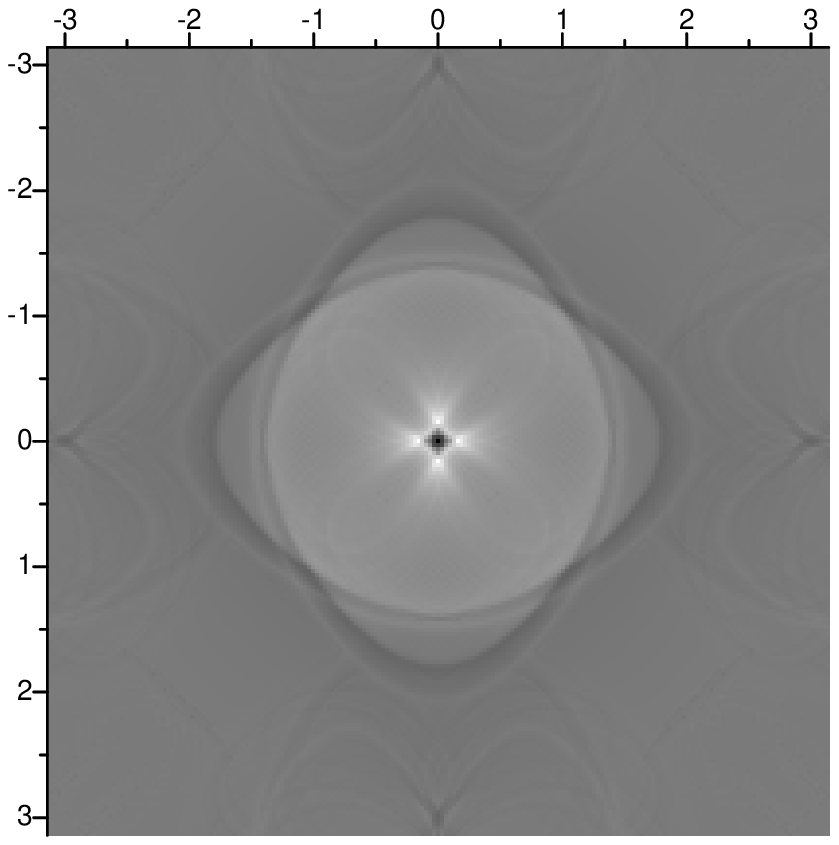}}
\subfigure[$\omega=-0.08$]{
\includegraphics*[bb=5 5 260 260,width=0.2\textwidth]{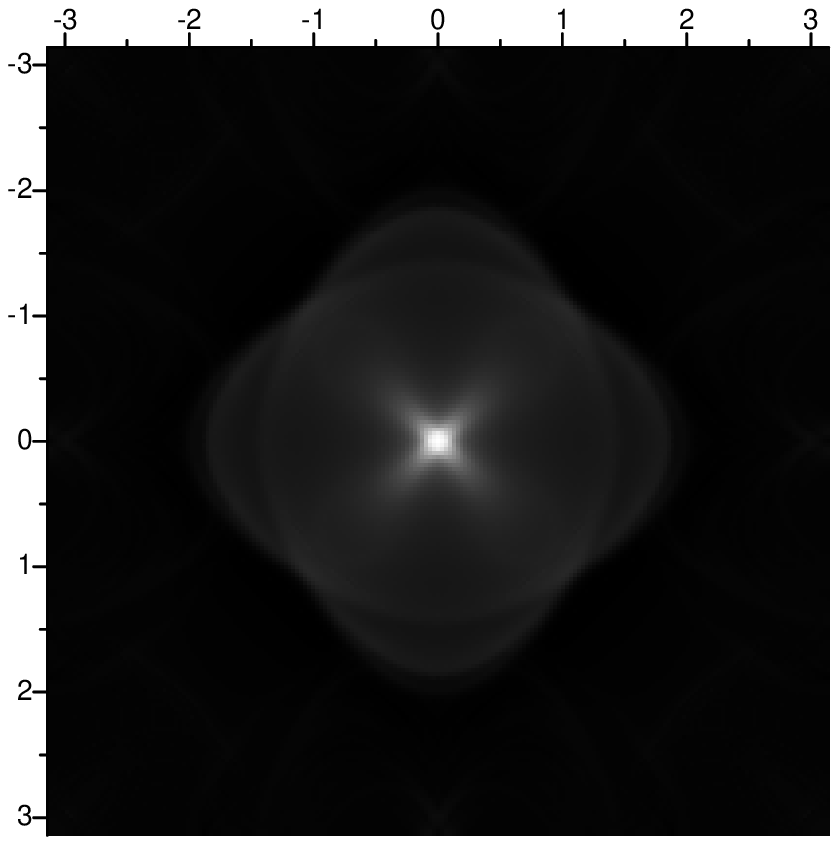}}
\subfigure[$\omega=-0.07$]{
\includegraphics*[bb=5 5 260 260,width=0.2\textwidth]{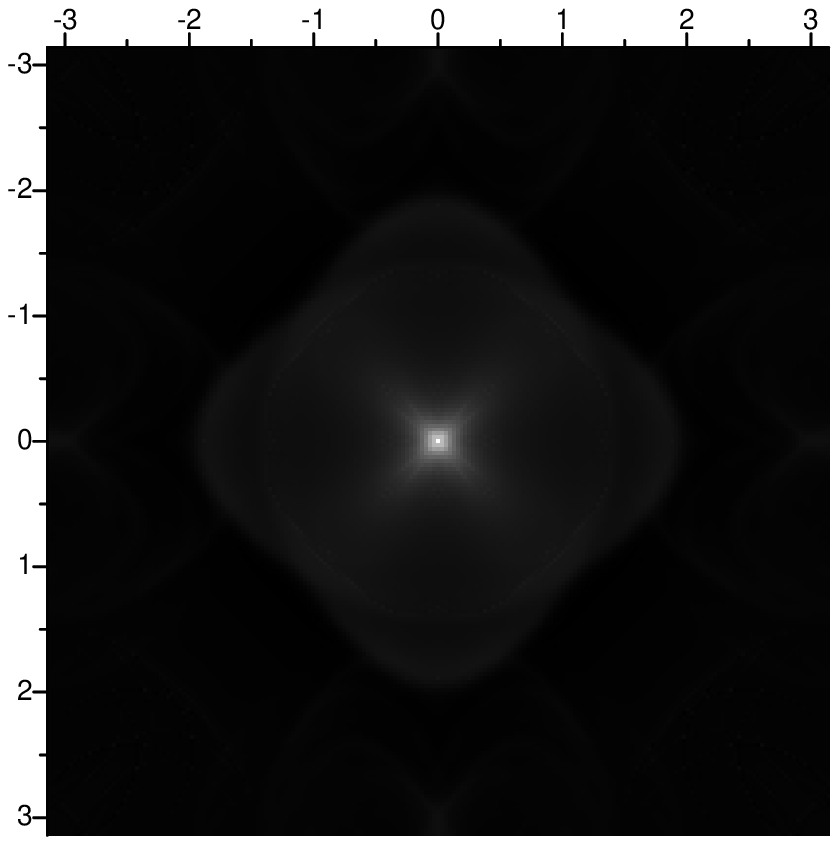}}
\subfigure[$\omega=-0.03$]{
\includegraphics*[bb=5 5 260 260,width=0.2\textwidth]{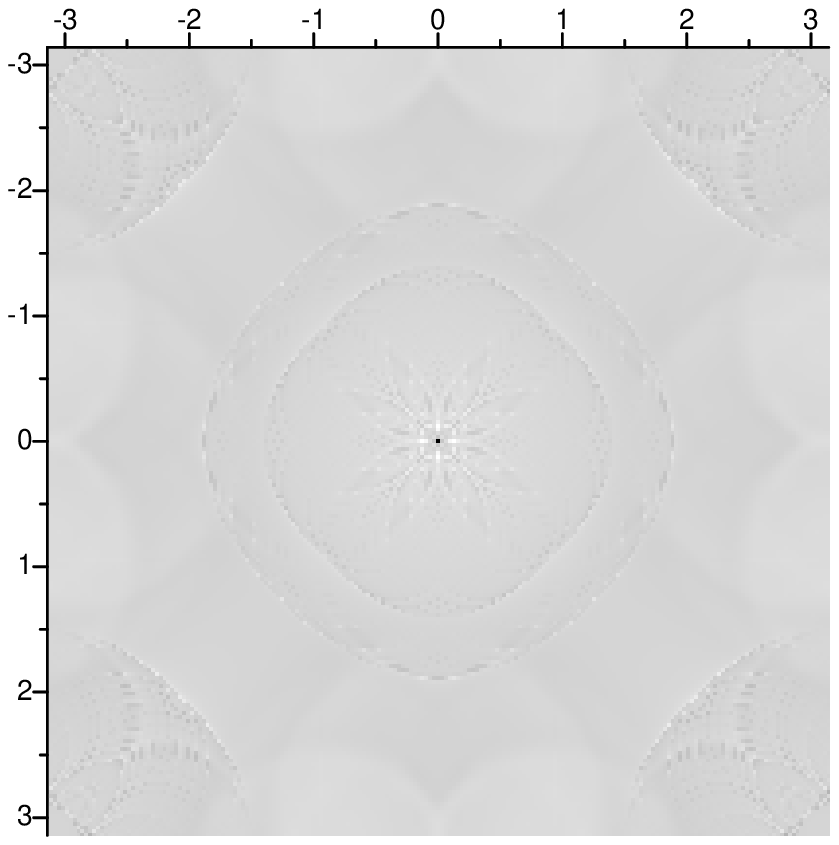}}
\subfigure[$\omega=0$]{
\includegraphics*[bb=5 5 260 260,width=0.2\textwidth]{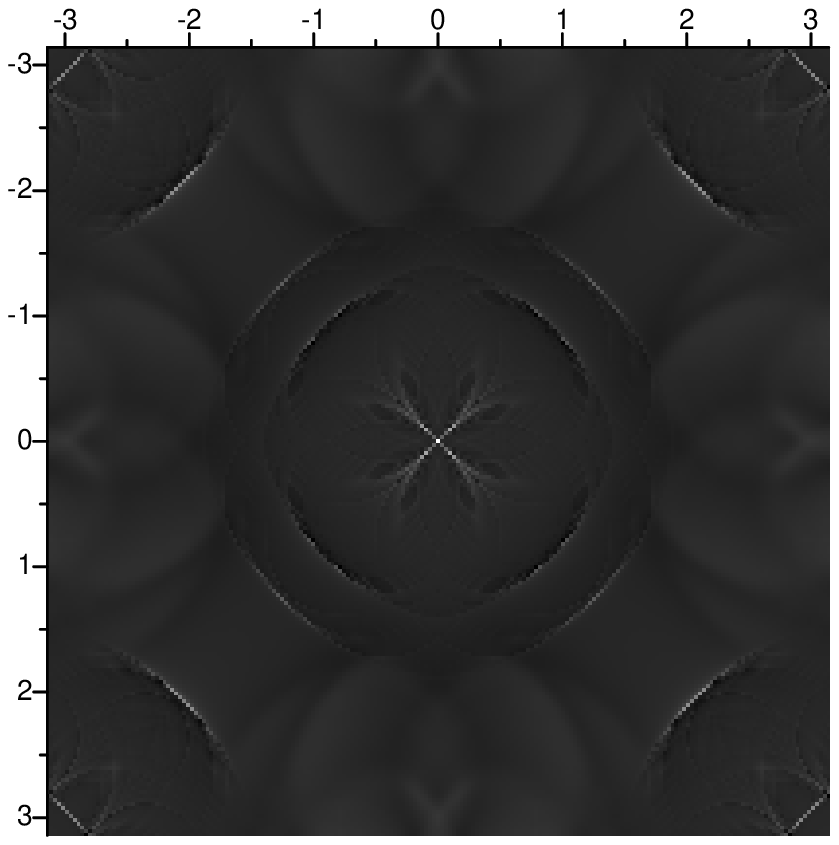}}
\subfigure[$\omega=0.03$]{
\includegraphics*[bb=5 5 260 260,width=0.2\textwidth]{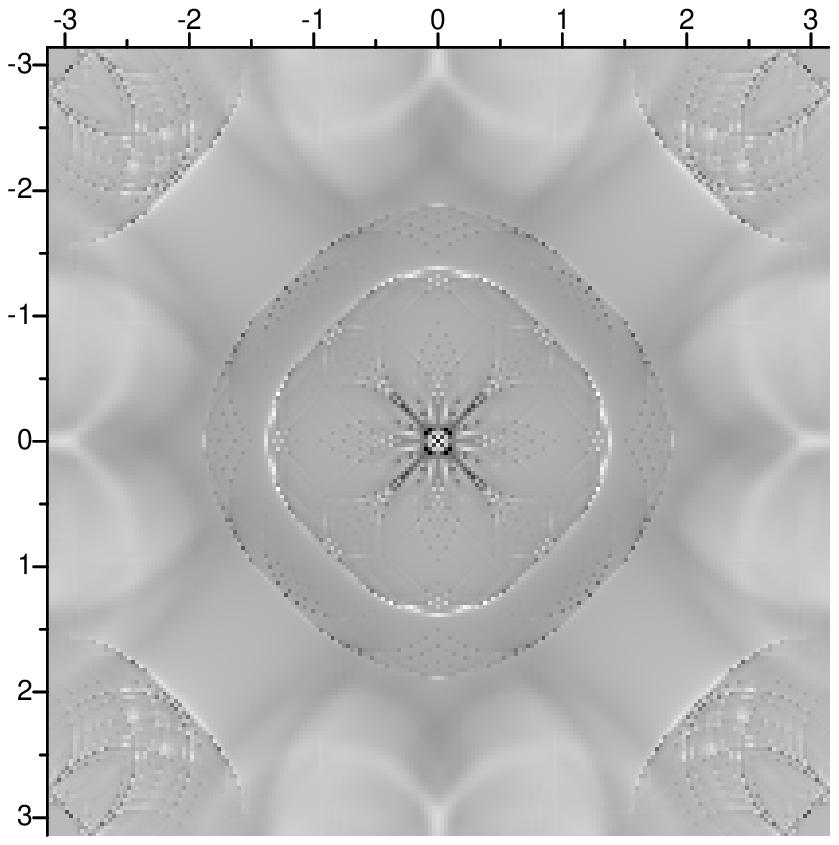}}
\subfigure[$\omega=0.07$]{
\includegraphics*[bb=5 5 260 260,width=0.2\textwidth]{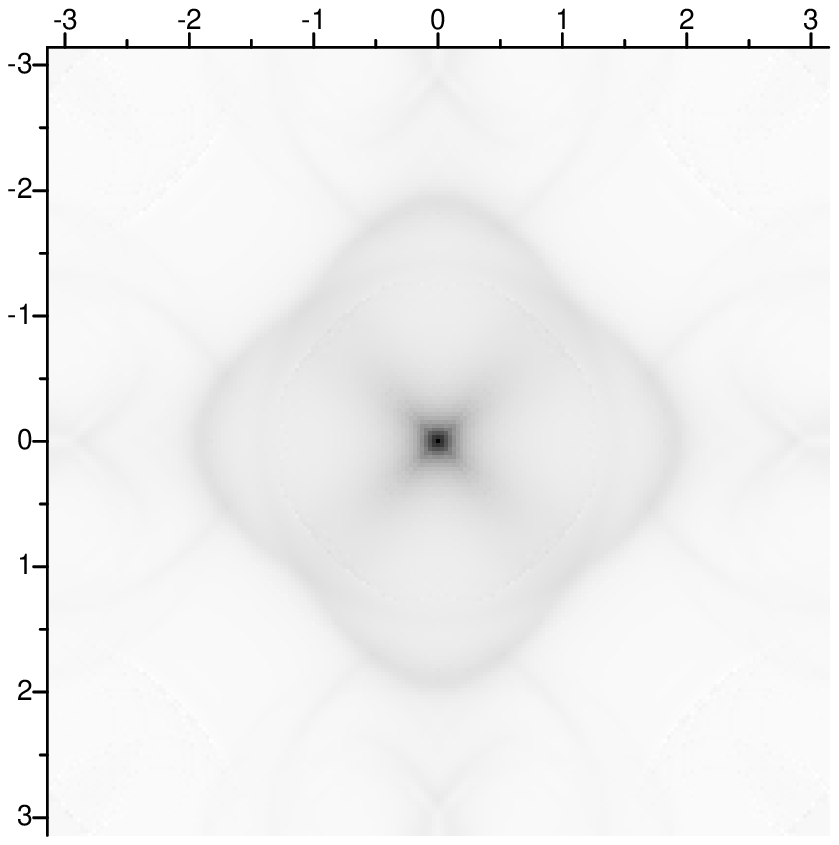}}
\subfigure[$\omega=0.08$]{
\includegraphics*[bb=5 5 260 260,width=0.2\textwidth]{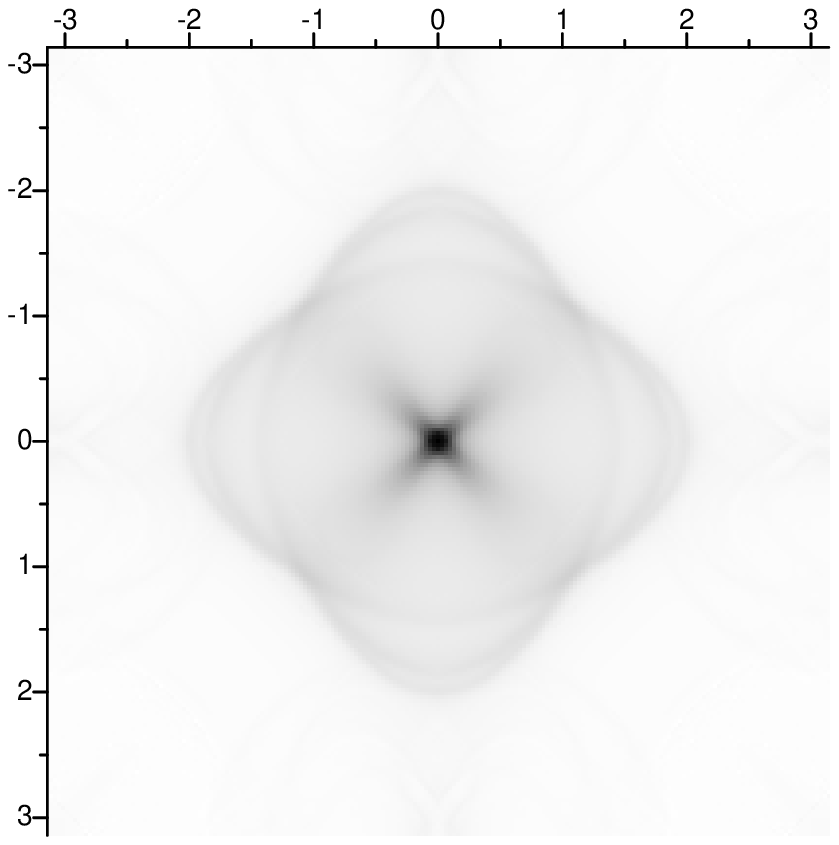}}
\subfigure[$\omega=0.09$]{
\includegraphics*[bb=5 5 260 260,width=0.2\textwidth]{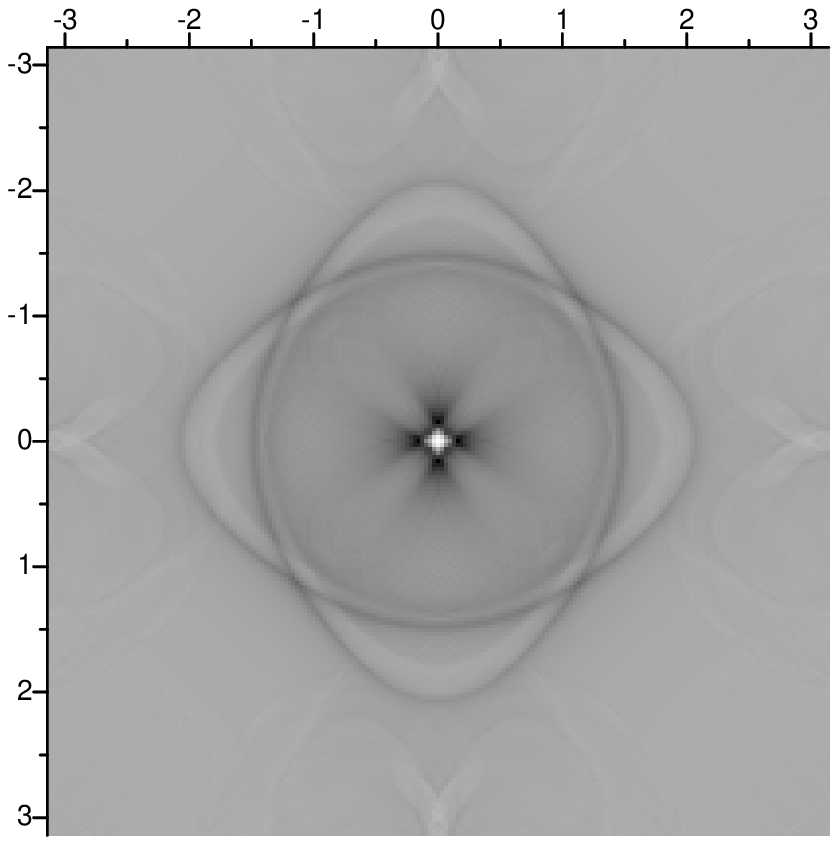}}
\caption{The same with Fig. \ref{FdNonM2D}, but for magnetic
impurity.} \label{FdM2D}
\end{figure}

We briefly discuss the $d_{x^2-y^2}$ pairing symmetry as an example
of a case with gapless nodal quasiparticles. The energy contours and
interference patterns are plotted in Fig. \ref{FdContour}, Fig.
\ref{FdNonM2D} and Fig. \ref{FdM2D}. Besides the robust intra-pocket
scattering around $\Gamma$ and the impurity-type sensitive peaks
near $(\pm\pi,0)$ and $(0,\pm\pi)$, the most specific feature is the
finite DOS within the pseudo-gap $\omega\in(-0.05,0.05)$, giving
rise to small but finite QPI in this region, as can be seen in (e),
(f) and (g) of Figs. \ref{FdNonM2D} and \ref{FdM2D}. This originates
from the intra-hole pocket scattering. When $|\omega|\rightarrow0$,
the QPI concentrates on the diagonal directions as expected from the
band structures in Fig \ref{FdContour}.

\subsection{Analytical analysis}
We now try to understand the above numerical
results analytically. The two-orbital Hamiltonian can be
diagonalized by a unitary transformation
\begin{eqnarray}
U^\dag(\mathbf{k})\left(%
\begin{array}{cc}
  \epsilon_x(\mathbf{k}) & \epsilon_{xy}(\mathbf{k}) \\
  \epsilon_{xy}(\mathbf{k}) & \epsilon_y(\mathbf{k}) \\
\end{array}%
\right)U(\mathbf{k})=\left(%
\begin{array}{cc}
  \epsilon_1(\mathbf{k}) & 0 \\
  0 & \epsilon_2(\mathbf{k}) \\
\end{array}%
\right), \end{eqnarray} where\bea U=\left(%
\begin{array}{cc}
  \cos(\theta_{\mathbf{k}}/2) & -\sin(\theta_{\mathbf{k}}/2) \\
  \sin(\theta_{\mathbf{k}}/2) & \cos(\theta_{\mathbf{k}}/2) \\
\end{array}%
\right),\eea and
$\tan(\theta_{\mathbf{k}})=\frac{2\epsilon_{xy}(\mathbf{k})}{\epsilon_x(\mathbf{k})-\epsilon_y(\mathbf{k})}$.
The integral over BZ in calculating the $T$-matrix in
Eq.~(\ref{eq6}) makes the analytical treatment untractable.
Fortunately we numerically verified that for the strength of
impurity in this work ($V_0=0.4$), the first order expansion in
Eq.~(\ref{eq5b}) is sufficiently precise (with error less than
$2\%$). In the following, we safely take $T=V$. In the
band-basis, the $V$-matrix for intra-orbital impurity is in the
following $\mathbf{k}$-dependent
form\bea V(\mathbf{k},\mathbf{k}')=\left(%
\begin{array}{cc}
  \cos(\frac{\theta_{\mathbf{k}}-\theta_{\mathbf{k}'}}{2}) & \sin(\frac{\theta_{\mathbf{k}}-\theta_{\mathbf{k}'}}{2}) \\
  -\sin(\frac{\theta_{\mathbf{k}}-\theta_{\mathbf{k}'}}{2}) & \cos(\frac{\theta_{\mathbf{k}}-\theta_{\mathbf{k}'}}{2}) \\
\end{array}%
\right)\otimes \left(%
\begin{array}{cc}
  V_0 & 0 \\
  0 & \pm V_0 \\
\end{array}%
\right),\eea where the upper (lower) sign is for the magnetic
(non-magnetic) impurity, as in Eqs.~(\ref{eq9a}) and (\ref{eq9b}). The
induced Green's function in Eq.~(\ref{eq4}) can now be transformed to the
band representation. After a lengthy but straightforward calculation,
we obtain
\begin{eqnarray}
&&\delta\rho(\mathbf{q})=\left[P_1(\mathbf{k},\mathbf{q})+P_2(\mathbf{k},\mathbf{q})\right] V_0\cos^2(\frac{\theta_{\mathbf{k}}-\theta_{\mathbf{k}+\mathbf{q}}}{2})\nonumber\\
&&+\left[Q_1(\mathbf{k},\mathbf{q})+Q_2(\mathbf{k},\mathbf{q})\right]V_0\sin^2(\frac{\theta_\mathbf{k}-\theta_{\mathbf{k}+\mathbf{q}}}{2}),
\label{eq14}
\end{eqnarray}
where $P_i$ and $Q_i$ denote the contributions from {\it
intra-pocket} and {\it inter-pocket} scattering respectively, and
take the form
\begin{widetext}
\begin{eqnarray}
P_{1(2)}(\mathbf{k},\mathbf{q})&=&\frac{(\omega+\epsilon_{1(2)}(\mathbf{k}))(\omega+\epsilon_{1(2)}(\mathbf{k}+\mathbf{q}))\pm\Delta(\mathbf{k})\Delta(\mathbf{k}+\mathbf{q})}{(\omega^2-\Delta(\mathbf{k})^2-\epsilon_{1(2)}(\mathbf{k})^2)(\omega^2-\Delta(\mathbf{k}+\mathbf{q})^2-\epsilon_{1(2)}(\mathbf{k}+\mathbf{q})^2)},\nonumber\\
Q_{1(2)}(\mathbf{k},\mathbf{q})&=&\frac{(\omega+\epsilon_{1(2)}(\mathbf{k}))(\omega+\epsilon_{2(1)}(\mathbf{k}+\mathbf{q}))\pm\Delta(\mathbf{k})\Delta(\mathbf{k}+\mathbf{q})}{(\omega^2-\Delta(\mathbf{k})^2-\epsilon_{1(2)}(\mathbf{k})^2)(\omega^2-\Delta(\mathbf{k}+\mathbf{q})^2-\epsilon_{2(1)}(\mathbf{k}+\mathbf{q})^2)}.
\label{eq15}
\end{eqnarray}
\end{widetext}
From these expressions we see that, for
$\omega\sim\pm\Delta(k_f)$, and on the energy contour $\epsilon_i(k)\sim0$, the magnetic impurity
contribution is almost zero if $q\sim(\pi,0)$  for the sign-changing
s-wave. The counterpart  is  true if
$\Delta$ does not change sign: the contribution by magnetic impurity
is now much larger than that by non-magnetic impurity. Furthermore,
at $q\sim0$ the contribution by magnetic impurity is much larger
than the non-magnetic impurity. These are consistent with our
numerical results and previous theoretical argument\cite{Wang2009}.

Now let us turn to the inter-orbital case. After the same process, one obtains\bea V(\mathbf{k},\mathbf{k}')=\left(%
\begin{array}{cc}
  \cos(\frac{\theta_{\mathbf{k}}+\theta_{\mathbf{k}'}}{2}) & \sin(\frac{\theta_{\mathbf{k}}+\theta_{\mathbf{k}'}}{2}) \\
  -\sin(\frac{\theta_{\mathbf{k}}+\theta_{\mathbf{k}'}}{2}) & \cos(\frac{\theta_{\mathbf{k}}+\theta_{\mathbf{k}'}}{2}) \\
\end{array}%
\right)\otimes\left(%
\begin{array}{cc}
  V_0 & 0 \\
  0 & \pm V_0 \\
\end{array}%
\right).\eea The result is
\begin{eqnarray}
&&\delta\rho(\mathbf{q})=\left[P_1(\mathbf{k},\mathbf{q})+P_2(\mathbf{k},\mathbf{q})\right] V_0\frac{\cos(\theta_\mathbf{k})+\cos(\theta_{\mathbf{k}+\mathbf{q}})}{2}\nonumber\\
&&+\left[Q_1(\mathbf{k},\mathbf{q})+Q_2(\mathbf{k},\mathbf{q})\right]V_0\frac{\cos(\theta_\mathbf{k})-\cos(\theta_{\mathbf{k}+\mathbf{q}})}{2},
\end{eqnarray}
We notice that around the $\Gamma$ point Fermi
surface, $\theta_{\mathbf{k}}$ changes from 0 to $4\pi$; around the M
point, $\theta_{\mathbf{k}}$ changes from 0 to $\pi$ and then back to 0.
In either case, we have $\int_{k_f}\cos(\theta_k)d^2k\sim0$, and
therefore the inter-orbit scattering is always much smaller than the
intra-orbit scattering, which is also seen in Fig.
\ref{FIntraInter}.

\section{Five-orbital model and numerical results}

\begin{figure} [htbp]
\begin{center}
\includegraphics[bb=20 10 310 240,width=0.4\textwidth]{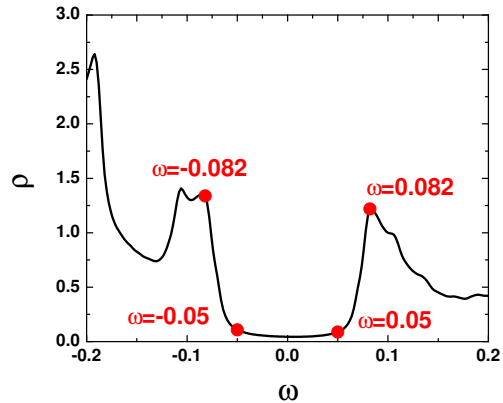}
\end{center}
\caption{ Bulk density of states $\rho$ for five-orbital model with
$s_{x^2y^2}$ pairing as a function of $\omega$, $\Delta_0=0.1$ and
no impurities.  The energy broadening width $\delta=0.002$.}
\label{FiveDOS}
\end{figure}

A better fit to the LDA band structure in iron-based superconductors
is given by  a  five orbital proposed in \cite{Kuroki2008}. To
investigate the model-dependence of the scattering patterns, we now
perform all the above calculations employing this five-orbital model
augmented by an intra-orbital $s_{x^2y^2}$ pairing symmetry.

\begin{figure} [htbp]
\centering \subfigure[$\omega=-0.20$]{
\includegraphics*[bb=5 5 260 260,width=0.2\textwidth]{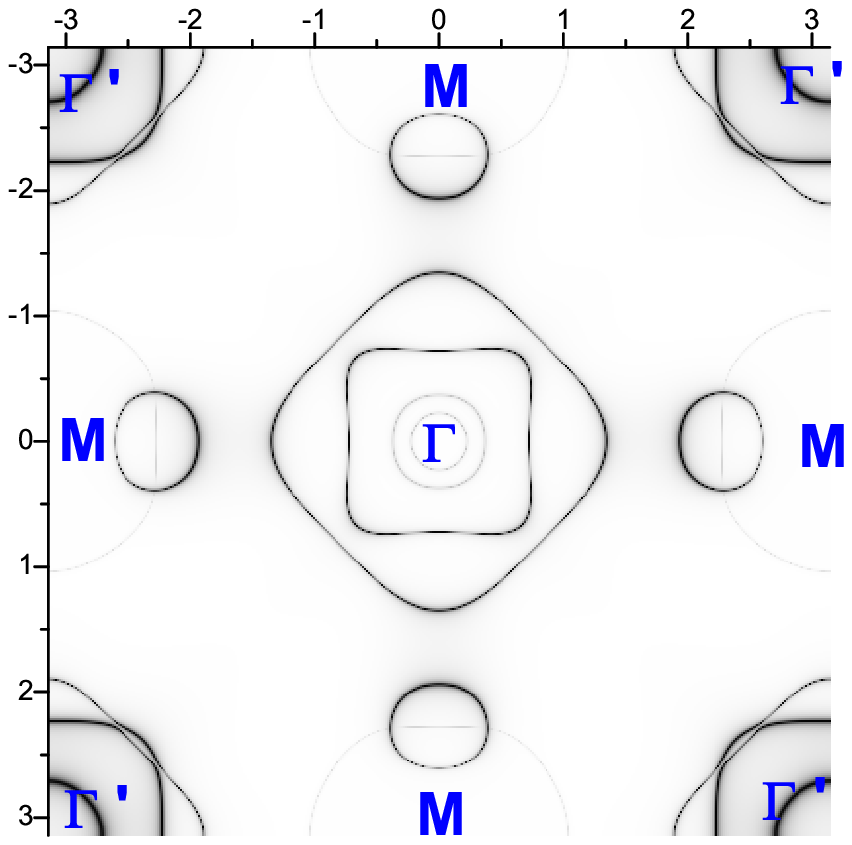}}
\subfigure[$\omega=-0.09$]{
\includegraphics*[bb=5 5 260 260,width=0.2\textwidth]{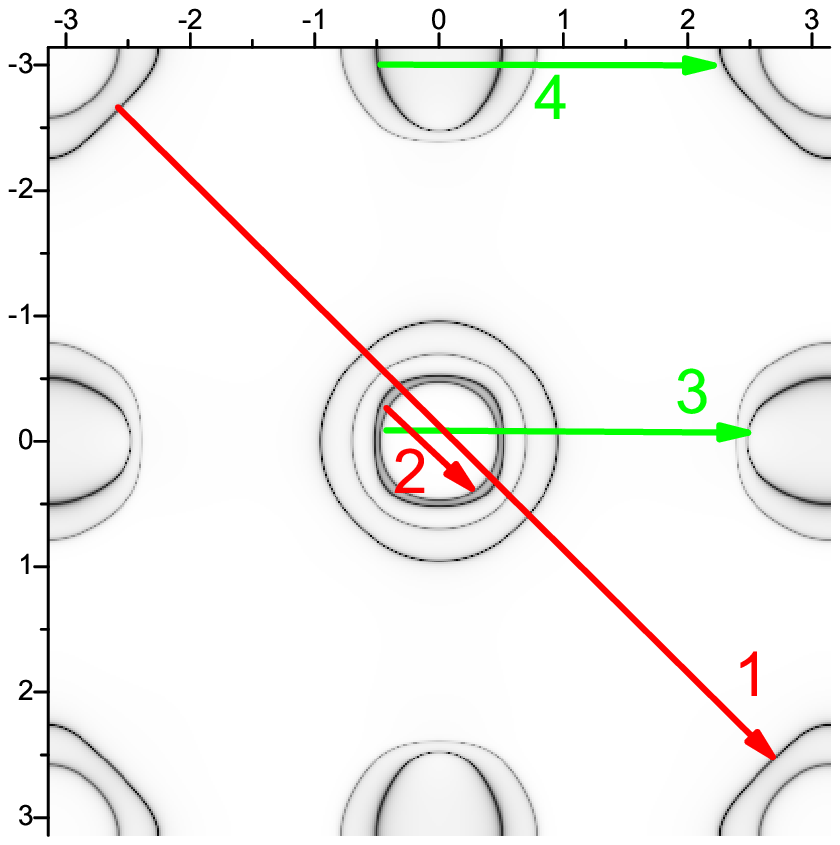}}
\subfigure[$\omega=-0.08$]{
\includegraphics*[bb=5 5 260 260,width=0.2\textwidth]{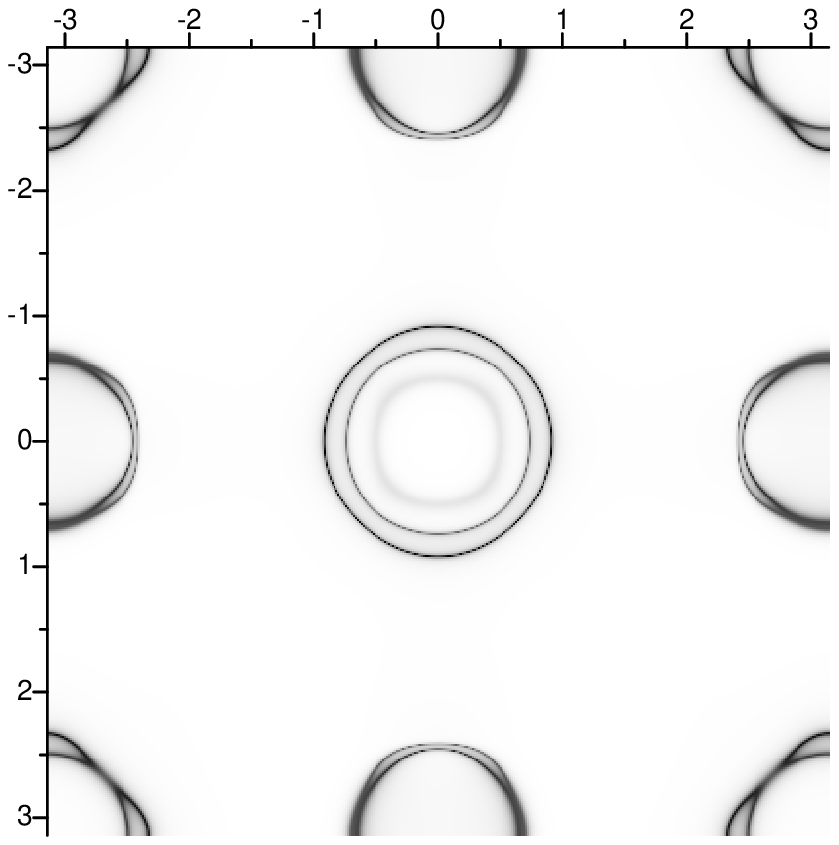}}
\subfigure[$\omega=-0.07$]{
\includegraphics*[bb=5 5 260 260,width=0.2\textwidth]{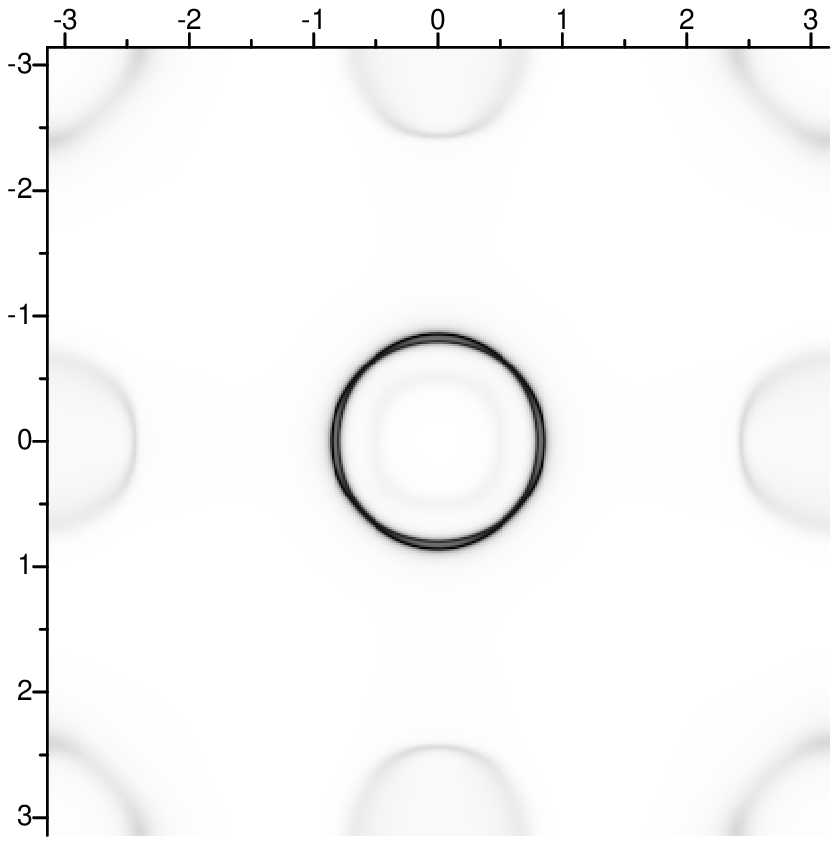}}
\subfigure[$\omega=0.07$]{
\includegraphics*[bb=5 5 260 260,width=0.2\textwidth]{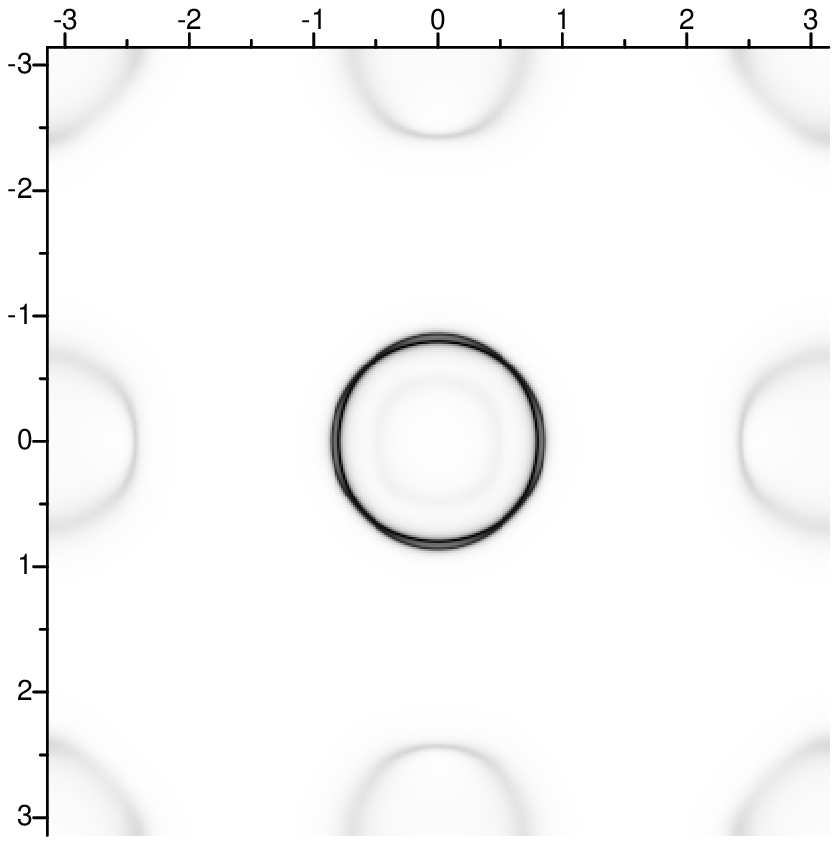}}
\subfigure[$\omega=0.08$]{
\includegraphics*[bb=5 5 260 260,width=0.2\textwidth]{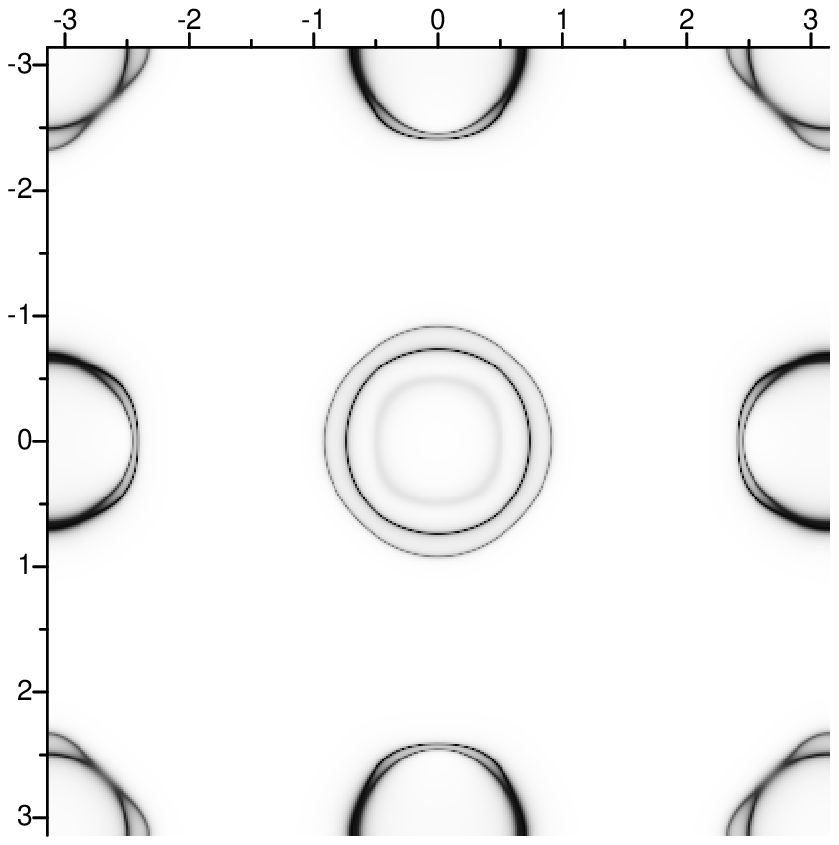}}
\subfigure[$\omega=0.09$]{
\includegraphics*[bb=5 5 260 260,width=0.2\textwidth]{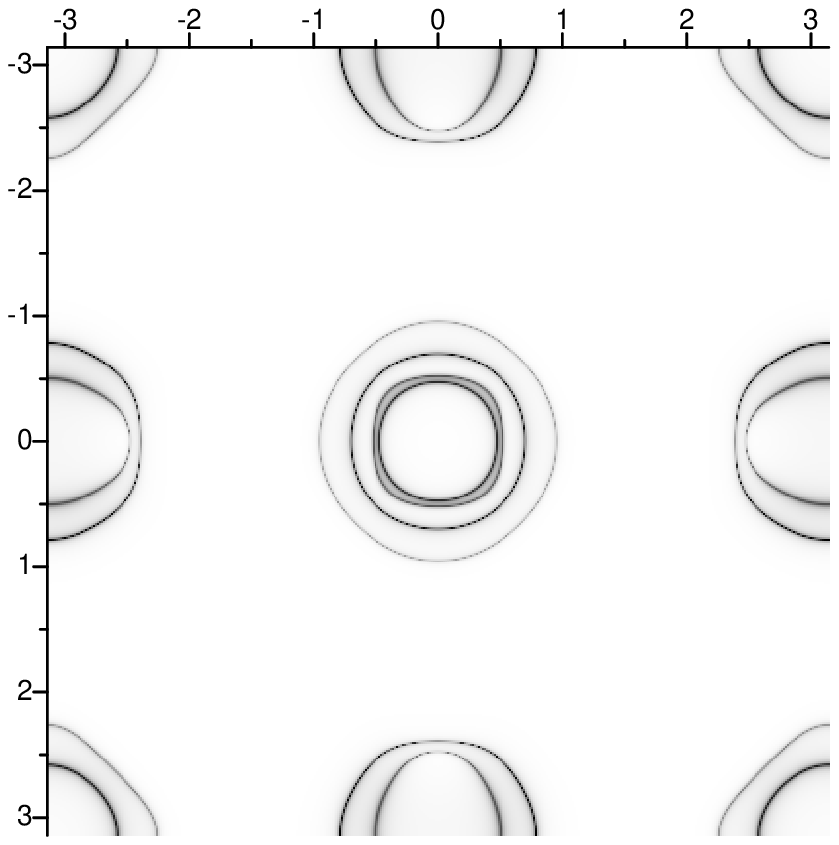}}
\subfigure[$\omega=0.20$]{
\includegraphics*[bb=5 5 260 260,width=0.2\textwidth]{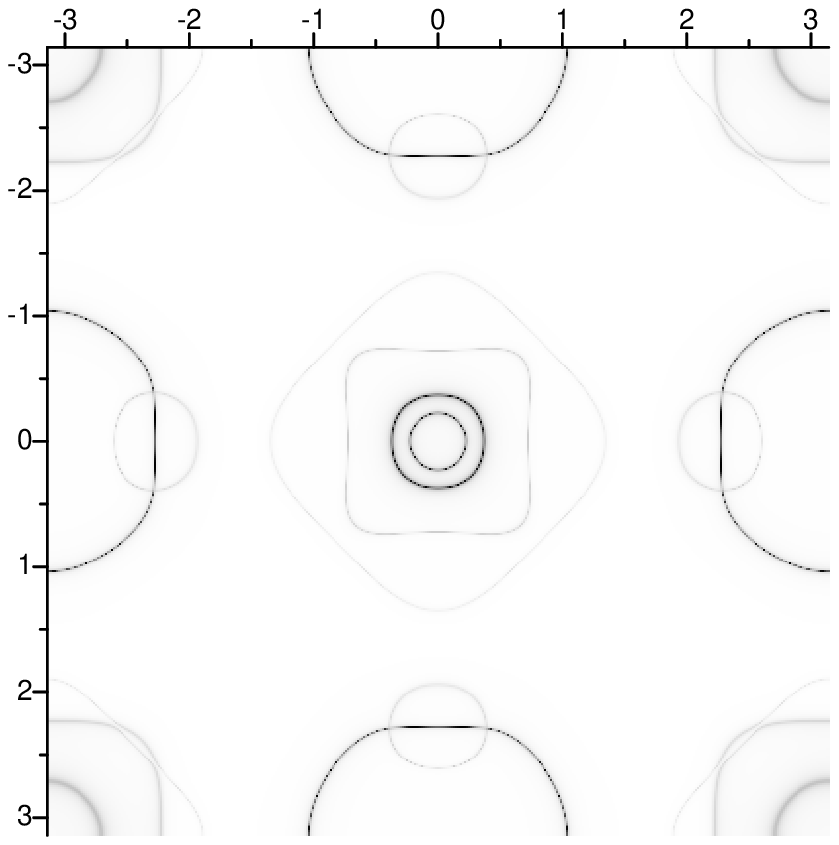}}
\caption{The spectral function $\mathcal{A}(\mathbf{k},\omega)$ in
the unfolded Brillioun zone, for five-orbital model with
$\Delta_0=0.1$. Darker regions correspond to larger values of
$\mathcal{A}$ hence larger DOS in $\mathbf{k}$ space.}
\label{FContourFive}
\end{figure}

\begin{figure} [htbp]
\centering \subfigure[$\omega=-0.2$]{
\includegraphics*[bb=5 5 260 260,width=0.2\textwidth]{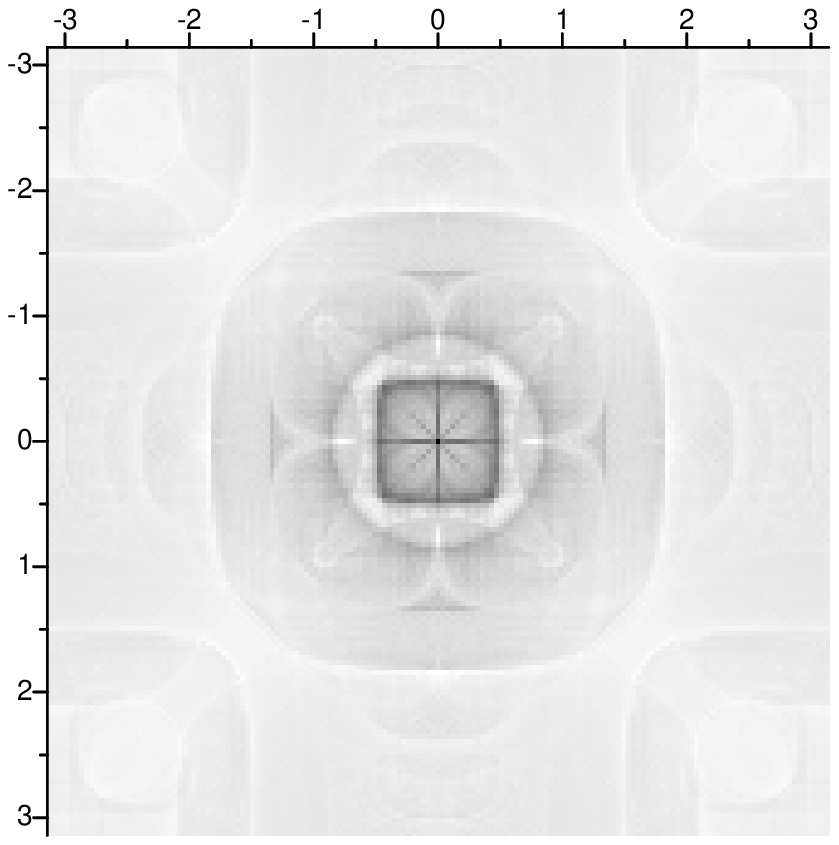}}
\subfigure[$\omega=-0.09$]{
\includegraphics*[bb=5 5 260 260,width=0.2\textwidth]{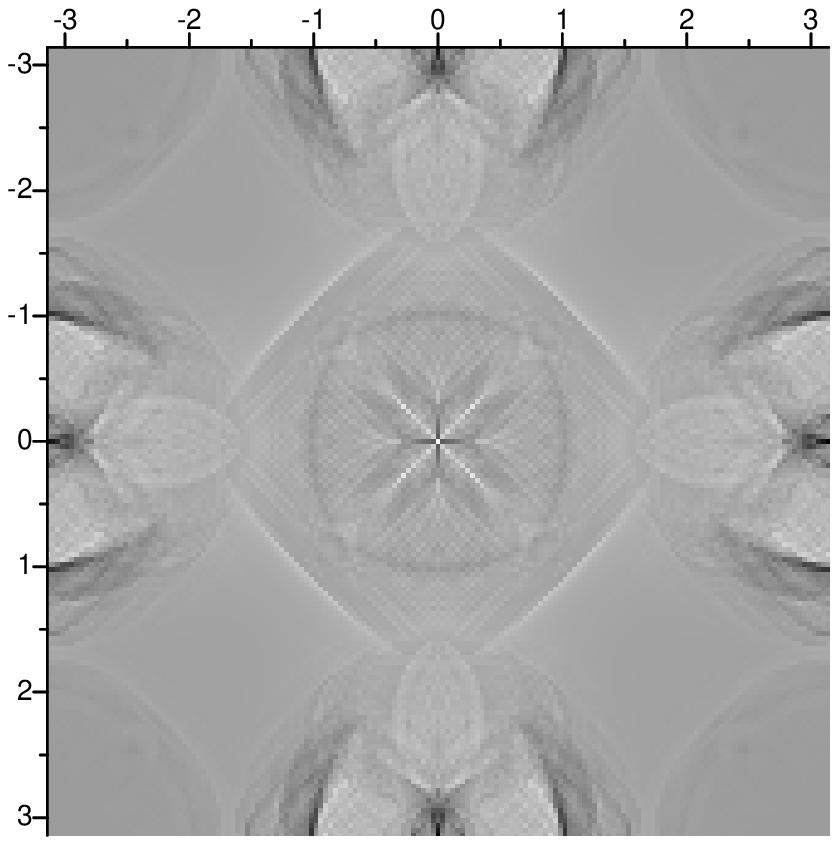}}
\subfigure[$\omega=-0.08$]{
\includegraphics*[bb=5 5 260 260,width=0.2\textwidth]{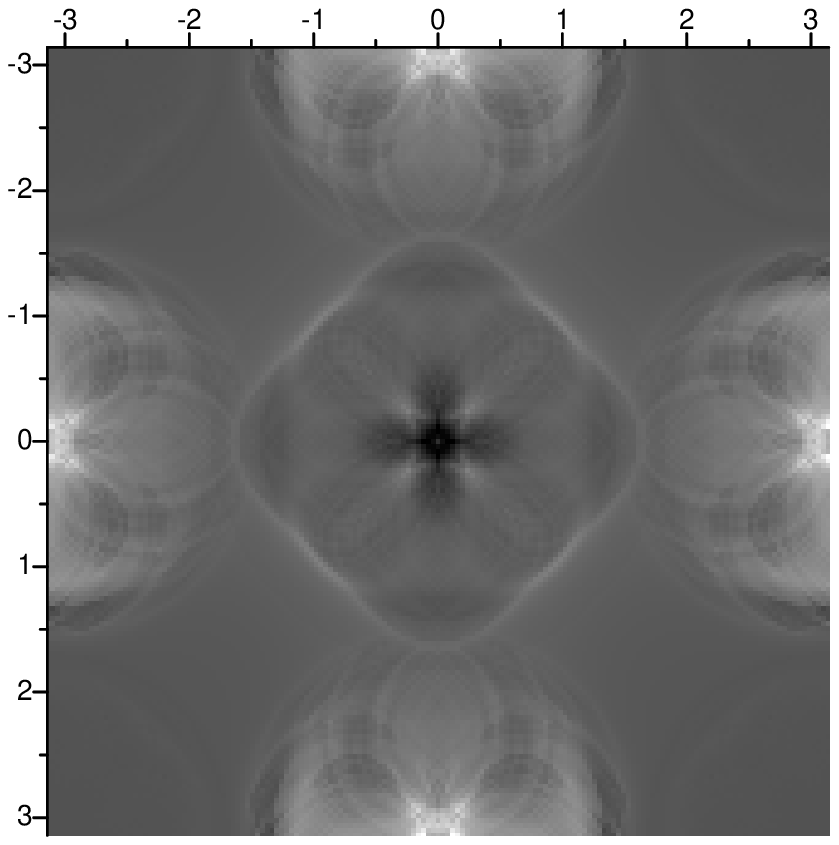}}
\subfigure[$\omega=-0.07$]{
\includegraphics*[bb=5 5 260 260,width=0.2\textwidth]{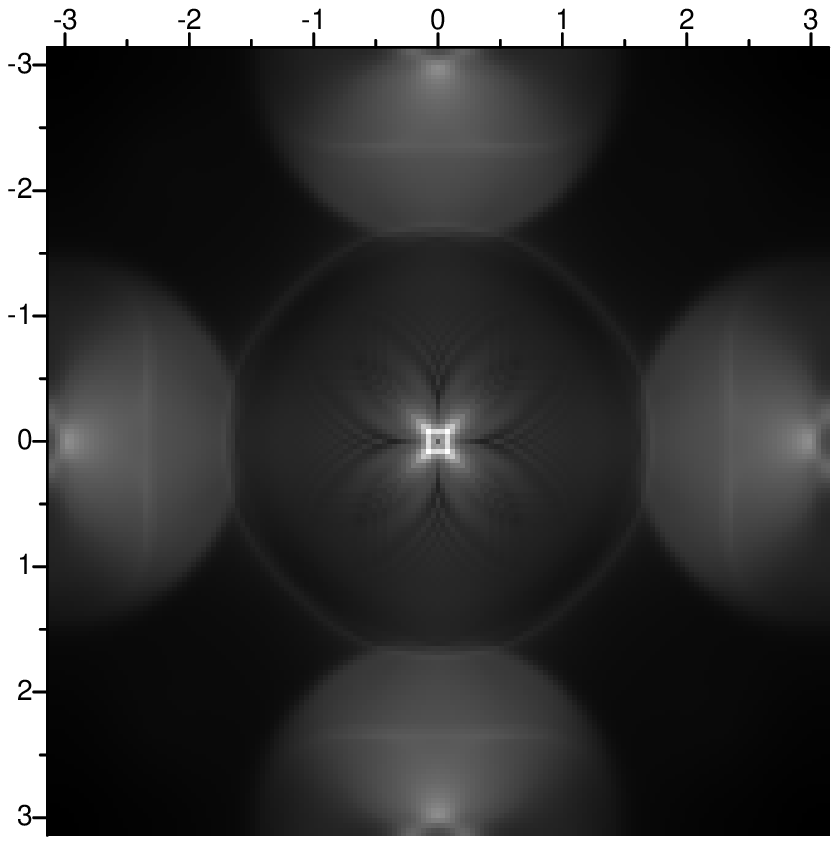}}
\subfigure[$\omega=0.07$]{
\includegraphics*[bb=5 5 260 260,width=0.2\textwidth]{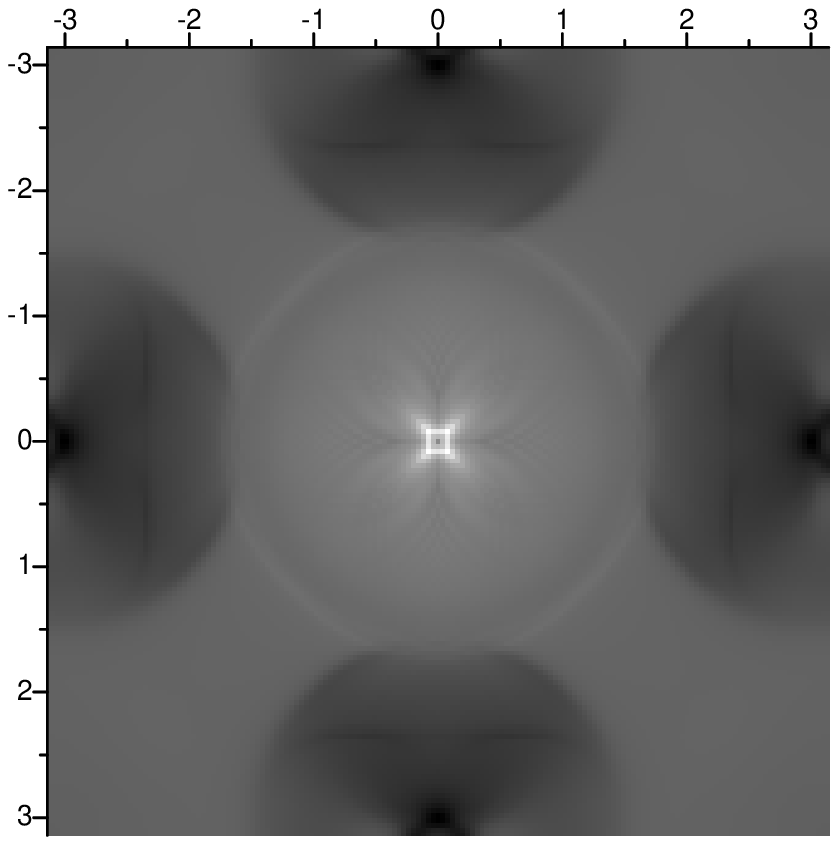}}
\subfigure[$\omega=0.08$]{
\includegraphics*[bb=5 5 260 260,width=0.2\textwidth]{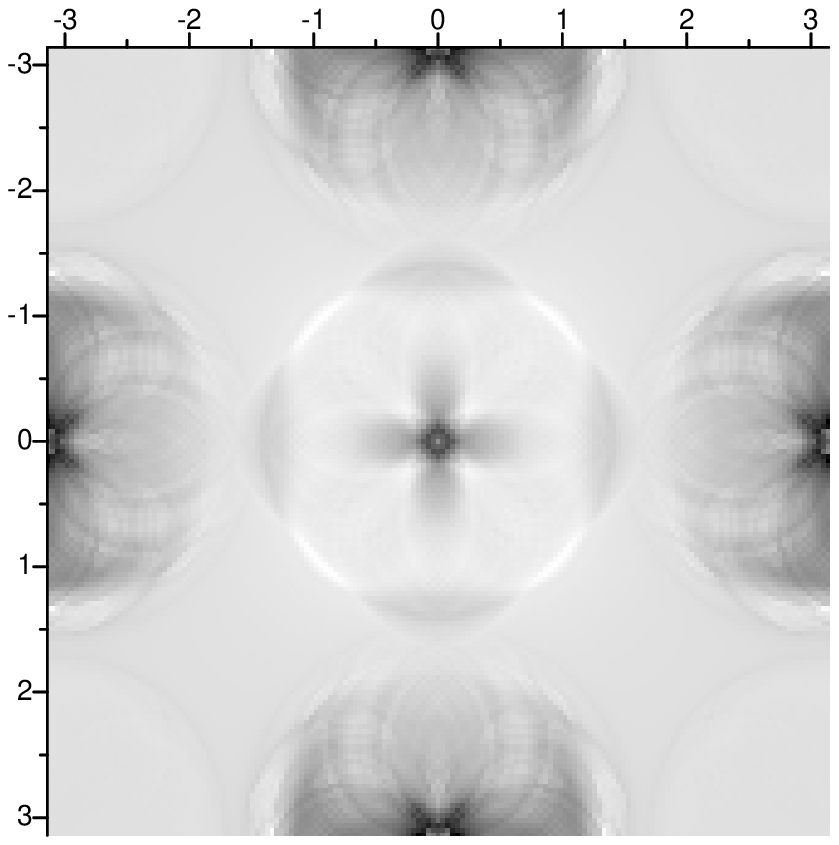}}
\subfigure[$\omega=0.09$]{
\includegraphics*[bb=5 5 260 260,width=0.2\textwidth]{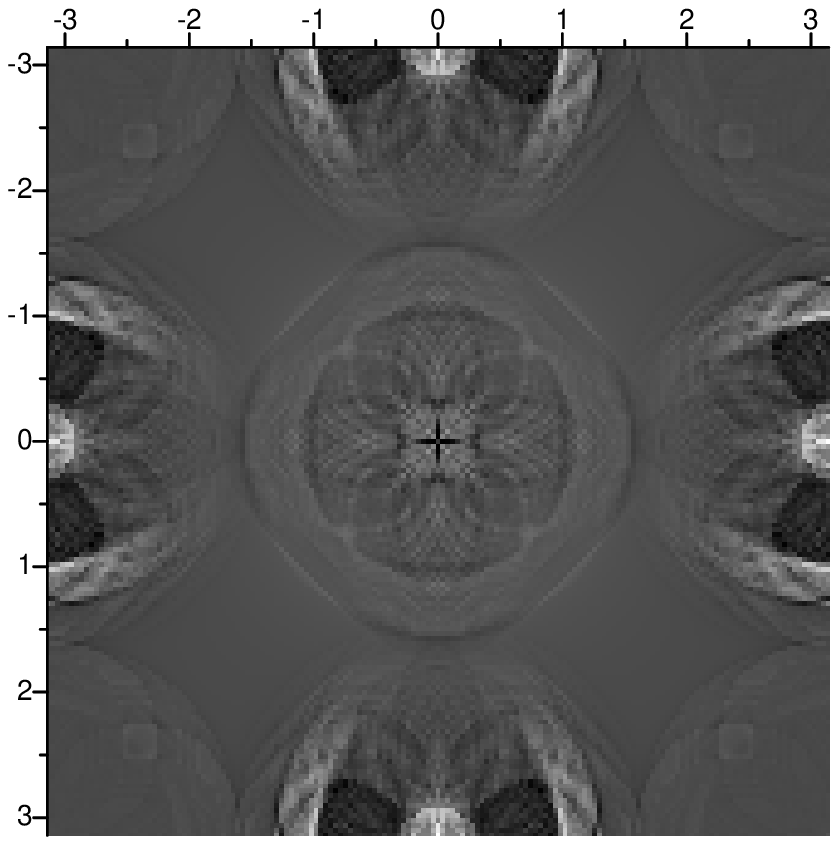}}
\subfigure[$\omega=0.2$]{
\includegraphics*[bb=5 5 260 260,width=0.2\textwidth]{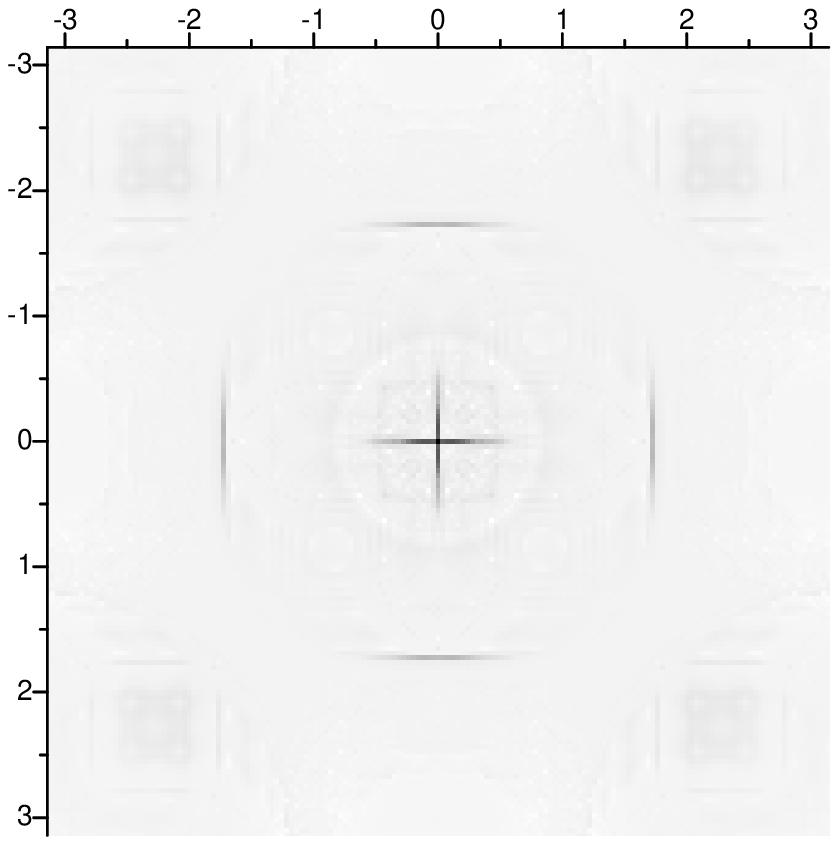}}
\caption{$\delta\rho(\mathbf{q})$ for five-orbital model,
non-magnetic impurity with intra-orbital scattering, $V_0=0.4$. A
$160\times160$ lattice in $\mathbf{k}$-space is used in numerical
integration of equation (\ref{eq6}) and the energy broadening width
$\delta=0.001$.} \label{FNonMFive}
\end{figure}

\begin{figure} [htbp]
\centering \subfigure[$\omega=-0.2$]{
\includegraphics*[bb=5 5 260 260,width=0.2\textwidth]{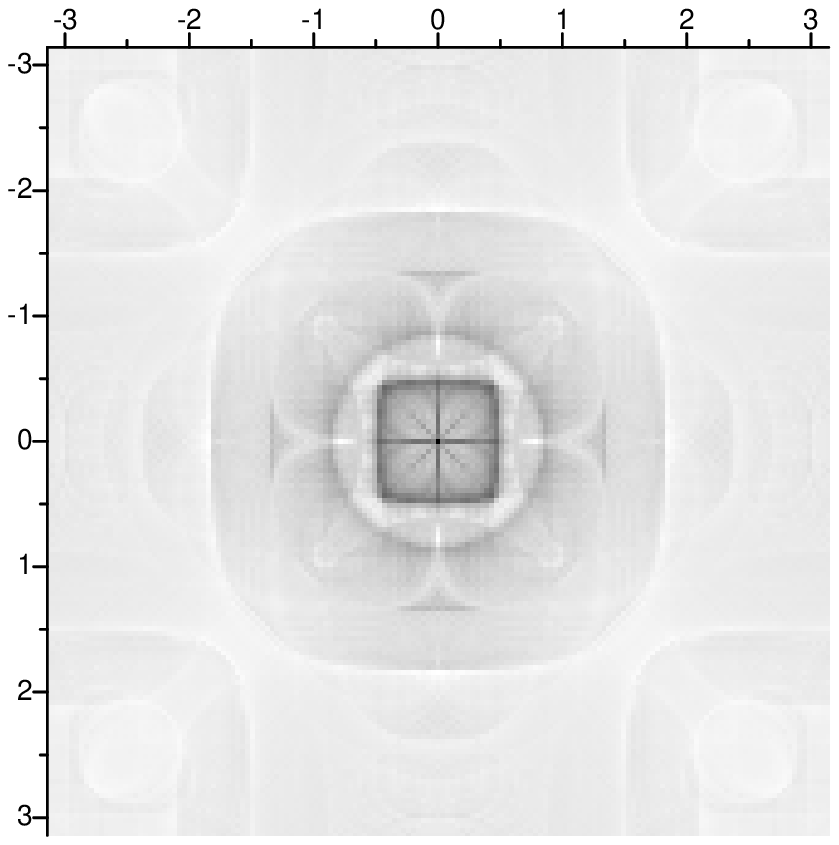}}
\subfigure[$\omega=-0.09$]{
\includegraphics*[bb=5 5 260 260,width=0.2\textwidth]{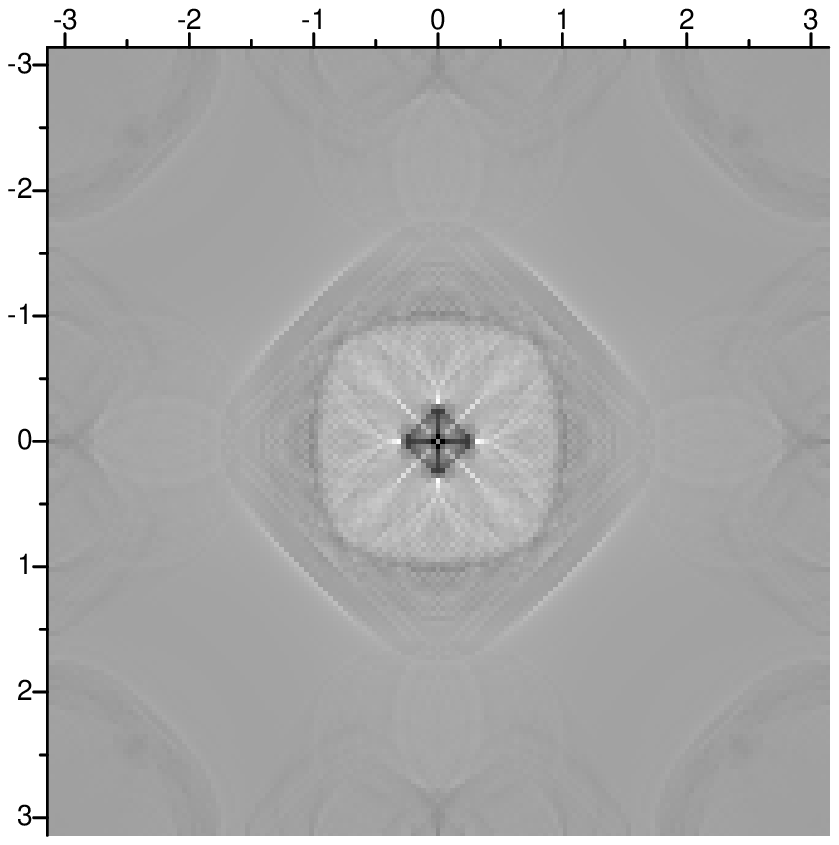}}
\subfigure[$\omega=-0.08$]{
\includegraphics*[bb=5 5 260 260,width=0.2\textwidth]{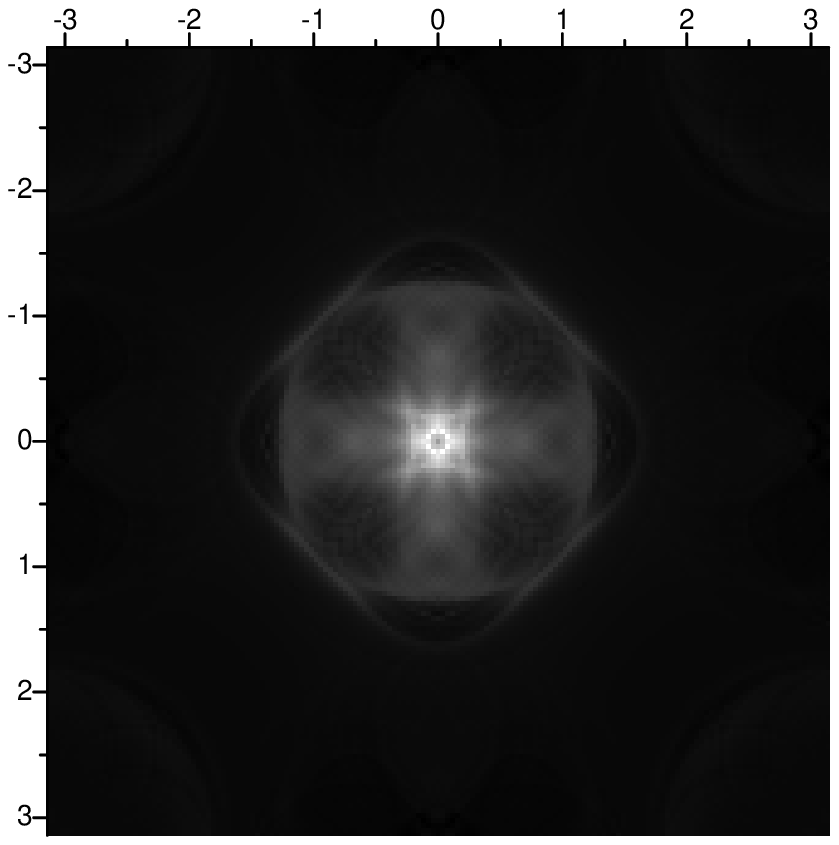}}
\subfigure[$\omega=-0.07$]{
\includegraphics*[bb=5 5 260 260,width=0.2\textwidth]{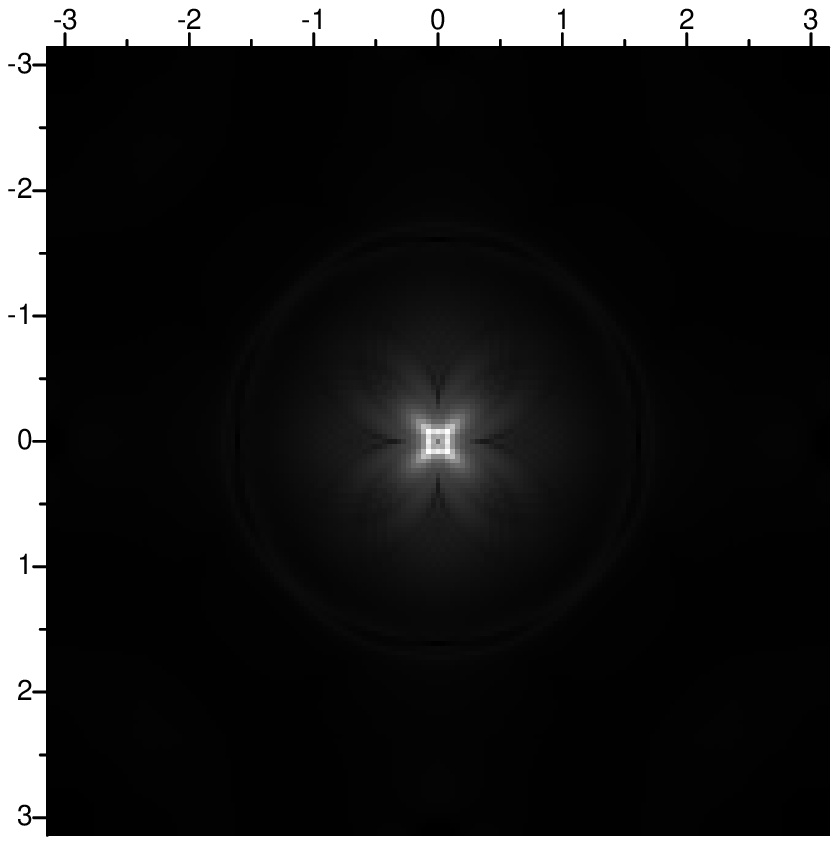}}
\subfigure[$\omega=0.07$]{
\includegraphics*[bb=5 5 260 260,width=0.2\textwidth]{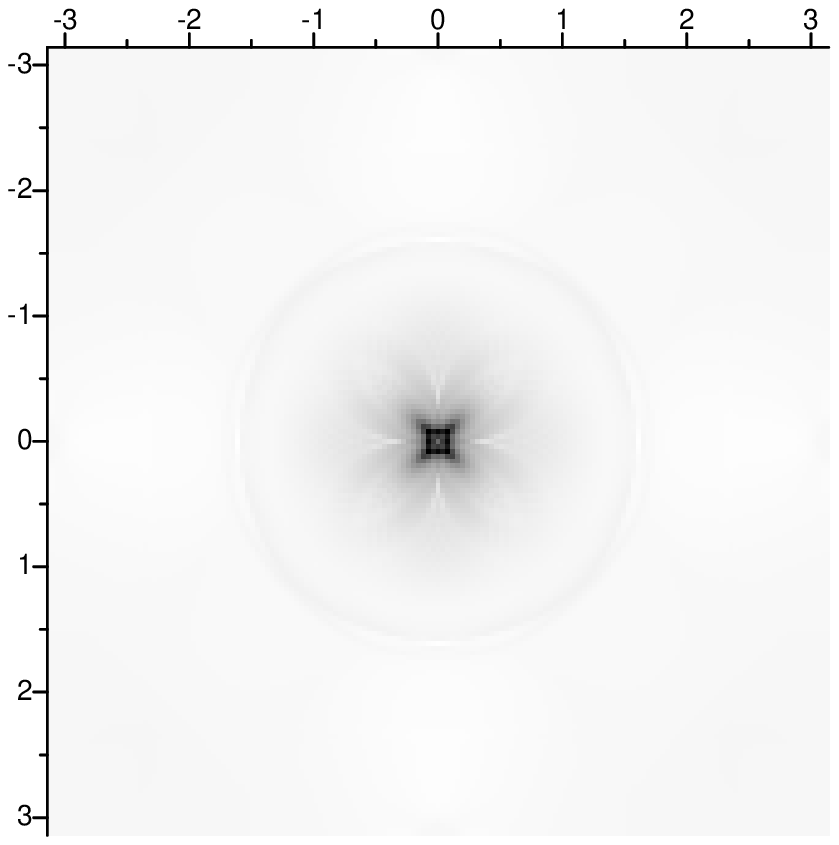}}
\subfigure[$\omega=0.08$]{
\includegraphics*[bb=5 5 260 260,width=0.2\textwidth]{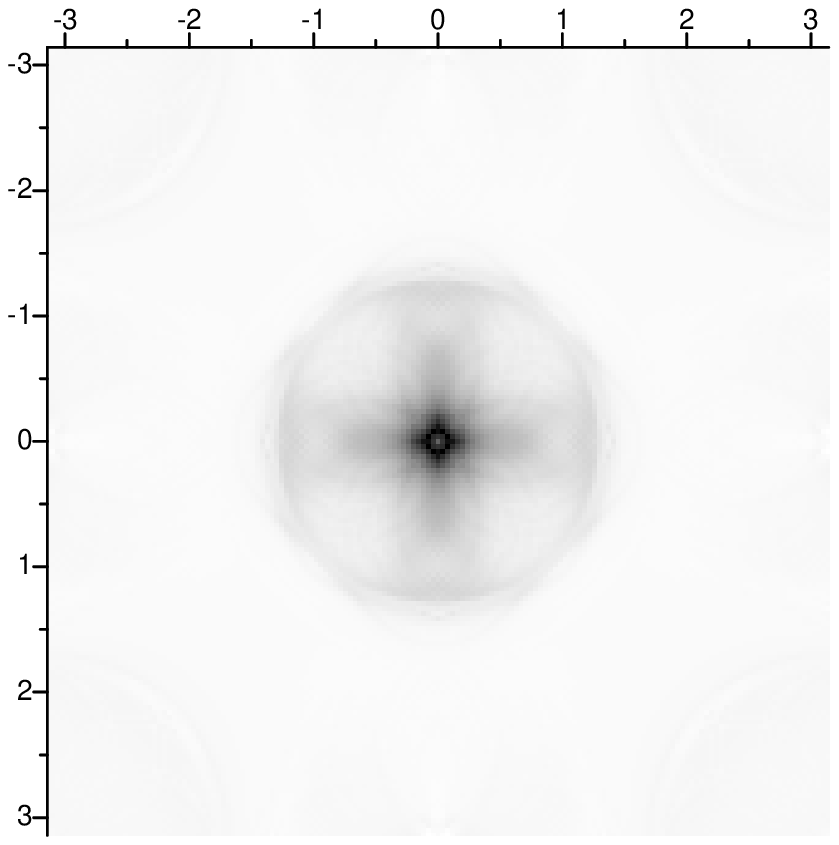}}
\subfigure[$\omega=0.09$]{
\includegraphics*[bb=5 5 260 260,width=0.2\textwidth]{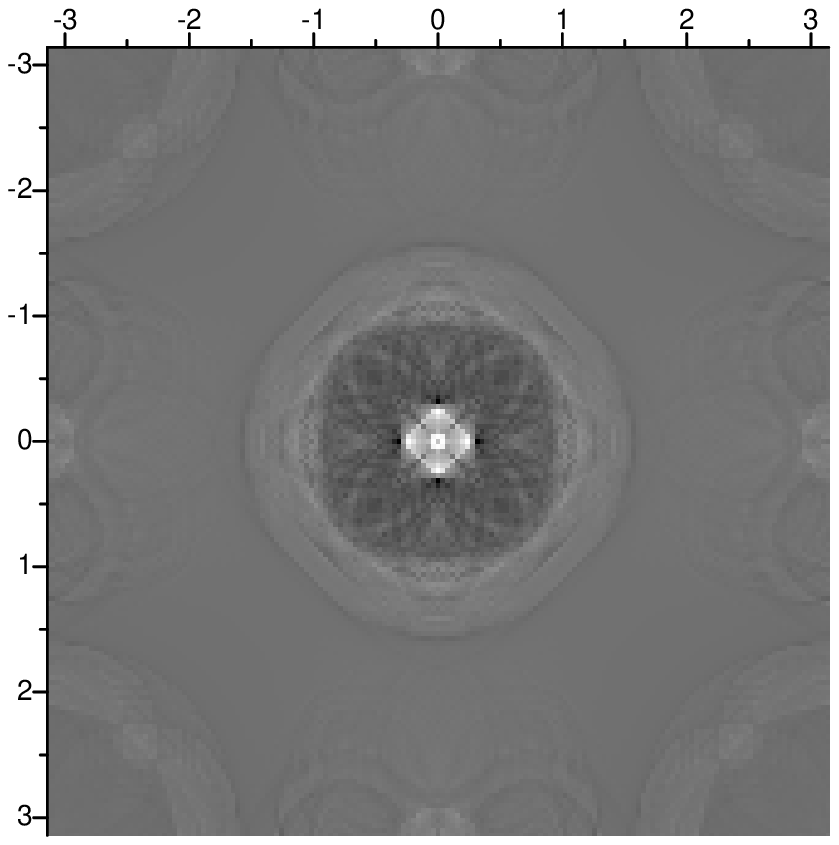}}
\subfigure[$\omega=0.2$]{
\includegraphics*[bb=5 5 260 260,width=0.2\textwidth]{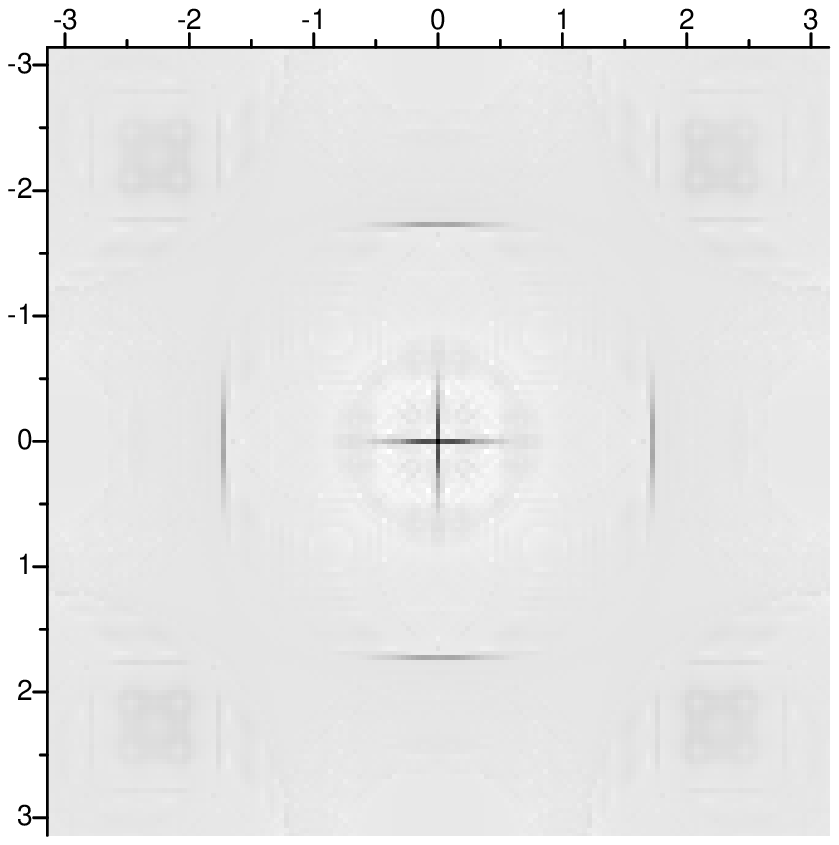}}\caption{Same as Fig. \ref{FNonMFive} but for magnetic impurity, $V_0=0.4$.}
\label{FMFive}
\end{figure}

\begin{figure} [htbp]
\begin{center}
\includegraphics*[width=0.4\textwidth]{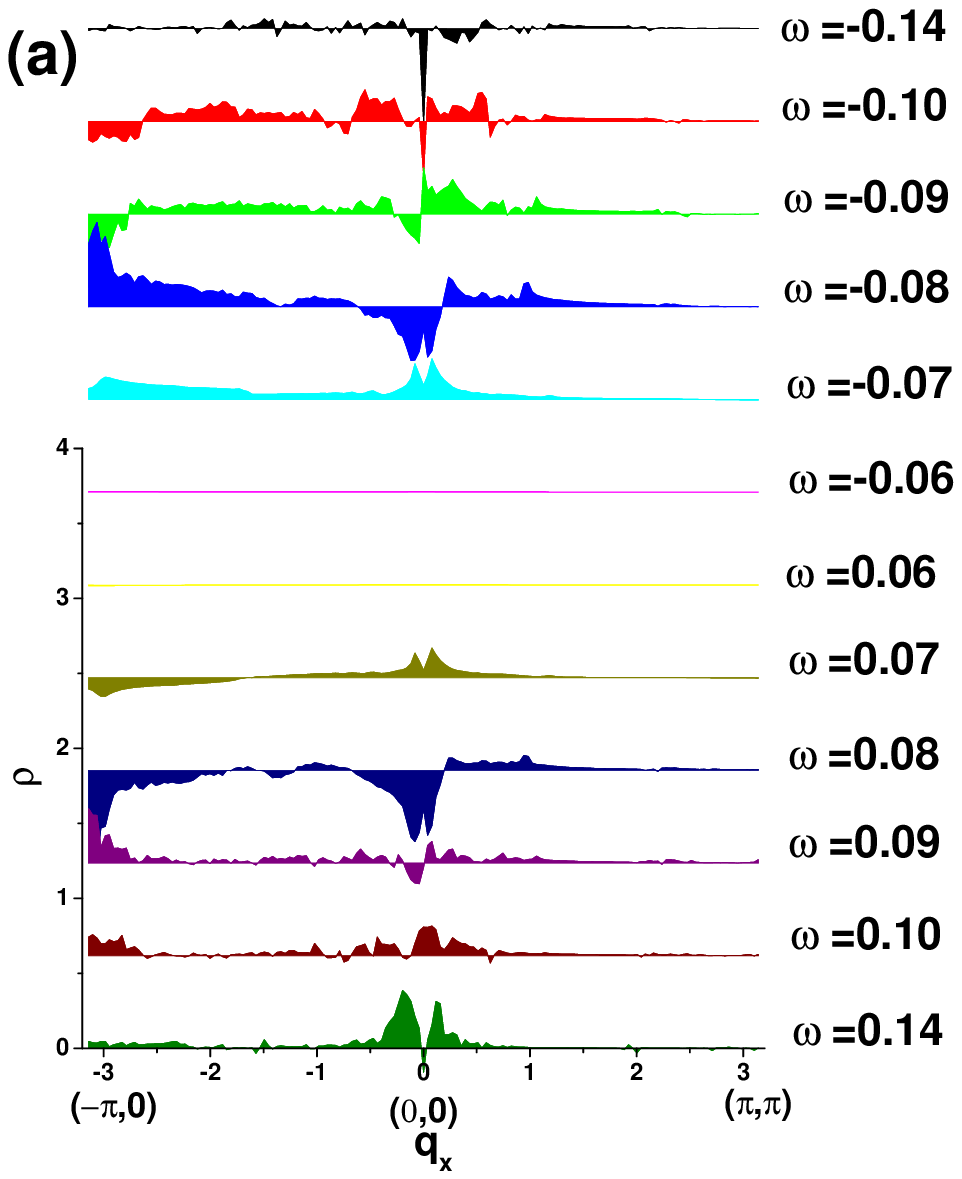}
\includegraphics*[width=0.4\textwidth]{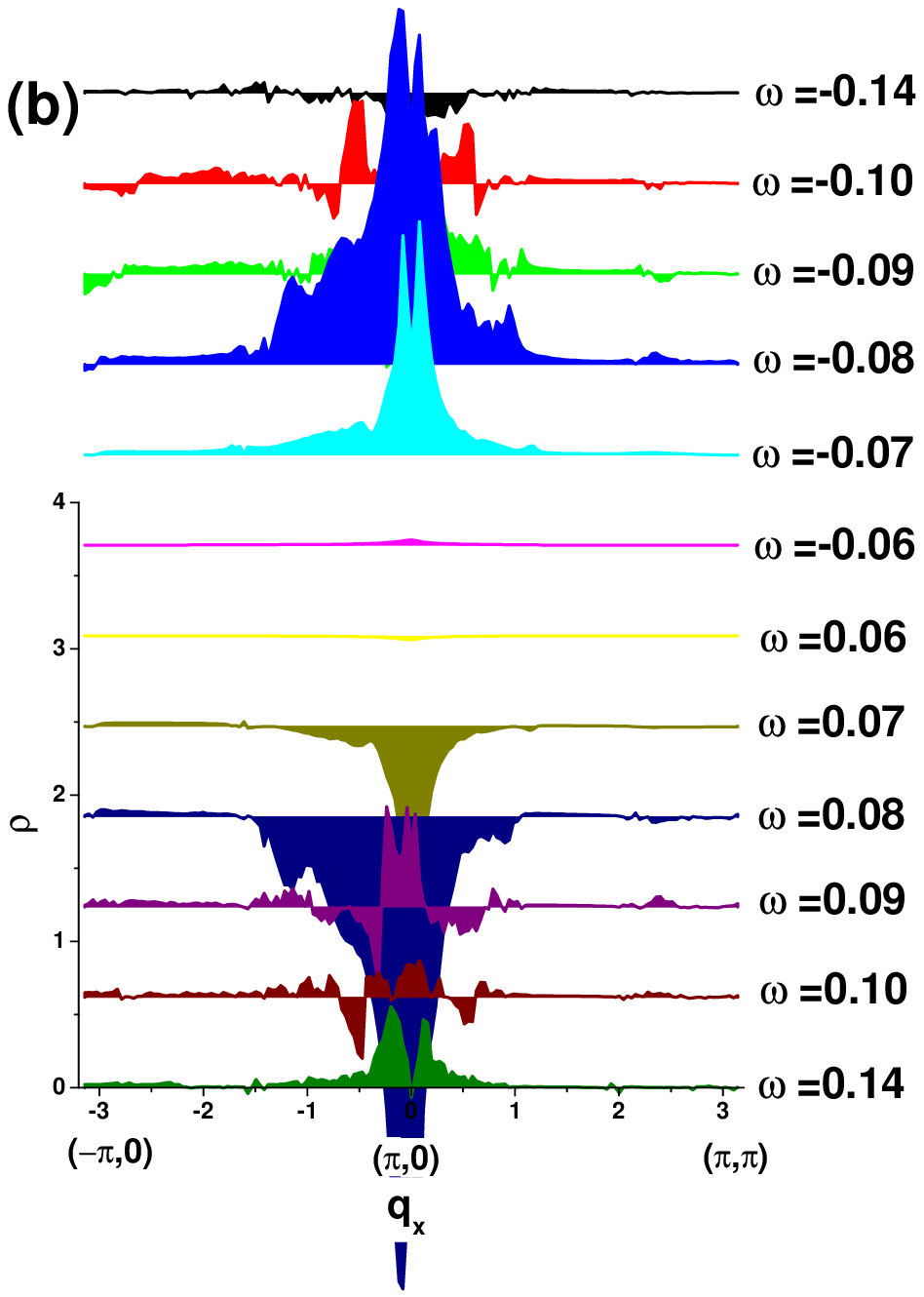}
\end{center}
\caption{(color online) Profiles of $\delta\rho(\mathbf{q},\omega)$
along $\mathrm{M}\rightarrow\Gamma\rightarrow\Gamma'$ for (a)
non-magnetic impurity and (b) magnetic impurity in five-orbital
model.} \label{Fiveab}
\end{figure}

\begin{figure} [htbp]
\begin{center}
\includegraphics*[width=0.4\textwidth]{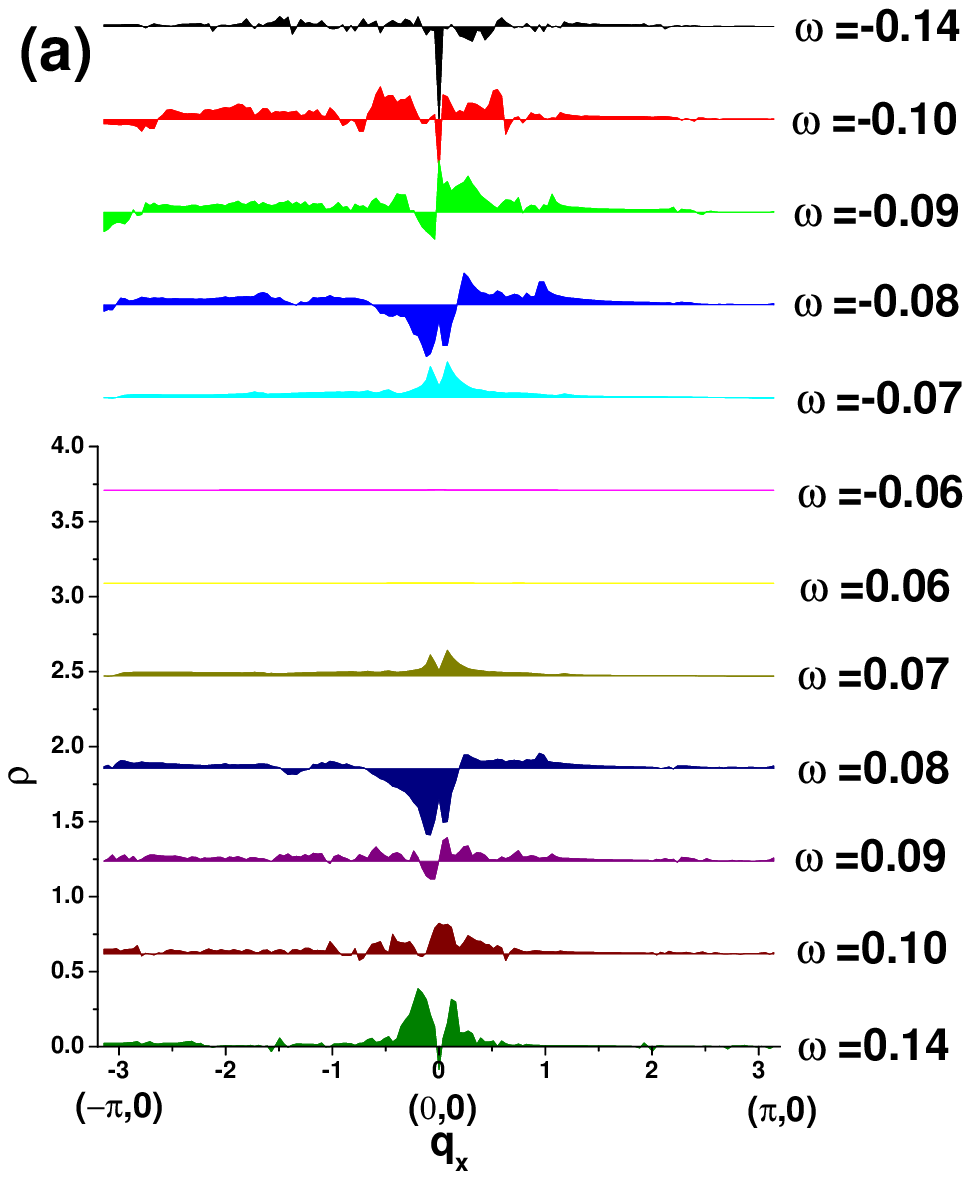}
\includegraphics*[width=0.4\textwidth]{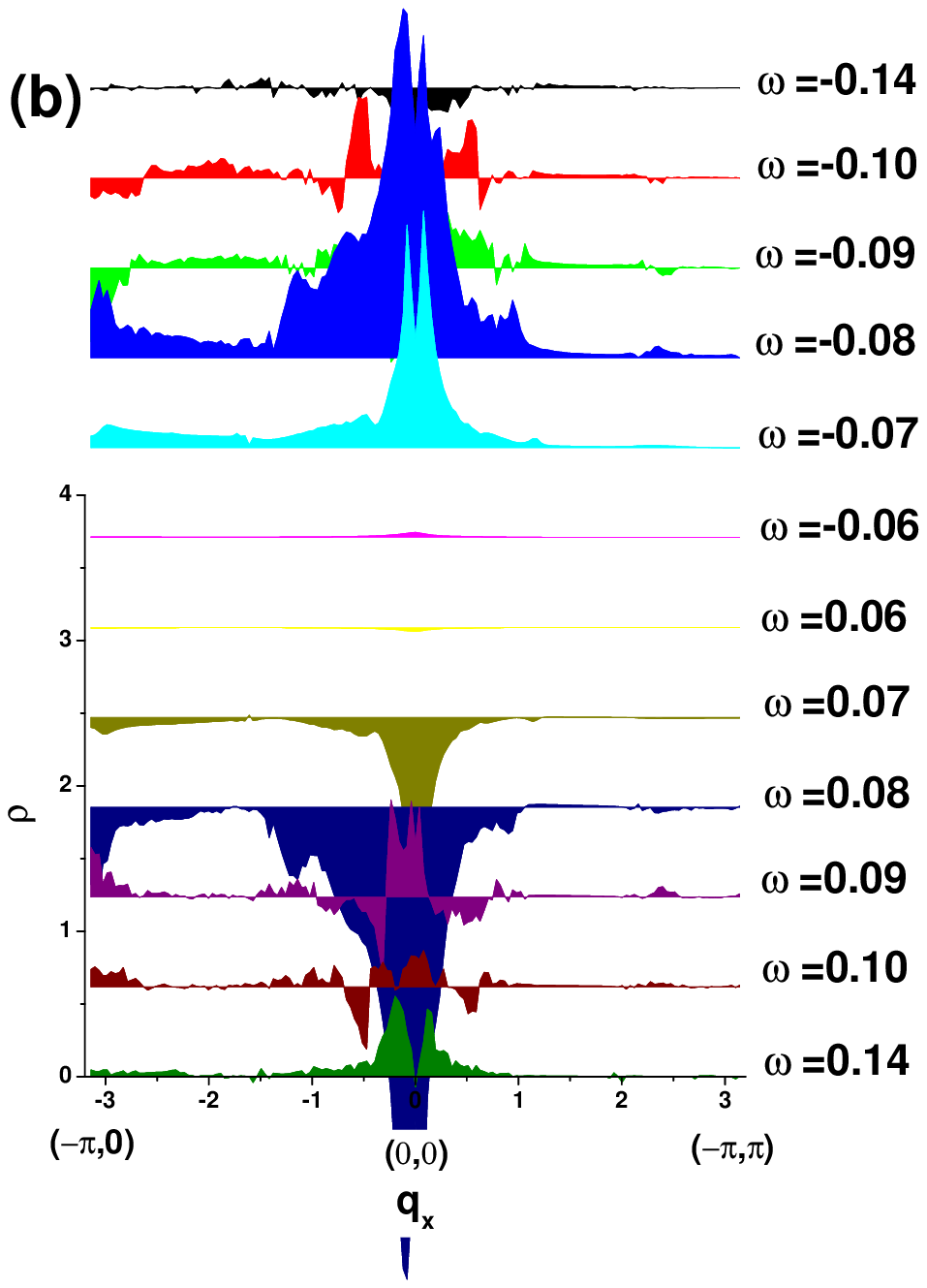}
\end{center}
\caption{(color online) Same as Fig. \ref{Fiveab} but for the case
without sign change of $\Delta$.} \label{FiveNSab}
\end{figure}

In Fig. \ref{FiveDOS}, we show the bulk DOS for the clean system.
The coherence peaks appear at $\pm0.082$ and the system is fully
gapped within $\sim(-0.05,0.05)$. In Fig. \ref{FContourFive}, we
plot the spectral function $\mathcal{A}(\mathbf{k},\omega)$ in the
unfolded Brillioun zone, where the energy contours and their weight
can be clearly seen. In Fig. \ref{FNonMFive} and Fig. \ref{FMFive},
we plot the interference pattern $\delta\rho(\mathbf{q})$ for
non-magnetic and magnetic impurities, respectively. Their profiles
in the direction $\mathrm{M}\rightarrow\Gamma\rightarrow\Gamma'$ are
plotted in Fig. \ref{Fiveab}. Finally, for comparison, we also plot
the results for order parameter without sign change in Fig.
\ref{FiveNSab}.

It is clear that the QPI in the five orbital model is quite
different from that in the two orbital model.
This major difference comes from the distribution of density of
states at the Fermi surfaces. For example, compared with the QPI in the
two-orbital model, scattering around $\Gamma$ now has its origin in the
intra-pocket scattering within the hole pockets - in the two orbital model it originates from
intra-electron pocket scattering.  The density of states is higher in the
electron than in the hole pockets in the two band model.
The opposite is true in the five orbital model. Another clear difference
is that due to the existence of additional orbitals, there exist
square shaped profiles in Fig. \ref{FNonMFive} (b) and (c)  which
correspond to the scattering process labeled by arrow 1 in Fig.
\ref{FContourFive} (b); and the circle shaped profile in Fig.
\ref{FNonMFive} (d) corresponds to arrow 2. These features are
absent in the two orbital model.

However, there are also common features in both models. The broad
and large peaks at $\mathbf{q}=(0,0)$ for magnetic impurity appear
in both models. More importantly, the $(\pm\pi,0)/(0,\pm\pi)$
sensitiveness on magnetic or non-magnetic impurity, and sign change
remains the same. For example, when $\Delta$ changes sign, the peak
around $\mathbf{q}=(\pm\pi,0)/(0,\pm\pi)$ for non-magnetic impurity
(Fig. \ref{Fiveab} (a)) disappears in the case of a magnetic
impurity (Fig. \ref{Fiveab} (b)). On the contrary, when $\Delta$
does NOT change sign (Fig. \ref{FiveNSab}), the peak at
$(\pm\pi,0)/(0,\pm\pi)$ is related to magnetic impurity. The
sensitiveness of the interference pattern around
$(\pm\pi,0)/(0,\pm\pi)$  corresponds to the inter-pocket scattering
labeled by arrows 3 and 4 in Fig. \ref{FContourFive} (b) and has
been explained explicitly in the two-orbital model.
Eqs.~(\ref{eq14}) and (\ref{eq15}) and the arguments following them
do not depend on the number of bands and therefore these features
are rather universal. Moreover, the $(\pm\pi,0)/(0,\pm\pi)$
sensitiveness on the order parameter sign change (Fig.
\ref{FiveNSab}) is quite similar with the two-orbital model.

\section{Conclusion} In summary, we have investigated in
detail the structures of the QPI in iron-based superconductors within
the current available two orbital and five orbital models. The
results obtained here suggest that the QPI can be used to determine
the band structure and orbital degrees of freedom in these materials
and can also provide  evidence of the SC pairing symmetries. In this
calculation, we have ignored possible three dimensional effects, more relevant for the 122 materials \cite{Zhao2008c, Yuan2009}. The physics
associated with the third dimension and possible competing orders or
coexistence states will be addressed in the
future.

\paragraph*{Acknowledge:} JPH and BAB thanks Dunghai Lee  for useful
discussions.  JPH, YYZ, WFT, XTZ,  KJS and CF were supported by the
NSF under grant No. PHY-0603759.

\end{document}